\documentclass{aa}  
\usepackage{placeins}
\usepackage{multirow}
\usepackage{verbatim}
\usepackage{amsmath}
\usepackage{adjustbox}
\usepackage{longtable}
\usepackage{graphicx}
\usepackage{gensymb}
\usepackage{subcaption}
\usepackage{lscape}
\usepackage{url}
\usepackage{longtable} 
\usepackage{tabularx}  
\newcommand{\AP}[1]{\textcolor{green}}
\usepackage[hidelinks,colorlinks=true,linkcolor=blue,citecolor=blue]{hyperref}
\usepackage{txfonts}
\def\code#1{\texttt{#1}}
\begin{document} 

   \title{TEGLIE: Transformer encoders as strong gravitational lens finders in KiDS}
     \subtitle{From simulations to surveys}

   \author{M.~Grespan\inst{\ref{ncbj}}
          \and
          H.~Thuruthipilly\inst{\ref{ncbj}}
          \and A.~Pollo\inst{\ref{ncbj}, \ref{ju}}
           \and M.~Lochner\inst{\ref{uwc}, \ref{radio}}  
           \and M.~Biesiada\inst{\ref{ncbj}}
           \and V.~Etsebeth \inst{\ref{uwc}}
          }

   \institute{National Centre for Nuclear Research, 
            ul. Pasteura 7, 02-093 Warsaw, Poland \\
              \email{margherita.grespan@ncbj.gov.pl} \label{ncbj}
              \and
              Astronomical Observatory of the Jagiellonian University, Orla 171, 30-001 Cracow, Poland\label{ju}
              \and
                Department of Physics and Astronomy, University of the Western Cape, Bellville, Cape Town, 7535, South Africa \label{uwc}
                \and
                South African Radio Astronomy Observatory, 2 Fir Street, Black River Park, Observatory, 7925, South Africa \label{radio}
              }
   \date{}

 
  \abstract
   {With the current and upcoming generation of surveys, such as the Legacy Survey of Space and Time (LSST) on the Vera C. Rubin Observatory and the Euclid mission,  tens of billions of galaxies will be observed, with a significant portion ($\sim$10$^5$) exhibiting lensing features. 
    To effectively detect these rare objects amidst the vast number of galaxies, automated techniques like machine learning are indispensable.}
   {We apply a state-of-the-art transformer algorithm to 221 deg$^2$ of the Kilo Degree Survey (KiDS) to search for new strong gravitational lenses (SGL). }
   {We test four transformer encoders trained on simulated data from the Strong Lens Finding Challenge on KiDS survey data. The best performing model is fine-tuned on real images of SGL candidates identified in previous searches. To expand the dataset for fine-tuning, data augmentation techniques are employed, including rotation, flipping, transposition, and white noise injection. The network fine-tuned with rotated, flipped, and transposed images exhibited the best performance and is used to hunt for SGL in the overlapping region of the Galaxy And Mass Assembly (GAMA) and KiDS surveys on galaxies up to $z$=0.8.
   Candidate SGLs are matched with those from other surveys and examined using GAMA data to identify blended spectra resulting from the signal from multiple objects in a fiber.}
   {We observe that fine-tuning the transformer encoder to the KiDS data reduces the number of false positives by 70\%.
   Additionally, applying the fine-tuned model to a sample of $\sim$  5\,000\,000 galaxies results in a list of $\sim$ 51\,000 SGL candidates. Upon visual inspection, this list is narrowed down to 231 candidates.
  Combined with the SGL candidates identified in the model testing, our final sample includes 264 candidates, with 71 high-confidence SGLs of which 44 are new discoveries. }
   {We propose fine-tuning via real augmented images as a viable approach to mitigating false positives when transitioning from simulated lenses to real surveys. While our model shows improvement, it still does not achieve the same accuracy as previously proposed models trained directly on galaxy images from the KiDS survey with added simulated lensing arcs. This suggests that a larger fine-tuning set might be necessary for competitive performance.
   Additionally, we provide a list of 121 false positives that exhibit features similar to lensed objects, which can benefit the training of future machine learning models in this field.}

   \maketitle
%

\section{Introduction}

Strong gravitational lensing (SGL) is a general relativity (GR) predicted phenomenon in which the potential of a massive foreground object, curving space-time, deflects the trajectory of the signal coming from a background source into separate multiple images.  When the lens and the foreground object align along the line of sight, the common observation is either two or four images, forming the iconic Einstein cross around the deflector. Alternatively, arc-like structures may manifest, creating nearly a complete ring around the lens, commonly referred to as the Einstein ring. 

SGL is rarely observed as it depends on two objects aligning along the line of sight. This is reflected in the limited number of candidates that have been discovered to date. A few thousand have been identified, but only a few hundred have been confirmed \citep[e.g.][]{Bolton_2008, Brownstein_2012,Tran_2022}.   

Gravitational lensing is a valuable tool with a number of applications. For example, it can be used to validate GR \citep{Cao2017,Wei2022}, to constrain dark matter models \citep[e.g.][]{Dye_2005, Hezaveh_2016}, to study the distribution of dark energy in the universe \citep[e. e.g.][]{Collett_2014, Cao_2015}, and study the formation and evolution of galaxies \citep[e.g.][]{Barnabe_2012,Nightingale_2019}.
 
Wide-area surveys like the Legacy Survey of Space and Time \citep[LSST;][]{lsstsciencecollaboration2009lsst} and the Euclid mission \citep{laureijs2011euclid} will collect an unprecedented volume of data, mapping about half of the sky ($\sim$ 20,000 deg$^2$). LSST is expected to observe $\sim$20 billion galaxies, including $\sim$10$^5$ galaxy-galaxy lenses and $\sim$10$^4$ strongly lensed quasars \citep{verma2019strong}.  Euclid is expected to detect more than $10^5$ SGL systems \citep{Collett_2015}.
The sheer volume of galaxies observed in these surveys, coupled with the relatively low occurrence of SGLs, render manual identification of SGLs prohibitively time-consuming and inefficient, necessitating the development of automated detection algorithms. Nonetheless, every SGL detected by an automated technique requires visual inspection by an expert to confirm its validity as a candidate. For conclusive confirmation, spectroscopic follow-up observations are required. 
Although crowd-sourcing initiatives such as Space Warps have proved useful in identifying SGL candidates in the past \citep{More_2015,Marshall_2016, Sonnenfeld_2020_sugohi, Garvin_2022}, the ever-increasing volume of observational data renders the approach of presenting all sources to volunteers unsustainable.  
Some successful automated methods for the detection of SGL-like features have been proposed; such as \code{ARCFINDER} \citep{alard2006automated} or \code{RingFinder}. More recently these algorithms, although shown to be effective in SGL detection, have been outperformed by more sophisticated machine learning (ML) methods in recent SGL detection challenges \citep{Metcalf_2019}. 
ML algorithms have thus appeared as a promising tool to automatically detect SGL candidates. Their application has already led to the identification of thousands of SGL candidates in surveys such as the Kilo Degree Survey \citep[KiDS;][]{Petrillo_2017, Petrillo_2018, links, Li_2020,He_2020, Li_2021}, Dark Energy Camera Legacy Survey \citep[DECaLS;][]{Huang_2020, Huang_2021,  Stein_2022, storfer_2023}, VLT Survey Telescope 
 \citep[VST][]{Gentile_2021}, Panoramic Survey Telescope and Rapid Response System \citep[Pan-STARRS;][]{Canameras_2021}, Dark Energy Survey \citep[DES;][]{Diehl_2017,Jacobs_2019,rojas2021strong,O_Donnell_2022, Zaborowski_2023}, Low-Frequency Array \citep[LOFAR;][]{Rezaei_2022_ML} and Hyper Suprime-Cam \citep[HSC;][]{ Canameras_2021, Shu_2022}.

In much of the aforementioned research, convolutional neural networks (CNNs) have emerged as the preferred architecture for SGL identification.
Notably, a CNN model secured the top performance in the SGL Finding Challenge \citep{Metcalf_2019}.
CNNs are supervised ML gradient-based algorithms first introduced by \cite{Lecun} for handwritten digit recognition. Since their debut, CNNs have revolutionized the field of image recognition \citep{Alexnet_2012, simonyan_2015,he_2015_resnet,  Szegedy_2015,huang_2018, tan_2020_efficientnet}. CNNs have been the state-of-the-art for object recognition for years, until the recent breakthrough in natural language processing (NLP) with the introduction of a new self-attention-based architecture, known as transformers \citep{vaswani2017attention}.
This architecture's significant impact on NLP is evidenced by models like BERT,  developed by researchers at Google AI \citep{devlin_2019_bert}, which uses a bidirectional encoding approach, considering both the left and right context of a word during training, unlike its predecessors. 
At the time of its release, BERT achieved state-of-the-art performance on a wide range of NLP tasks, including text classification, question answering and text summarization \citep[e.g.][]{wang2019multipassage}. At the core of transformers lies the attention mechanism, which dynamically assigns varying weights to different parts of the input data, enabling the model to prioritize relevant information and model long-range dependencies.

The attention mechanism has been successfully extended to image analysis \citep{carion2020endtoend,dosovitskiy2020image,Jiahui, Mitchell} proving transformers as more robust learners than CNNs \citep{Paul_Chen_2022}.  Beyond their success in image classification and object detection, transformers have also proven adept at various image processing tasks, including image segmentation, image super-resolution, and image restoration \citep{Khan_2022,aslahishahri2023darts,li2023transformerbased}.
The astronomical community has rapidly recognized the potential of transformers, leading to a variety of applications \citep{Merz_2023_trans,Donoso_2023_trans,allam2023paying, hwang2023universe,haressh_lsbs} including strong lens detection \citep{Hareesh, Huang_transforemr, Hareesh2, Jia_Transformers}.

Supervised ML models require a substantial amount of labeled data, typically in the range of tens of thousands of samples, to attain optimal performance. However, for rare phenomena such as SGLs, where only a few thousand lens candidates have been identified and observations are restricted to specific instruments, large labeled datasets are not readily available.
To address this challenge, artificial datasets with lens systems as realistic as possible are generated employing various techniques. 
Mimicking the complex morphology of lens systems is a challenging task, and creating realistic examples of non-lenses is equally difficult. Irregular galaxies, mergers, and ring galaxies can all exhibit features that resemble those of SGL.
For these reasons, transitioning from a simulated dataset to real-world SGL searches invariably results in an increase in false positives compared to the numbers encountered during training with artificial data.
However, when moving from simulated datasets to real-world data, where non-lenses vastly outnumber lenses, final visual inspection remains indispensable for validation.

This study examines the potential of transformers to detect SGLs, the role of sample selection by galaxy type, and the impact of fine-tuning in reducing false positives. We also investigate different data augmentation techniques to increase the fine-tuning dataset size.
Examining false positives, the primary reason models achieve lower accuracy, is crucial for building robust training sets. This analysis helps us understand model shortcomings and guides the development of more accurate training data, especially considering the vast number of galaxies that next-generation surveys will detect.  

The transformers we use are described in \cite{Hareesh} and were trained on the simulated Bologna Strong Lens Finding Challenge dataset \citep{Metcalf_2019}. 
The best-performing fine-tuned model is supplied with images from where KiDS and Galaxy And Mass Assembly (GAMA) footprints overlap.

The paper is structured as follows:  Sec. \ref{sec:prev_ml} examines the previous SGL searches in KiDS. Sec. \ref{sec:data} introduces the data and outlines the preprocessing criteria used to select the sample. The methodology adopted in this work is detailed in Sect. \ref{sec:methodology}. Section \ref{sec:test} presents a comprehensive evaluation of various transformers to assess their performance in detecting strong lenses on KiDS survey images. 
Section \ref{sec:220_Deg} characterizes the newly discovered lenses, introducing the acronym TEGLIE (Transformer Encoders as strong Gravitational Lens finders In the kilo-degreE survey) for this sample.
Section \ref{sec:candidates_propriety} delves into the properties of the TEGLIE lenses. Section \ref{sec:discussion} provides an analysis of the results, while Sect. \ref{sec:conclusions} summarises the main findings and highlights the potential of ML for SGL detection.

\section{Previous Strong Lens Searches in KiDS with CNNs}
\label{sec:prev_ml}
The KiDS survey, with its extensive size and depth, provides an ideal testing ground for the application of ML techniques to the detection of SGLs, as demonstrated by previous studies \citet{Petrillo_2017, Petrillo_2018, links, Li_2020, Li_2021}. All of these searches have been conducted using CNN architectures. For this reason, we investigate the effectiveness of the current state-of-the-art for image classification on terrestrial images, namely self-attention-based models, in detecting rare astronomical objects such as SGLs within the KiDS survey. 

In order to provide an adequate context for our subsequent analyses, we present a brief overview of the SGL searches that have been conducted as part of the KiDS survey. This comparative perspective will prove useful when evaluating different approaches and sample selection strategies in the following sections.

Initial studies by \citet{Petrillo_2017,Petrillo_2018} used a CNN to identify SGLs within luminous red galaxies (LRGs) extracted from KiDS DR3. 
The network was trained on simulated SGL images created by superimposing simulated lensed images on real LRGs observed in the survey. Subsequent studies by \citet{links, Li_2020} and \citet{Li_2021} further refined the approach by extending the target set to include Bright Galaxies (BGs) in addition to LRGs from DR4 and DR5 (not yet publicly available).
Both LRGs and BGs are massive galaxies with a large lensing cross-section, making them prime candidates for SGL detection \citep{Oguri_2010}.
Since LRGs are a subset of BGs by definition, the target dataset undergoes a preliminary selection process to identify specific BGs. In all the KiDS SGL searches, the galaxies meeting the following criteria were considered as BGs.
\begin{itemize}
    \item \code{SExtractor} $r$-band \code{Flag} $<$ 4, to eliminate objects with incomplete or corrupted photometry, saturated pixels, extraction, or blending issues.
    \item \code{IMA\_FLAGS} = 0,  excluding galaxies in compromised areas.
    \item \code{SExtractor} Kron-like magnitude  \code{MAG\_AUTO}, in the $r$-band, $r_{auto}\leq 21$ mag, maximizing the lensing cross-section \citep{Schneider_1992}.
    \item KiDS star-galaxy separation parameter \code{SG2DPHOT} = 0, selecting galaxy-like objects. 
    This parameter has values of 1 for stars, 2 for unreliable sources, 4 for stars based on star/galaxy separation criteria, and 0 for non-stellar objects. For further details refer to \citet{La_Barbera_2008, deJong2015,Kuijken_2019}.
\end{itemize}
From the BG sample,  LRGs are extracted by imposing two further criteria, (1) $r_{auto}\leq 20$ mag  and (2) color-magnitude selection at low redshift ($z<0.4$) based on \citet{Eisenstein_2001}, with some modifications to include fainter and bluer sources:
\begin{equation}
\label{eq:color_cut}
\begin{split}
  &  |c_{perp}|<0.2 \\
  &  r_{auto} < 14 + c_{par}/0.3
\end{split}
\end{equation}
where
\begin{equation}
\begin{split}
   & c_{par} = 0.7(g-r) +1.2[(r-i) - 0.18] \\
   & c_{perp} = (r-i) - (g-r)/4.0 - 0.18
 \end{split}
\end{equation}

In \cite{Petrillo_2017}, a CNN was used to identify SGLs from a sample of 21\,789 LRG in KiDS DR3, yielding 761 SGL candidates. After visual inspection, 56 of these candidates were deemed reliable.
Similarly, \cite{links} used two CNNs to search for SGLs between the LRG galaxies in KiDS DR4. The CNNs, using $r$-band images and $g$, $r$, and $i$ color-composited images as input, detected 2\,510 and 1\,689 candidates, respectively, from a sample of 88\,327 LRG. 
After visual inspection, these candidates were narrowed down to 1\,983 potential SGLs and 89 strong high-quality (HQ) candidates. This catalog of Lenses in the KiDS survey (LinKS) is publicly available \footnote{\url{https://www.astro.rug.nl/lensesinkids/}}. 
  
\cite{Li_2020, Li_2021} extended the target sample to include both BGs and LRGs from DR4 and DR5, respectively. 
In their study, \cite{Li_2020} employed the same CNN architecture as in \cite{links}, taking as input only $r$-band images, to a sample of 3\,808\,963 BGs and 126\,884 LRGs. This CNN successfully identified 2\,848 SGL candidates among the BGs and 3\,552 candidates among the LRGs. After visual inspection, 133 BG and 153 LRG candidates were classified as high-probability SGLs.

In subsequent work, \cite{Li_2021} employed two different CNN architectures to analyze the 341 deg$^2$ of the KiDS survey final unpublished release (DR5). The first CNN, which takes as input $r$-band images, identified 1\,213 SGL candidates, and the second CNN, which takes as input $g$, $r$, and $i$ color-composited images, identified 1\,299 candidates. Visual inspection deemed the candidates to be 487, with 192 BG and 295 LRG candidates classified as high-probability SGLs. Among these high-probability SGLs, 97 were selected as the most likely SGL candidates.

The full HQ SGL catalog from these studies is available on the KiDS DR4 website \footnote{\url{https://kids.strw.leidenuniv.nl/DR4/hqlenses.php}}.

\section{Data}
\label{sec:data}

\subsection{Kilo-Degree Survey}
\label{sec:kids}

The Kilo-Degree Survey (KiDS, \cite{dejong2013, deJong2015, deJong2017, Kuijken_2019}) is a European Southern Observatory (ESO) public wide-field medium-deep optical four-band imaging survey with the main aim of investigating weak lensing.
It is carried out with an OmegaCAM camera \citep{Kuijken2011} mounted on the VLT Survey Telescope (VST, \cite{capaccioli2011vlt}) at the Paranal Observatory in Chile.
The OmegaCAM has one deg$^2$ of field of view and the pixels have an angular scale of 0.21 arcseconds.
The $r-$band has the best seeing, the point spread function's (PSF) full width-half maximum (FWHM) is 1.0, 0.8, 0.65, and 0.85 arcseconds in the $u,g,r,i$  -bands, respectively \citep{deJong2015, deJong2017}.
The source extraction and the associated photometry have been obtained by the KiDS collaboration using \code{SExtractor} \citep{Bertin_1996} and the multi-band colors are measured by the Gaussian Aperture and PSF (GAaP) code \citep{gaap}.

The latest publicly available release of the KiDS survey is Data Release 4  \citep{Kuijken_2019}, encompassing 10,006 tiles covering approximately 1\,000 deg$^2$ of the sky. With the upcoming final data release (DR5) becoming public, the KiDS dataset will cover approximately 1\,350 deg$^2$ of the sky in the four  $u,g,r,i$ optical bands.

\subsection{KiDS survey sample selection}
\label{sec:sample}
For this study, we utilize a subset of KiDS DR4, comprising 221 tiles ($\sim$ 220 deg$^2$) which all overlap with the GAMA spectroscopic survey footprint. 
The GAMA regions overlapping with the KiDS survey are detailed in Tab.~\ref{tab:ganma_survey}. Within the survey regions G09, G12, G15, and G23, there are 56, 56, 60, and 49 tiles in KiDS DR4, respectively.
Future studies could model the newly discovered lenses by leveraging a combination of GAMA spectroscopic redshifts and other measurements with the KiDS images, adopting an approach similar to that employed in \cite{Knabel2023}.

\begin{table}[h]
\centering
\caption{GAMA survey regions.\protect\footnote{The survey details are taken from the following GAMA website: \url{http://www.gama-survey.org}}}
\begin{tabular}{lll}
\hline\hline\noalign{\smallskip}
Survey region & R.A. J2000       & Dec. J2000    \\
                & (deg) & (deg) \\
\hline\noalign{\smallskip}
G09    & 129.0 to 141.0 & -2 to +3   \\
G12    & 174 to 186     & -3 to +2   \\
G15    & 211.5 to 223.5 & -2 to +3   \\
G23    & 339.0 to 351.0 & -35 to -30 \\ 
\hline
\end{tabular}

\label{tab:ganma_survey}
\end{table}

In line with the approach employed in previous SGL searches, as discussed in Sect. \ref{sec:prev_ml}, we restrict our target to sources with an $r$-band \code{Flag} value less than 4. 
Additionally, to minimize contamination from galaxies in compromised regions such as reflection halos and diffraction spikes, we only consider sources with a \code{ima\_flags} value of zero in all bands. 
In contrast to previous studies that employed a Kron-like magnitude threshold of $r_{auto}<$ 21 mag \citep{Li_2020, Li_2021,Petrillo_2017, Petrillo_2018, links}, we extend our target sample to include all galaxies with photometric redshifts up to $z$=0.8, without putting any limit on $r_{auto}$. This allows for a more comprehensive exploration of the lensing population and potentially identifies fainter SGLs. 

The photometric redshifts for the target sample are obtained from the KiDS DR4 catalog using the Bayesian Photometric Redshift Code (\code{BPZ}) \citep{Benitez_2011}. For redshift selection, the focus is on \code{Z\_B}, an estimate of the most probable redshift from the 9-band images (peak of the posterior probability distribution), \code{Z\_ML}, the 9-band maximum likelihood redshift, and the lower and upper bounds of the 68\% confidence interval of \code{Z\_B}. Objects with estimated redshifts \code{Z\_B} and \code{Z\_ML} within the 0.0 - 0.8 redshift range are included, and additionally \code{Z\_ML} is required to fall within the lower and upper 1$\sigma$ bounds of \code{Z\_B}, ensuring that the redshift estimates are consistent and reliable.

We aim to identify lenses that may have been overlooked in previous searches, including those where the lens and lensed objects are in close proximity.  Therefore we refrain from imposing any constraints on the star-galaxy separation parameter \code{SG2DPHOT} which flags celestial objects as stars based on the source $r$-band morphology.
However, in future studies, it may be crucial to impose constraints on this parameter to optimize computational efficiency.

This sample selection limits the sample to 5\,538\,525 elements out of 21\,950\,980 within the 221 KiDS tiles.

For the image cutouts, we select dimensions of $101 \times 101$ pixels, corresponding to $20 \times 20$ arcsec$^2$ in the KiDS survey. These cutouts are centered on the centroid sky position derived from the $r$-band coadded image. The cutout size corresponds to $90 \times 90$ kpc at $z$=0.3 or $120 \times 120$ kpc at $z$=0.5 \citep{Li_2020}.

\subsection{Synthetic KiDS-like data}
\label{sec:synthetic}
The Bologna Strong Gravitational Lens Finding Challenge\footnote{\url{http://metcalf1.difa.unibo.it/blf-portal/gg_challenge.html}} \citep{Metcalf_2019}, a now-concluded open challenge, tasked participants with developing automated algorithms to identify strong gravitational lenses among other types of sources. 
The mock galaxies were generated within the Millennium simulation \citep{Boylan_2009} and lensed using the GLAMER lensing code \citep{Metcalf_2014_GLAMER, Petkova_2014_glamer}.
The challenge dataset consisted of 85\% purely simulated images and 15\% real images of bright galaxies from the KiDS survey. Lensing features were added to both samples with a rate of 50\%. 
The challenge dataset comprises 100\,000 images while the training set consists of 20\,000 images, each measuring 101 $ \times $ 101 pixels. These images have been created to mimic the noise levels, pixel sizes, sensitivities, and other parameters of KiDS.  The images encompass four bands, designated as  $(u, g, r,$ and $i)$, with the $r$ band serving as the reference band. The image resolution is 0.2 arcseconds, which results in a 10 x 10 arcsec$^2$  image.

\subsection{GAMA spectroscopy and \code{AUTOZ} redshifts}
\label{sec:gama}

The Galaxy And Mass Assembly survey \citep{Driver_2009, Driver_2011, Liske_2015, Driver_2022} is a multi-wavelength spectroscopic survey conducted using the Anglo-Australian Telescope's spectrograph.
GAMA Data Release 4 \citep{Driver_2022} provides redshifts for over 330\,000 targets spanning over $\sim$ 250 deg$^2$.
These redshifts are determined through a fully automated template-fitting pipeline known as \code{AUTOZ} \citep{Baldry_2014}. \code{AUTOZ} is capable of identifying spectral template matches that may contain contributions from two distinct redshifts within a single spectrum.  In addition to the redshift, the algorithm retrieves cross-correlation redshifts against a library of stellar and galaxy templates.  The galaxy templates (numbered 40 to 47 in order of increasing emission-line strength) are broadly categorized as passive galaxies (PG) for templates 40-42 and emission-line galaxies (ELG) for templates 43-47, with lower-numbered templates corresponding to stars \citep{Holwerda_2015}.
Based on the best-fit galaxy type at each redshift component, it is possible to classify the blended galaxies.  These classifications include two passive galaxies (PG+PG), two emission-line templates (ELG+ELG), a passive galaxy at a lower redshift and an emission-line galaxy at a higher redshift (PG+ELG), or the inverse scenario (ELG+PG).

The public online catalog \code{AATSpecAutozAllv27\_DR4}\footnote{\url{https://www.gama-survey.org/dr4/schema/dmu.php?id=91}} provides the outputs of the \code{AUTOZ} algorithm. The catalog includes the best-fitting galaxy template and the four flux-weighted cross-correlation peaks ($\sigma$) of redshift matches with galaxy templates. The highest peak  ($\sigma_1$) corresponds to the best-fit redshift, while subsequent peaks have progressively lower correlation values with $\sigma_4$ having the weakest correlation.

\citet{Holwerda_2015} used the outputs from \code{AUTOZ} to compile a catalog of blended galaxy spectra within GAMA DR2 \citep{Liske_2015}. Blended spectra arise when the light from two distinct galaxies falls within the same GAMA fiber. Consequently, cases where two separate galaxy templates provide high-fidelity matches, yet yield significantly different redshifts, are indicative of potential blended sources. These blended sources can be attributed to either strong gravitational lensing events or serendipitous overlaps of two galaxies along the line of sight.
To quantify the strength of the secondary redshift candidates, \citet{Holwerda_2015} selected  double-z candidates via the ratio R: 
\begin{equation}
\label{eq:R}
    R =  \frac{\sigma_2}{\sqrt{\sigma_3/2 + \sigma_4/2 }}
\end{equation}
where $\sigma_2$ represents the second highest redshift peak value, and $\sigma_3$ and $\sigma_4$ the third and fourth peaks, respectively. This ratio allows for the selection of candidate redshifts with prominent secondary peaks compared to the remaining two.

\citet{Holwerda_2015} adopted a threshold of $R>1.85$ along with the requirement of foreground+background galaxy pairs as PG+ELG for identifying SGL candidates, being the configuration most observed \citep{Bolton_2008}. After visually inspecting the images of candidates in SDSS and KiDS they obtained 104 lens candidates and
176 occulting galaxy pairs.

\citet{Knabel2023} employed a similar strategy for selecting strong lensing candidates by analyzing GAMA spectra. They examined the spectra of candidates previously identified during SGL searches within the KiDS survey  \citep{links, Li_2020}. 
Since these candidates were pre-selected by the KiDS collaboration, \citet{Knabel2023} used a less stringent threshold on R ($R\geq$1.2) compared to \citet{Holwerda_2015}. Additionally, they required a foreground source redshift exceeding 0.5.
Following this selection, 42 candidates were chosen for further analysis and modeling using the \code{PYAUTOLENS} software \citep{Nightingale_2019}.  Among these, 19 candidates yielded successful modeling results. While the majority exhibited PG+ELG galaxy configurations, other configurations were also present.

\section{Methodology}
\label{sec:methodology}

 \begin{figure*}[h]
\centering
\includegraphics[width=\textwidth,keepaspectratio]{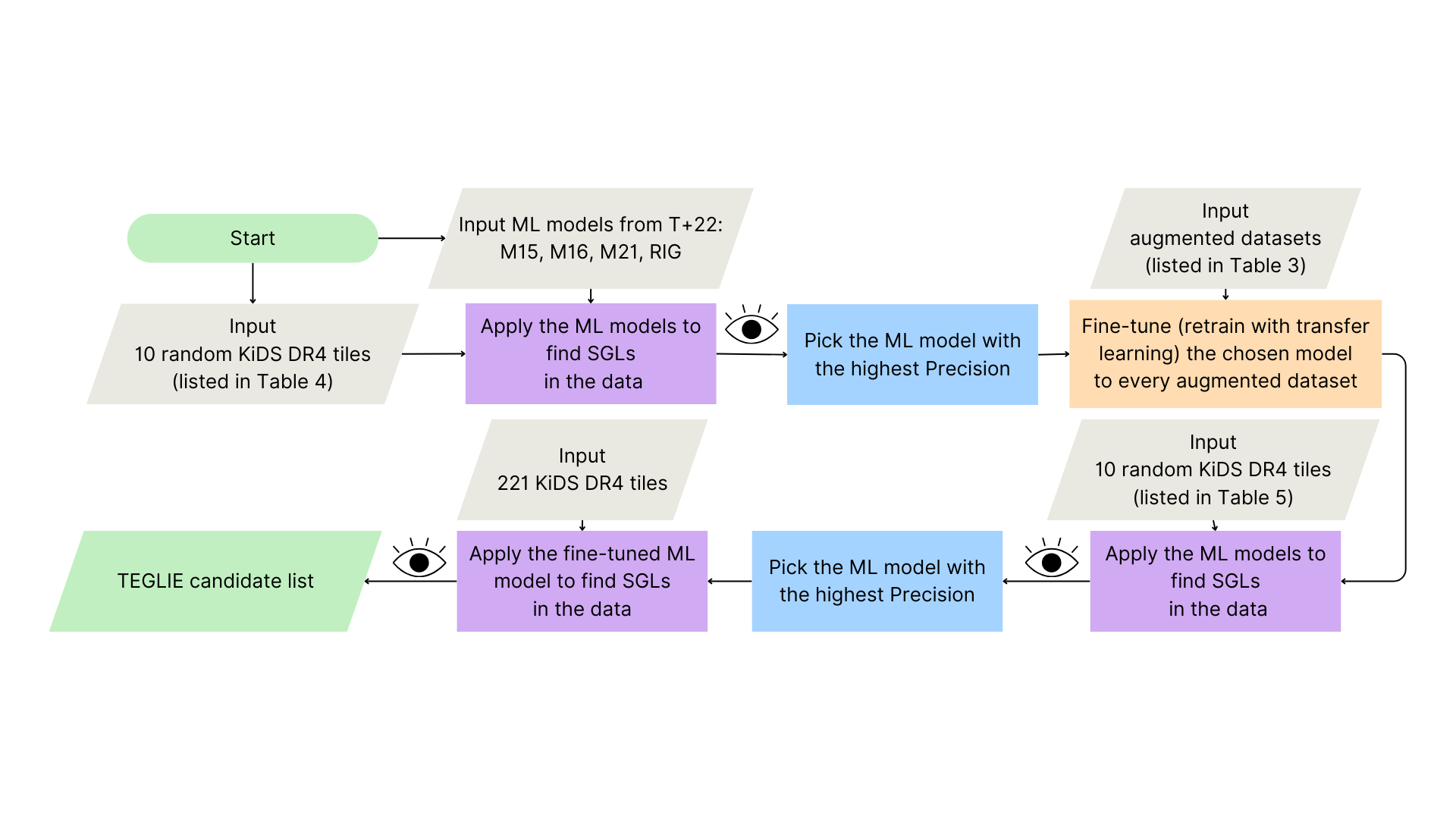}
\caption{Schematic view of this work's methodology.
The eye icon denotes the process of visually evaluating ML candidates and T+22 stands for \cite{Hareesh}. The comprehensive details of each step in this flowchart are presented in Sect. \ref{sec:methodology}.}
\label{fig:flow_chart}
\end{figure*}

In a recent publication, \cite{Hareesh} used, for the first time, transformer encoders to identify SGLs.  A total of 18\,000 images from the Lens Challenge dataset (see \ref{sec:synthetic}) were used to train 21 different transformer encoders (TE). This section outlines the transformer encoder architectures, performance metrics, and fine-tuning approach used in our study.
A comprehensive overview of our methodology can be found in Fig. \ref{fig:flow_chart}.
The best-performing weights (before and after fine-tuning) and architecture of the model are publicly available on GitHub\footnote{\url{https://github.com/margres/TEGLIE}}.

\subsection{The model - Transformer Encoder}
\label{sec:model}
\begin{figure}[h]
\centering
\includegraphics[width=250 pt,keepaspectratio]{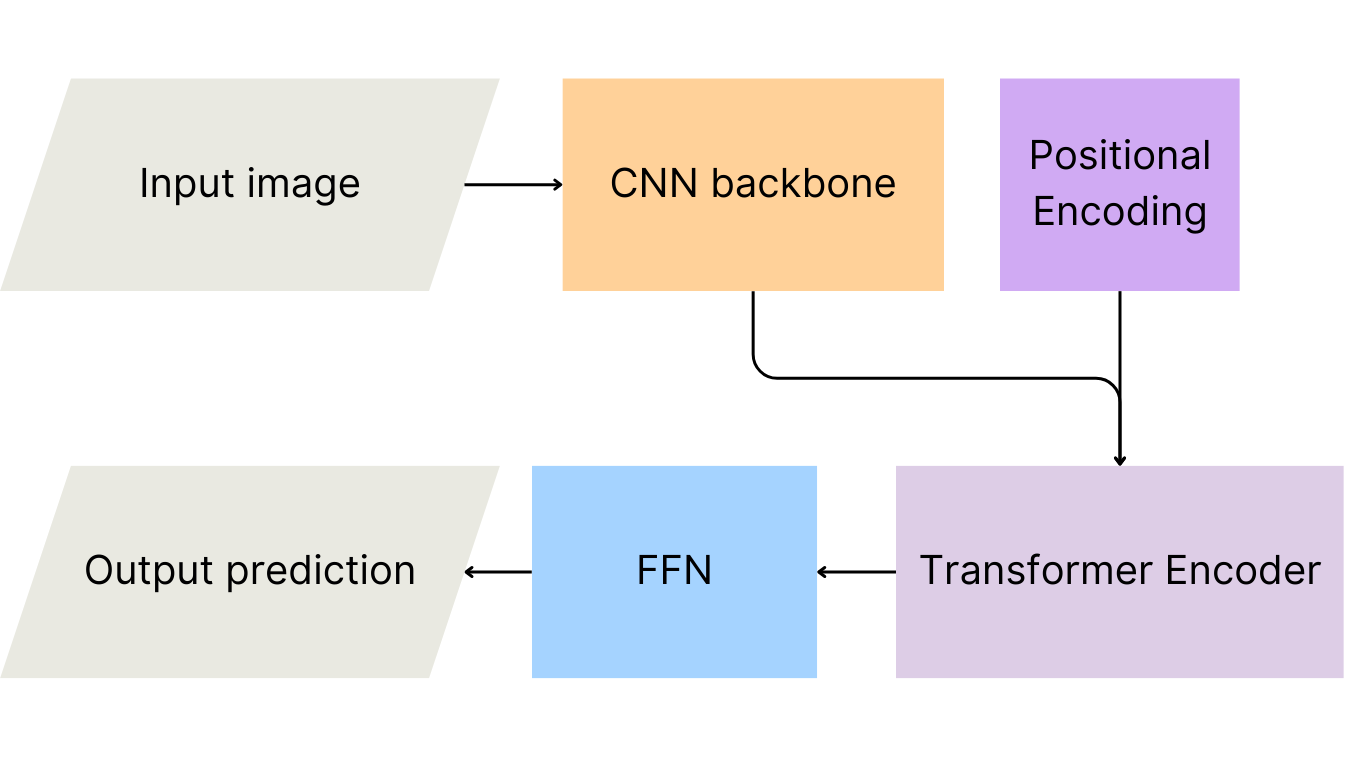}
\caption{Scheme of the general architecture of the transformer encoder. The extracted features of the input image by the CNN backbone are combined with positional encoding and passed on to the encoder layer to assign attention scores to each feature. The weighted features are then passed to the feed-forward neural network (FFN) to predict the probability.}
\label{fig:Transformer}
\end{figure}

Figure \ref{fig:Transformer} presents a schematic view of the transformer employed in this work.
The initial component is a simple CNN designed to extract pertinent features from the input image. Since the transformer architecture is permutation-invariant, a fixed positional encoding is incorporated into the output of the CNN backbone. The positional encoding is defined as \citep{vaswani2017attention}: 
 \begin{gather}
     PE_{(pos,2i)} =  \sin \bigg{(}pos/12800^\frac{2i}{d_{model}}\bigg{)}\\
     PE_{(pos,2i+1)} = \cos \bigg{(}pos/12800^\frac{2i}{d_{model}}\bigg{)}
 \end{gather}
where $pos$ is the position, $i$ is the dimension of the positional encoding vector, and $d_{model}$ is the dimension of the input feature vector. 
The positional encoding layer embeds the positional information of each feature into the representation. This information is essential for the model to understand the spatial context of the features.

The features, along with their positional encoding, are then fed into the transformer encoder layer, which comprises a multi-head self-attention module and a feed-forward network (FFN). Within this layer, self-attention is applied to the input. This process involves treating each point in the feature map as a random variable and determining the pairwise covariances. The value of each prediction is enhanced or diminished based on its similarity to other points in the feature map. The FFN learns the weighted features filtered by the encoder layers.

The model output is a single neuron with a sigmoid activation function that predicts the likelihood of the input image being a lens. This value, known as the prediction probability, ranges from 0 to 1, with 1 representing the maximum probability of the image containing a lens. We consider all elements with a prediction probability exceeding 0.8 as lens candidates.

In this work, we employ the three best-performing models trained on 4-band simulated images by \citet{Hareesh} and one model specifically trained on 3-band images. The incorporation of the 3-band model stems from the observation that the $u$-band data in the real KiDS images exhibit significantly poorer seeing conditions. Consequently, a 3-band network could potentially show superior performance on real data. Furthermore, previous searches in the KiDS survey primarily employed 3- or 1-band networks.

Table~\ref{tab:best_model} summarizes the model architectures. In a transformer encoder, the notation X H$_{Y}$ indicates the presence of $X$ independent attention heads, each processing a hidden representation of dimension Y. These heads act as multiple independent units focusing on different aspects of the data.  The dimension, Y, refers to the size of the hidden representation processed by each head and determines the model's complexity and capacity.
Furthermore, Z(E) denotes the inclusion of Z, the main processing units that extract information from the input data and pass it through subsequent layers. The AUROC, TPR and FPR metrics definitions are provided in the Appx. \ref{sec:metric_labeled}.

\begin{table}[h]
\caption{Table incorporating the name, model structure, AUROC, TPR, and FPR of the models from \citet{Hareesh} used in this study.
}
\centering
\begin{adjustbox}{width=1\columnwidth}
\begin{tabular}{lllll}
\hline\hline
\noalign{\smallskip}
Name       & Model structure                   & AUROC & TPR & FPR \\
\hline
\noalign{\smallskip}
M15 & 8 CNN Layers + 8 H$_{128}$ + 4 (E)      & 0.978 & 0.92  &  0.06      \\
M16 & 16 CNN Layers + 8 H$_{128}$ + 8 (E)   & 0.962 & 0.90  & 0.10        \\
M21 & 8 CNN Layers + 8 H$_{128}$ + 4 (E)     & 0.980  &  0.93  & 0.08     \\
RIG & 8 CNN Layers + 8 H$_{128}$ + 4 (E)  & 0.974   &  0.91   &  0.07   \\
\hline
\end{tabular}
\end{adjustbox}
\label{tab:best_model}
\end{table}

\subsection{Metrics}
\label{sec:metric}
Astronomical surveys like KiDS lack precise counts of genuine lenses and non-lenses within the data. While visual inspection allows us to identify correctly classified lenses (true positives, TP) and non-lenses misidentified as lenses (false positives, FP), it is difficult to determine the exact number of missed lenses as well as the correctly classified non-lenses. This limitation renders commonly used evaluation metrics, such as AUROC, TPR, and FPR (defined in Appx. \ref{sec:metric_labeled}), inapplicable.

To assess our models' performance we use, instead, the precision metric:
\begin{equation}
\label{eq:P}
   P = \frac{TP}{TP + FP} .
\end{equation}

\subsection{Visual inspection}
\label{sec:visual_insp}

Visual inspection is a valuable, albeit time-consuming, step for preliminary quality assessment of candidate lenses identified by the machine learning model. This step is essential for obtaining a clean training dataset and estimating model precision.

The individual tasked with evaluating the candidate images, referred to as the inspector, is presented with four bands of 101 x 101 pixel images and a composite RGB image constructed from the $r$, $i$, and $g$-bands, as outlined in the methodology presented in  \cite{Lupton_2004}.
The inspector is then required to assign a grade to the object displayed on the screen. The available options are as follows:
 
\begin{itemize}
    \item 0: No lensing features are present
    \item 1: Most likely a SGL. The presence of clear lensing features, such as arcs or multiple distorted images, strongly suggests strong gravitational lensing. Accordingly, grade 1 was assigned to candidates with blue arcs and a red source or when multiple images were positioned in a manner consistent with strong gravitational lensing.
    \item 2: Possibly a SGL. 
    The presence of a single image or a weakly curved arc-like structure raises the possibility of gravitational lensing. However, the absence of a distinct ring or the ambiguity between an Einstein ring and a ring galaxy hinders conclusive classification. This grade was awarded when only a single faint blue arc or image was visible and the configuration of the lens and source remained unclear.
    \item 3: Object exhibiting lensing-like features (arc-like structures) without being a lens: This grade was assigned to candidates that exhibited arc-like structures but were not caused by gravitational lensing. Such objects could be merging galaxies or unusual galaxies with peculiar structures.

\end{itemize}

The utility of a model hinges on the balance between TP and FP that is deemed acceptable by the inspectors. 
In the context of wide-area surveys, visual inspection represents a bottleneck in candidate identification, necessitating the implementation of a model with a relatively low FPR. Despite the unknown total number of strong lenses (TPs + FPs), the model's performance can be assessed by evaluating its ability to identify lenses discovered by other surveys.

As previously stated, visual inspection is a time-consuming task. Due to the broad redshift range and lack of specific galaxy population restrictions in this study, only one expert initially assesses the candidates. Subsequently, a group of experts scrutinizes the more promising candidates (labeled 1, 2, or occasionally 3).
The lenses retained with a grade of 1 after this final round are designated as high-quality (HQ) or high-confidence candidates, while all candidates (labeled 1 and 2) are referred to as TEGLIE candidates.

\subsection{Fine-tuning and Data Augmentation}
\label{sec:augmented_dataset}

The Lens Finding Challenge dataset has been created for the purpose of testing different automated techniques for the identification of SGL. Consequently, the dataset was not intended to precisely replicate real observations; rather, KiDS was employed as a guideline for image creation. For this reason, for instance, \cite{Davies_2019} identified inconsistencies, such as an over-representation of large Einstein rings and swimulated lensed galaxies fainter than the COSMOS lenses \citep{Faure_2008}.
Given the discrepancies observed, it is anticipated that models trained exclusively on the simulated dataset will exhibit suboptimal performance when presented with actual KiDS observations. Consequently, in order to mitigate this issue, the model with the highest performance is selected by evaluating it on a representative subset of KiDS data (10 deg$^2$). Furthermore, transfer learning and data augmentation techniques are employed to enhance the training data, with the objective of fine-tuning the model to the characteristics of the KiDS survey.

Transfer learning involves fine-tuning a model that has already been trained on a large, diverse dataset to adapt it to a specific task or domain. Previous studies have demonstrated the effectiveness of transfer learning in improving the performance of machine learning models on small datasets \citep{Pan_2010, Yosinski_2014}. 
Instead of simply fine-tuning specific layers, we load the pre-trained weights and fine-tune the entire model on a representative subset of real KiDS data augmented to address the model's limitations. This approach aims to use the knowledge learned from the simulated data while adapting to the unique features and characteristics present in real KiDS observations, potentially overcoming the challenge of identifying FP. 

Data augmentation refers to techniques that artificially expand the training set size by generating new images from the existing ones. Common approaches include geometric transformations (scaling, rotation, cropping, flipping), random erasing, noise injection, color space modifications, and many more. For an in-depth overview of image augmentation techniques, refer to  \citet{Mumuni_2022,Xu_2023}. 
However, the ideal augmentation strategy depends heavily on the specific model and task. Therefore, we use the best-performing model on the small KiDS data sample to test different combinations of augmentation techniques specifically for SGL in KiDS. This process ensures we tailor the augmented data to the model's needs, optimizing its ability to generalize to real-world observations. Ultimately, the model fine-tuned on the dataset yielding the best performance with data augmentation is selected to search for lenses in a larger KiDS area of $\sim$ 200 deg$^2$.

\begin{figure}[h]
\includegraphics[width=8cm]{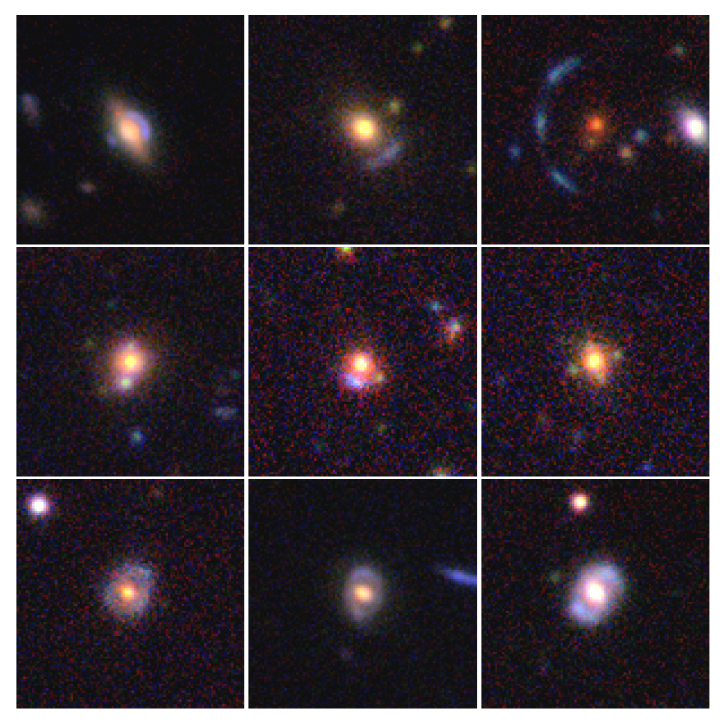}
\centering
\caption{Selection of the exemplary strong lens candidates found by the KiDS collaboration in DR4 \citep{Li_2020, links} and implemented in the fine-tuning dataset.  }
\label{fig:lenses_kids}
\end{figure}

The dataset used for fine-tuning includes a balanced quantity of real observations of both SGLs and non-lenses, the SGL class being the only one with few available examples being the one augmented. For the SGL class, 169 high-quality strong lens candidates from KiDS DR4, identified by \cite{links, Li_2020, Li_2021}  and publicly available on the KiDS DR4 website\footnote{\url{https://kids.strw.leidenuniv.nl/DR4/hqlenses.php}}, are employed. Some of these previously discovered lenses used for fine-tuning are shown in Fig. \ref{fig:lenses_kids}. 

\begin{figure}[!htbp]
\includegraphics[width=8cm]{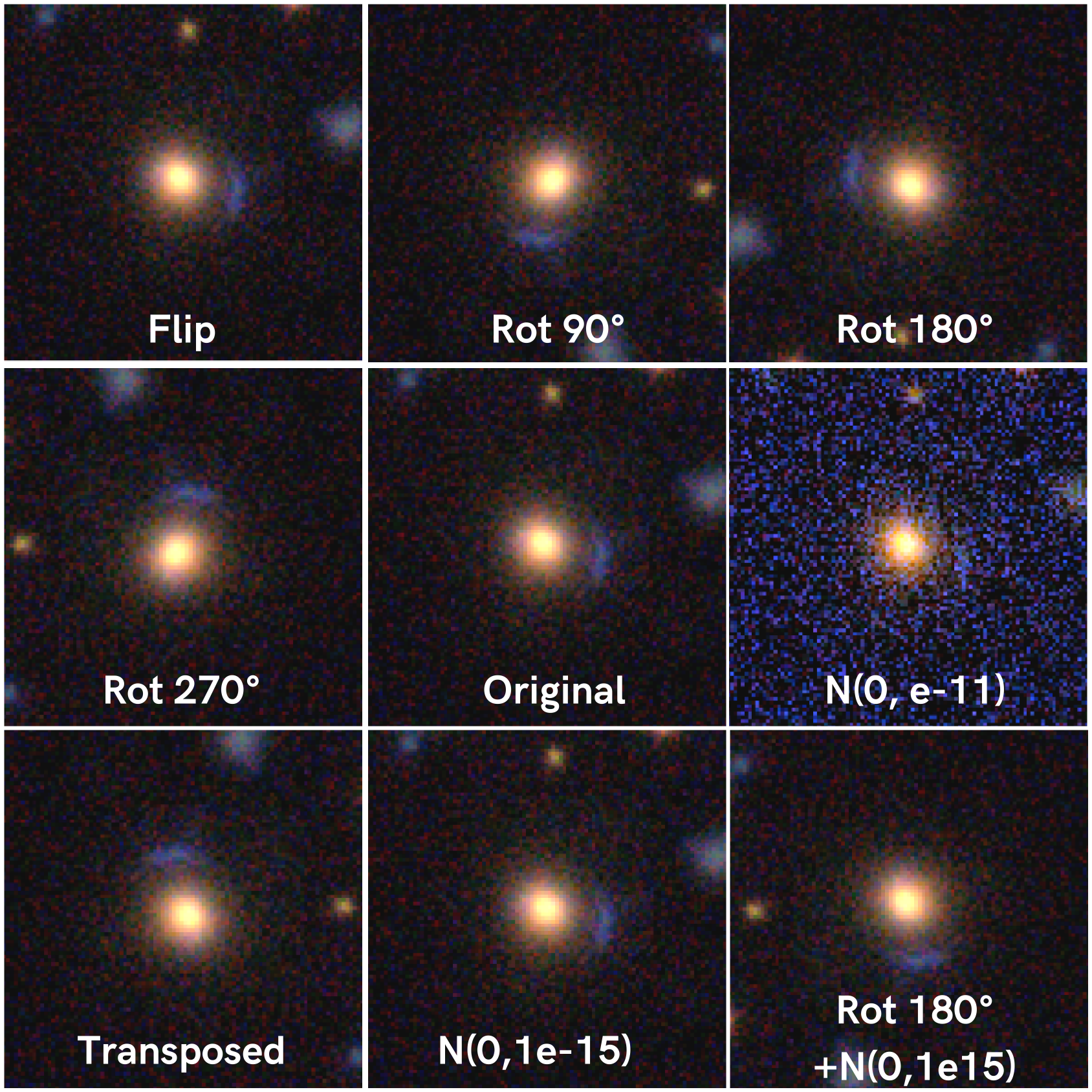}
\centering
\caption{High-quality KiDS SGL candidate (ICRS coordinates J030540.105-293416) cutout with different data augmentations. }
\label{fig:data_augmentation}
\end{figure}

In this work, we use straightforward image augmentation techniques, including rotation, flipping, transposing, and noise injection. To preserve the intrinsic properties of the observed objects, we avoid transformations that alter the image's color or zoom ratio.
Specifically, we use rotation angles of 90\textdegree, 180\textdegree, and 270\textdegree,  which are necessary to avoid empty corners. For noise injection, we introduce white Gaussian (${\cal N}$) noise with a mean $\mu=0$ and standard deviations of $\sigma=10^{-11}$ and $10^{-15}$. 
To determine these values, we randomly select 1\,000 cutouts that pass our sample selection. For each cutout, we calculate the average, minimum, and maximum pixel values. We then calculate the average of these three measurements, resulting in a minimum average value of -5.767$\times$10$^{-11}$, a maximum average value of 2.644$\times$10$^{-9}$, and an average of 1.846$\times$10$^{-11}$. We choose a noise level of the same order of magnitude as the average pixel value and a significantly smaller value to assess the network's response to both magnitudes.
An example of SGL with different data augmentations is shown in Fig. \ref{fig:data_augmentation}. 

\begin{figure*}
\includegraphics[width=\textwidth]{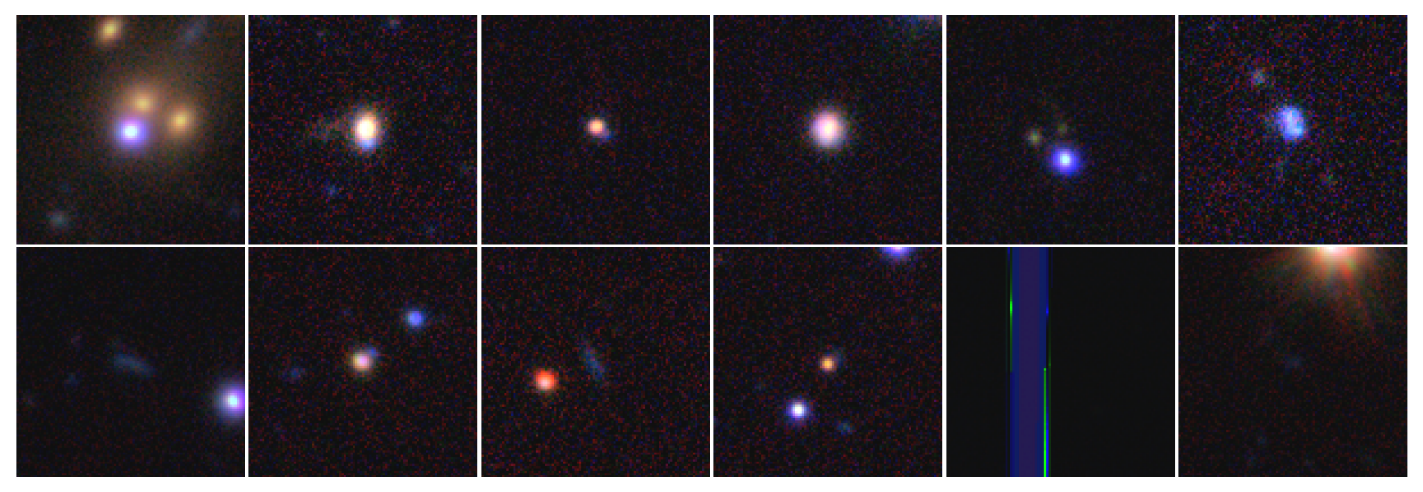}
\centering
\caption{Examples of cutouts employed as non-lenses in the fine-tuning dataset}
\label{fig:grade_0_0}
\end{figure*}

The non-lens class in our dataset consists of images misclassified as strong lenses by the model. To improve its performance, we prioritize images with the highest probability of being misclassified lenses, a process known as uncertainty sampling \citep{lewis_1994_uncert_sampl}.
This technique identifies examples where the model exhibits the most uncertainty, allowing for efficient learning with minimal labeling effort.
In our case, we have a fixed number of SGL candidates but a larger pool of non-lens examples. Therefore, we utilize all SGL candidates while selecting the most uncertain non-lens images based on their prediction error. This least confident sampling strategy prioritizes the most informative examples for model improvement.
Specifically, we gather the 150 most uncertain FP images from each of 10 randomly selected KiDS DR4 tiles, ensuring equal representation from each tile.
Figure \ref{fig:grade_0_0} showcases some of these particularly uncertain non-lenses.

The datasets investigated in the fine-tuning process are
\begin{itemize}
    \item No Augmentation: cutouts of SGL candidates found by the KiDS collaboration
    \item Rotated (R): 4 images for every SGL, the original plus the three rotations
    \item Rotated + Gaussian Noise $\mu$=0, $\sigma$=10$^{-11}$:  the rotated images described above but with some additional white noise drawn by 
${\cal N}(0, 10^{-11})$ 
    \item Rotated + Gaussian Noise $\mu$=0, $\sigma$=10$^{-15}$: similar to the dataset described above, rotated images with additive white noise drawn from ${\cal N}(0, 10^{-15})$ 
    \item Rotated, Rotated+Gaussian Noise $\mu$=0, $\sigma$=10$^{-15}$: 
    rotated images and rotated images with additive noise drawn from ${\cal N}(0, 10^{-15})$ 
    \item Rotated, Flipped (RF): rotated images, and for every cutout, the flipped SGL with respect to the x-axis       
    \item Rotated, Flipped, Transposed (RFT): rotated images, flipped SGL with respect to the x-axis, and transposed original image                 
\end{itemize}

For each dataset, we randomly partition the data into training, testing, and validation sets, maintaining the equal representation of lenses and non-lenses classes. The test set comprises 20\% of the overall augmented dataset, while the validation set constitutes 10\% of the training set, leaving the remaining 72\% of the augmented dataset for the training set. The sizes of the augmented datasets are presented in Tab. \ref{tab:dataset_size}.

\begin{table}[h]
\caption{Test and training set sizes for the different kinds of data augmentations. The dataset size includes lenses and not lenses together.}
\label{tab:dataset_size}
\centering
\begin{tabular}{lll}
\hline\hline
\noalign{\smallskip}
Augmentation type                              & Test set & Training set \\
\hline
\noalign{\smallskip}
No Augmentation                                    & 68       & 243          \\
Rotated (R)                                         & 271      & 972          \\
Rotated+${\cal N}(0, 10^{-11})$           & 338      & 1\,216         \\
Rotated+${\cal N}(0, 10^{-15})$          & 338      & 1\,216         \\
Rotated, Rotated+${\cal N}(0, 10^{-11})$  & 541      & 1\,946         \\
Rotated, Flipped (RF)                                   & 340      & 1\,222         \\
Rotated, Flipped, Transposed (RFT)                        & 406      & 1\,459        \\
\hline
\end{tabular}
\end{table}


\begin{figure*}[!htbp]
\includegraphics[width=\textwidth]{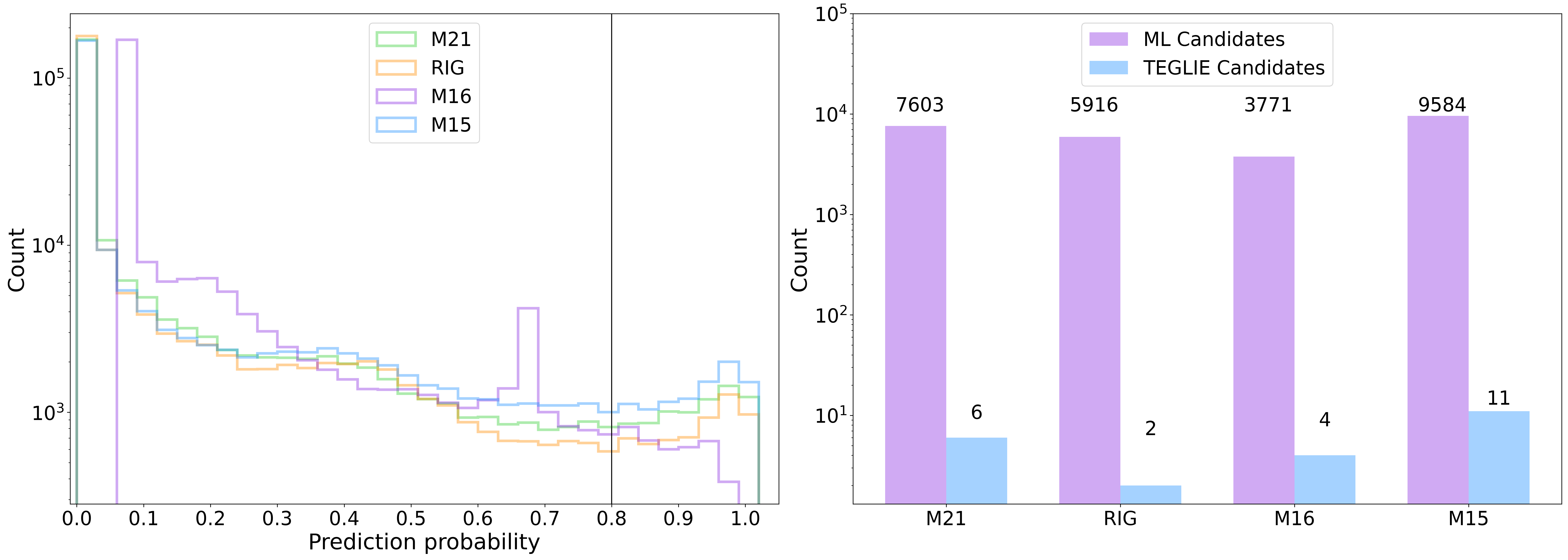}
\centering
\caption{Prediction probability sample extracted from the 10 tiles in Table \ref{tab:tiles_toaug}.
The left plot shows the prediction probability of the four models used in this work from \cite{Hareesh}, the black line shows the threshold probability, every object above 0.8 is considered an ML candidate.  The right plot shows the total number of ML candidates compared to the true positives with grades 1 and 2 (TEGLIE candidates) for every model. Bar element counts are shown above each bar. }
\label{fig:fp_before_retraining}
\end{figure*}

\section{Model evaluation}
\label{sec:test}

\begin{table}[!htbp]
\caption{List of the 10 randomly chosen KiDS tiles used for testing the transformers on the survey data. The columns from left to right include the tile name as found in the KiDS DR4 catalog and the coordinates of the tile center in J2000 degrees. }
\label{tab:tiles_toaug}
\centering
\footnotesize
\begin{tabular}{lll}
\hline\hline
\noalign{\smallskip}
Tile name   & R.A. J2000      & Dec. J2000    \\
\hline
\noalign{\smallskip}
KIDS\_0.0\_-28.2   & 0.0     & -28.192         \\
KIDS\_212.6\_2.5   & 212.591 & 2.478          \\
KIDS\_168.0\_1.5   & 168.0   & 1.489           \\
KIDS\_4.7\_-32.1   & 4.706   & -32.148       \\
KIDS\_185.0\_1.5   & 185.0   & 1.489        \\
KIDS\_156.0\_1.5   & 156.0   & 1.489        \\
KIDS\_339.3\_-30.2 & 339.297 & -30.17          \\
KIDS\_53.5\_-35.1  & 53.514  & -35.115   \\
KIDS\_138.0\_0.5   & 138.0   & 0.5         \\
KIDS\_353.1\_-30.2 & 353.099 & -30.17       \\
\hline
\end{tabular}
\end{table}

\subsection{Testing on KiDS DR4}
\label{sec:testing_real}

Table \ref{tab:best_model} summarizes the performance of the three best-performing models by \cite{Hareesh}. These models are tested on 10 randomly selected KiDS tiles, listed in Tab. \ref{tab:tiles_toaug}, none of these overlap with GAMA.  The cutouts fed into the models are preprocessed according to the procedure described in Sect. \ref{sec:sample}. 
The distribution of prediction probabilities for each model across the 10 tiles is depicted in the left plot of Fig. \ref{fig:fp_before_retraining}. The cutouts with a prediction probability exceeding the threshold of 0.8 (black vertical line in the figure) are referred to as ML candidates.
To illustrate the proportion of TP (blue bars) to FP (purple bars), the candidates confirmed after visual inspection (TEGLIE candidates) are represented in the right plot of Fig. \ref{fig:fp_before_retraining}.
The HQ (grade 1) candidates identified during this testing process have IDs ranging from 0 to 4 in Fig. \ref{fig:all_candidates} and Tab. \ref{tab:all_candidates_1}.

\begin{figure}[!htbp]
\includegraphics[width=8cm]{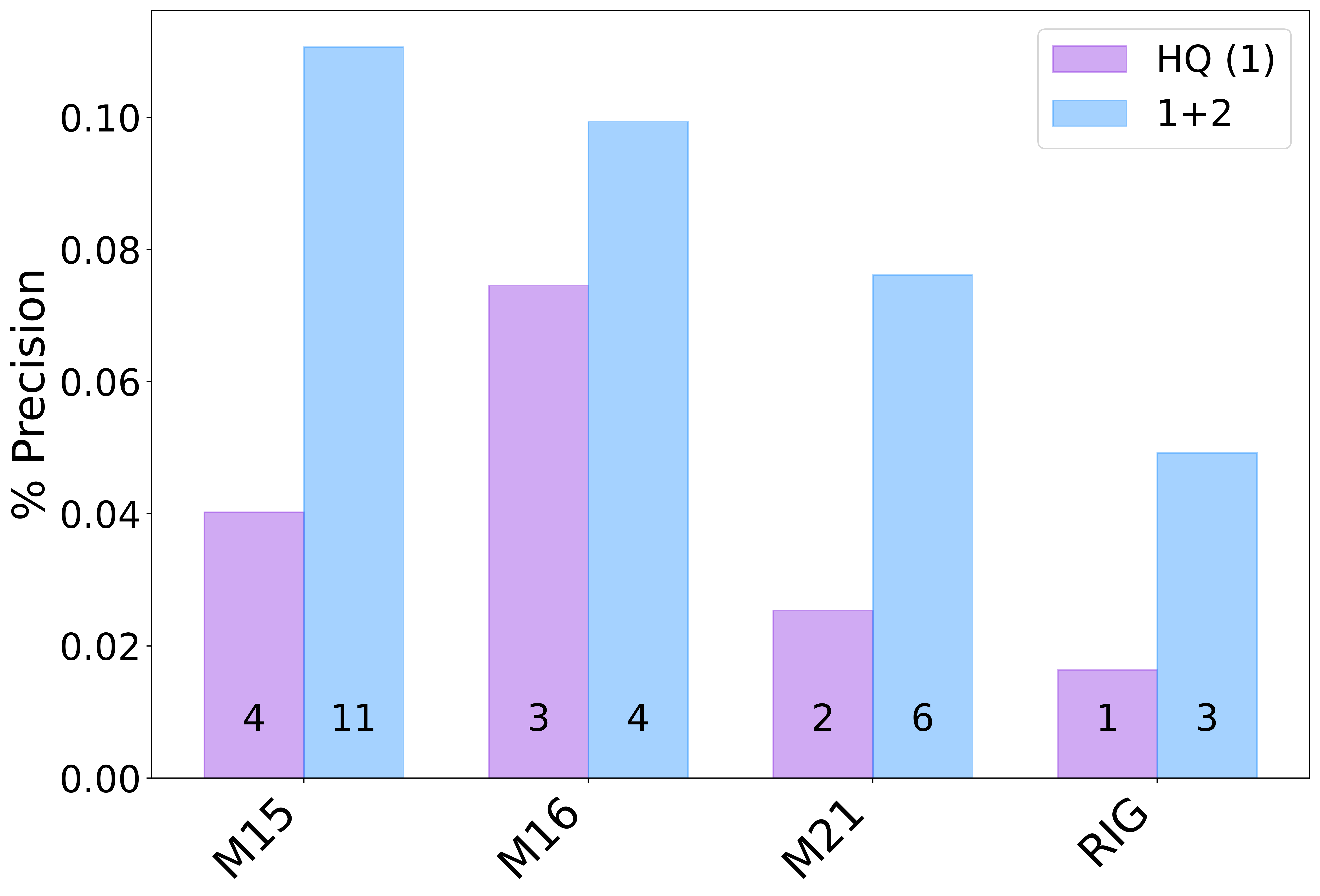}
\centering
\caption{Precision metric (Eq. \ref{eq:P}) of the models in Table \ref{tab:best_model} for KiDS sample from the tiles in Table \ref{tab:tiles_toaug}.
In blue we show the metric computed considering objects with labels 1 and 2 as true positives. In purple, the same metric is shown but for objects with a confidence label of 1. The number at the bottom indicates the total number of true positives found in the 10 tiles with each model.} 
\label{fig:Precision_before}
\end{figure}

In the 10 test tiles, 238\,028 elements pass the preprocessing stage and are presented to the models. The precision of the different models is illustrated in Fig. \ref{fig:Precision_before}.
Compared to the other models, Model 16 identifies significantly fewer candidates (ML{\footnotesize cand}$_{M16}$=4\,026), while Model 15 identifies the most candidates (ML{\footnotesize cand}$_{M15}$=9\,944). Models 21 and RIG identify ML{\footnotesize cand}$_{M21}$=7\,882 and ML{\footnotesize cand}$_{RIG}$=6\,100, respectively. Of these, TP$_{M21}$=6 and TP$_{RIG}$=3 are visually confirmed as SGL candidates.

Model 15 achieves the highest precision P$_{M15}$ = 0.11\% among the models, followed by model 16 of $P_{M16}$=0.099\%, model 21 with  $P_{M21}$=0.076 \%, and model RIG with $P_{RIG}$=0.049 \%.
Despite its high precision, Model 15 also generates double the number of FP compared to Model 16.  

\begin{figure}[!htbp]
\includegraphics[width=6cm]{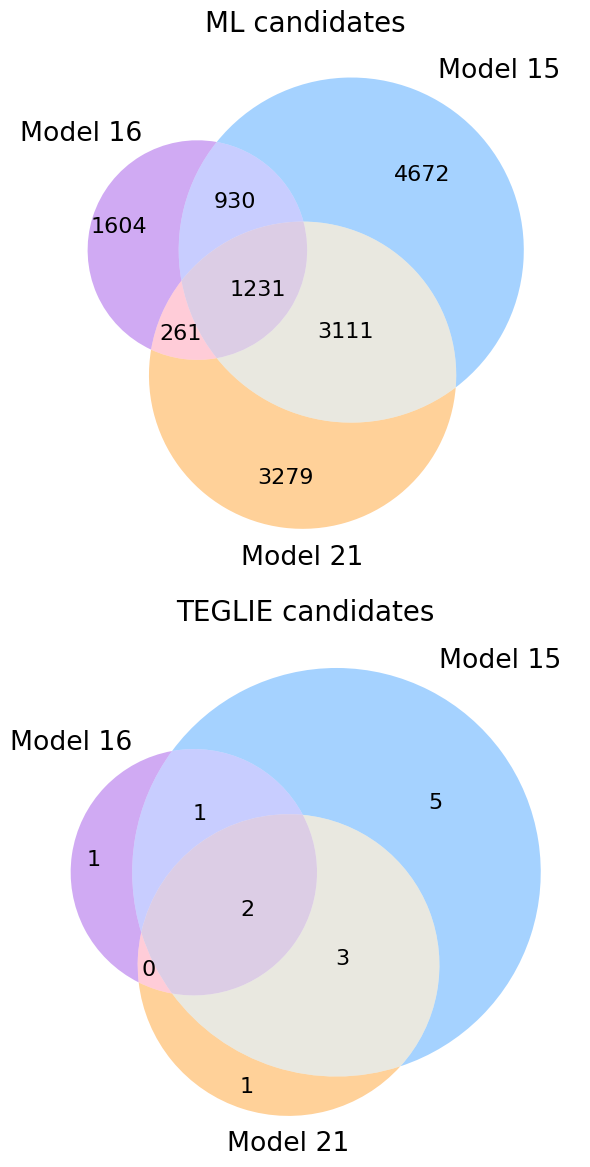}
\centering
\caption{Venn diagram exploring common elements found by different models. Numbers indicate the numbers of elements in common between datasets. Top panel: diagram of the ML candidates found by the three 4-bands models.
Bottom panel: diagram of the visually inspected candidates only.} 
\label{fig:ven_diag_test_tilesz0-0.8}
\end{figure}

Figure \ref{fig:ven_diag_test_tilesz0-0.8} illustrates the overlap between the three models in terms of the candidates detected. The top diagram depicts the overlap for all candidates, while the bottom diagram represents the overlap for confirmed candidates. Given model RIG's inferior performance, it has been excluded from this analysis.
The three models identify distinct candidates, yet Model 15 uncovers the majority of the candidates found by Models 21 and 16. This indicates that Model 15 possesses a more comprehensive grasp of the shared characteristics of strong gravitational lenses compared to the other two models.

The substantial prevalence of false positives renders the SGL detection impractical, necessitating fine-tuning. To accomplish this, one of the presented models must be selected.
Model 15 exhibits the highest P among all models, despite having the highest number of false positives. Model 15 also has a significant overlap with the other models in terms of the detected candidates. This implies that Model 15 is capable of identifying a broad spectrum of SGL candidates and does not simply rely on a small number of very specific features. Additionally, Model 15 has demonstrated effectiveness on simulated data \citep{Hareesh}, indicating that it is a robust model that is likely to generalize well to new data sets.
Based on this assessment, Model 15 stands out as the most appropriate candidate for fine-tuning for the KiDS survey.

\subsection{Testing of the fine-tuned models on KiDS DR4}
\label{sec:test_retrained}

Using each dataset specified in Tab. \ref{tab:dataset_size}, we individually fine-tune Model 15 for the KiDS survey. We adhere to the same procedure outlined in Section \ref{sec:testing_real}, randomly selecting 10 KiDS tiles. The tile names and characteristics are presented in Tab. \ref{tab:test_tiles} with two of these (KIDS$\_$340.8$\_$-28.2, KIDS$\_$131.0$\_$1.5) situated within the GAMA survey area.

\begin{figure}[!htbp]
\includegraphics[width=8cm]{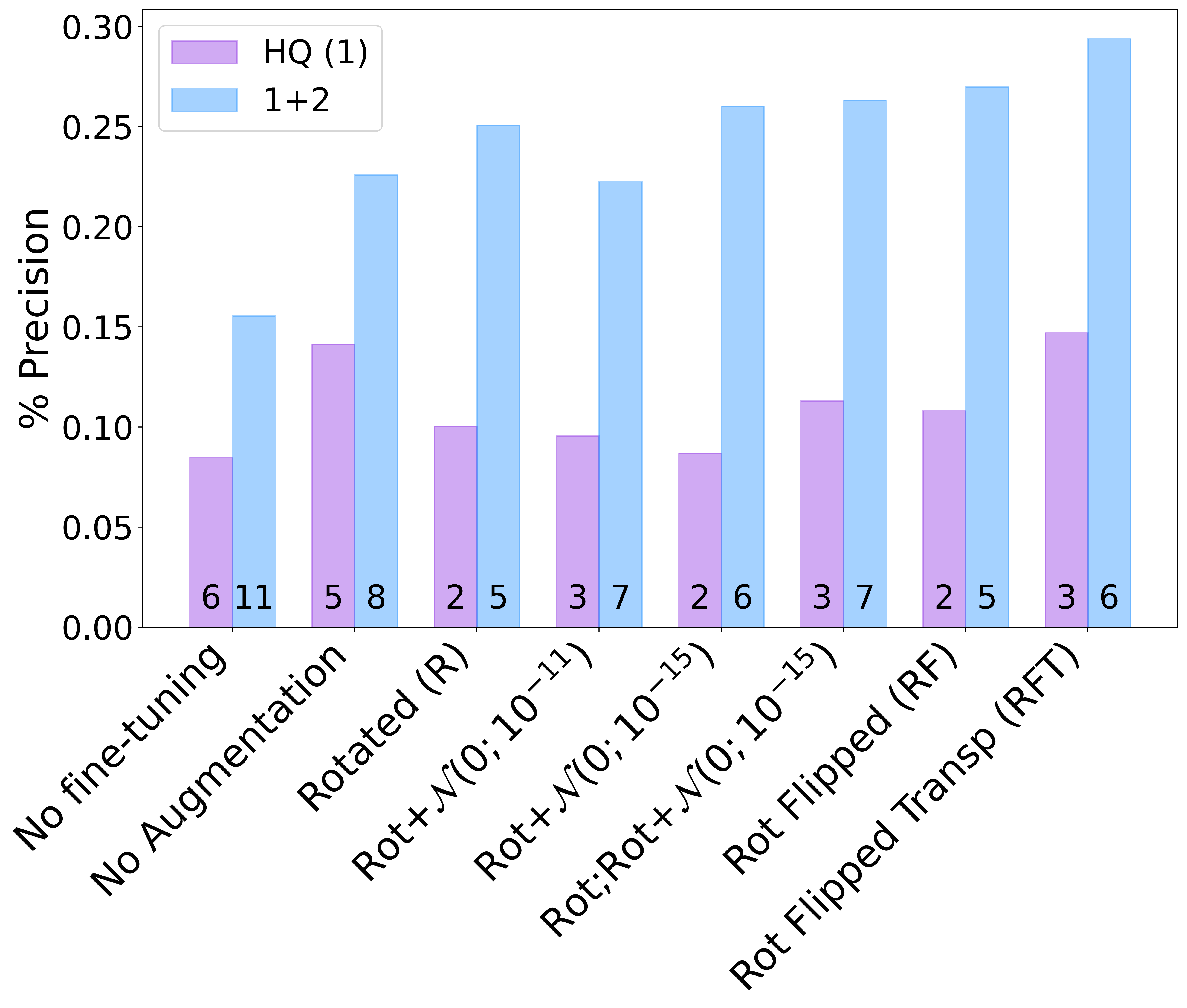}
\centering
\caption{Precision metric (Eq. \ref{eq:P}) for different types of augmented datasets. 
In blue we show the metric computed considering objects with labels 1 and 2 as true positives. In purple, the same metric is shown but for objects with a confidence label of 1. TP counts are shown at the bottom of every bar.}
\label{fig:diff_aug}
\end{figure}

Figure \ref{fig:diff_aug} presents the precision achieved on the 10 test tiles, for various augmentation techniques. The purple boxes represent the precision for only the high-quality candidates, the blue boxes represent the metric for all confirmed candidates and the numbers at the bottom of the boxes the SGL candidates found. Figure \ref{fig:diff_aug} demonstrates how the performance of the transformers varies with the fine-tuning dataset. As expected, the model with the most false positives is the one employing the original, not fine-tuned weights - "No fine-tuning".
The precision of the "No fine-tuning" model is approximately half of the best-performing fine-tuned model, highlighting the impact of transfer learning. The models with the least false positives are RFT, RF, and R. However, the number of true positives changes substantially between these models. The model fine-tuned with rotated flipped and transposed images, RFT, emerges as the best-performing model in this 10-tile evaluation. 
The number of FPs significantly impacts the precision of the "No Fine-Tuning" model. The model identified 7068 FPs, considerably higher than the 2035 FPs found by the RFT model. While the RF model achieved the lowest number of FPs (1847), it also detected one fewer SGL candidate compared to the RFT model. This trade-off between FPs and TP is well depicted by the precision metric.

In total, 23 objects have been labeled as SL candidates across the 10 test tiles, and the 10 high-quality (HQ) instances have IDs 5 to 14 in Fig. \ref{fig:all_candidates} and Tab. \ref{tab:all_candidates_1}. The RFT model exhibits an average of 70\% fewer false positives than "No fine-tuning".

\begin{table}[!htbp]
\caption{List of the 10 randomly chosen KiDS tiles used for testing the fine-tuned transformers. 
The columns from left to right include the tile name as found in the KiDS DR4 catalog and the coordinates of the tile center in J2000 degrees. \label{tab:test_tiles}}
\centering
\footnotesize
\begin{tabular}{lll}
\hline\hline\noalign{\smallskip}
Tile name & R.A. J2000  & Dec. J2000    \\
\hline
\noalign{\smallskip}
KIDS\_40.6\_-28.2 & 40.627 & -28.192   \\
KIDS\_331.5\_-29.2 & 331.519 & -29.181   \\
KIDS\_131.0\_1.5 & 131.0 & 1.489 \\
KIDS\_340.8\_-28.2 & 340.815 & -28.192  \\
KIDS\_32.7\_-28.2 & 32.727 & -28.192  \\
KIDS\_39.9\_-29.2 & 39.873 & -29.181  \\
KIDS\_52.4\_-29.2 & 52.405 & -29.181 \\
KIDS\_199.6\_2.5 & 199.554 & 2.478  \\
KIDS\_179.0\_-0.5 & 179.0 & -0.5  \\
KIDS\_167.0\_-0.5 & 167.0 & -0.5 \\
\hline
\end{tabular}
\end{table}

\begin{table*}[]
\centering
\caption{The complete catalog of the machine learning SGL candidates from the 221 deg$^2$ check, including the label given after visual inspection and the prediction probability given by the model. The full catalog is only available at the CDS.}
\label{tab:all_checked}
\begin{tabular}{llllllll}
\hline\hline \noalign{\smallskip}
ID    & KiDS Tile        & KiDS ID               & R.A. J2000 & Dec. J2000 & $z_{phot}$ & Grade & Prob. \\
\hline \noalign{\smallskip}
0     & KIDS\_213.0\_1.5 & J141138.303+010326.90 & 212.9096   & 1.0575     & 0.59       & 0   & 0.99  \\
1     & KIDS\_213.0\_1.5 & J141050.772+010425.21 & 212.7116   & 1.0737     & 0.55       & 0  & 0.99  \\
2     & KIDS\_213.0\_1.5 & J141034.169+010518.74 & 212.6424   & 1.0885     & 0.66       & 0   & 1.0   \\
\multicolumn{8}{c}{...}                                                                                 \\
51624 & KIDS\_131.0\_1.5 &  J084406.467+015347.21 & 131.026948 & 1.896449 & 0.16 & 0   & 0.94 \\
51625 & KIDS\_131.0\_1.5 &  J084544.554+015636.43 & 131.435644 & 1.943453 & 0.26 & 0   & 0.99 \\
51626 & KIDS\_131.0\_1.5 &  J084540.075+015844.75 & 131.416982 & 1.979098 & 0.63 & 0   & 0.99 \\
\hline 
\end{tabular}
\end{table*}

\begin{table*}[!htbp]
\footnotesize
\centering
\caption{High-quality candidates identified through visual inspection. 
    The first column `ID' indicates the index of the object in Fig. \ref{fig:all_candidates}. `KiDS Tile' and `KiDS ID' specify the name of the tile and ID (omitting the prefix KiDSDR4) as specified in the KIDS catalog, respectively. `R.A.' and `Dec.' are the object's coordinates in degrees; `z$_{phot}$', `z$_{spec}$' and `z$_{spec,2}$' are the photometric redshift of the object in the KiDS catalog, the two high-fidelity redshifts from the blended spectra in GAMA, respectively. `Test' specifies whether the object was found before (BF) or after (AF) fine-tuning of the model or during the search within the GAMA (G) footprint. Some candidates have been detected during the AF test as well as during the GAMA search.
    The `crosscheck' column includes any additional comments about the object crossmatch with other surveys within a 10 $arcsec$ range.}
\label{tab:all_candidates_1}
\begin{adjustbox}{width=0.93\textwidth}
\begin{tabular}{llllllllll}
\hline
\hline
  \noalign{\smallskip}
ID & KiDS Tile         & KiDS ID               & R.A. J2000 & Dec. J2000 & z$_{phot}$ & z$_{spec}$ & z$_{spec,2}$ & Test & Crosscheck                \\
\hline
   \noalign{\smallskip}
0  & KIDS\_0.0\_-28.2   & J235811.969-282916.97 & 359.5499 & -28.488  & 0.42 &      &      & BF   &                         \\
1  & KIDS\_0.0\_-28.2   & J000014.372-281135.77 & 0.0599   & -28.1933 & 0.31 &      &      & BF   &                         \\
2  & KIDS\_168.0\_1.5   & J111050.643+014544.91 & 167.711  & 1.7625   & 0.17 &      &      & BF   &                         \\
3  & KIDS\_212.6\_2.5   & J141208.101+021905.12 & 213.0338 & 2.3181   & 0.54 &      &      & BF   &                         \\
4  & KIDS\_339.3\_-30.2 & J223657.865-303421.92 & 339.2411 & -30.5728 & 0.45 &      &      & BF   &                         \\
5  & KIDS\_40.6\_-28.2  & J024129.936-282646.70 & 40.3747  & -28.4463 & 0.67 &      &      & AF   &                         \\
6  & KIDS\_331.5\_-29.2 & J220803.182-293206.60 & 332.0133 & -29.5352 & 0.24 &      &      & AF   &                         \\
7  & KIDS\_39.9\_-29.2  & J023831.797-291408.66 & 39.6325  & -29.2357 & 0.15 &      &      & AF   &                         \\
8  & KIDS\_39.9\_-29.2  & J023813.654-291219.81 & 39.5569  & -29.2055 & 0.3  &      &      & AF   & White Dwarf             \\
9  & KIDS\_52.4\_-29.2  & J032933.425-293534.26 & 52.3893  & -29.5929 & 0.77 &      &      & AF   &                         \\
10 & KIDS\_179.0\_-0.5  & J115657.416-004729.75 & 179.2392 & -0.7916  & 0.1  &      &      & AF   &                         \\
11 & KIDS\_40.6\_-28.2  & J024130.061-280752.35 & 40.3753  & -28.1312 & 0.49 &      &      & AF   &                         \\
12 & KIDS\_340.8\_-28.2 & J224144.305-282155.34 & 340.4346 & -28.3654 & 0.46 &      &      & AF   &                         \\
13 & KIDS\_179.0\_-0.5  & J115522.762-002930.42 & 178.8448 & -0.4918  & 0.5  & 0.58 & 0.43 & AF,G   &                         \\
14 & KIDS\_179.0\_-0.5  & J115637.743-000903.21 & 179.1573 & -0.1509  & 0.43 & 0.09 & 0.37 & AF,G   &                         \\
15 & KIDS\_130.0\_-1.5  & J083933.274-014038.45 & 129.8886 & -1.6773  & 0.33 &      &      & G    & HQ KiDS, SuGOHI (A)         \\
16 & KIDS\_131.0\_-0.5  & J084341.960-002715.05 & 130.9248 & -0.4542  & 0.28 & 0.23 & 0.23 & G    &                         \\
17 & KIDS\_131.0\_-0.5  & J084520.009-005454.78 & 131.3334 & -0.9152  & 0.47 & 0.36 & 0.4  & G    & HQ LinKS, SuGOHI (A)       \\
18 & KIDS\_131.0\_1.5   & J084449.044+010002.26 & 131.2044 & 1.0006   & 0.43 & 0.28 & 0.29 & G    & LinKS, SuGOHI (C)          \\
19 & KIDS\_131.0\_1.5   & J084434.974+011105.24 & 131.1457 & 1.1848   & 0.63 & 0.04 & 0.28 & G    &                         \\
20 & KIDS\_133.0\_1.5   & J085156.962+013016.65 & 132.9873 & 1.5046   & 0.44 &      &      & G    & HQ LinKS                \\
21 & KIDS\_134.0\_-1.5 & J085446.547-012137.03 & 133.6939 & -1.3603 & 0.43 & 0.35 & 0.36 & G & HQ LinKS, SuGOHI (A), SL2S    \\
22 & KIDS\_136.0\_-0.5  & J090248.973-000742.59 & 135.7041 & -0.1285  & 0.58 &      &      & G    &                         \\
23 & KIDS\_136.0\_-0.5  & J090507.336-001029.85 & 136.2806 & -0.175   & 0.71 & 0.03 & 0.49 & G    & HQ KiDS, SuGOHI (B)         \\
24 & KIDS\_136.0\_-0.5  & J090553.214-000319.05 & 136.4717 & -0.0553  & 0.46 &      &      & G    &                         \\
25 & KIDS\_140.0\_-1.5  & J092134.931-010343.72 & 140.3955 & -1.0621  & 0.54 & 0.01 & 0.36 & G    &                         \\
26 & KIDS\_140.4\_2.5   & J092208.588+021357.51 & 140.5358 & 2.2326   & 0.45 & 0.35 & 0.4  & G    &                         \\
27 & KIDS\_140.4\_2.5   & J092020.548+025412.84 & 140.0856 & 2.9036   & 0.33 & 0.03 & 0.28 & G    &                         \\
28 & KIDS\_140.4\_2.5   & J092136.654+021409.44 & 140.4027 & 2.236    & 0.45 & 0.22 & 0.32 & G    & SuGOHI (B), Pan-STARRS      \\
29 & KIDS\_174.5\_-2.5  & J113741.606-020559.02 & 174.4234 & -2.0997  & 0.32 & 0.63 & 0.22 & G    &                         \\
30 & KIDS\_175.0\_0.5   & J114113.002+001610.81 & 175.3042 & 0.2697   & 0.6  & 0.48 & 0.38 & G    &                         \\
31 & KIDS\_176.0\_-1.5  & J114329.639-014429.98 & 175.8735 & -1.7417  & 0.2  &      &      & G    & HQ LinKS, SuGOHI (A), SLACS \\
32 & KIDS\_176.0\_-1.5  & J114305.505-012316.27 & 175.7729 & -1.3879  & 0.3  & 0.26 & 0.42 & G    &                         \\
33 & KIDS\_176.0\_0.5   & J114444.815+001346.70 & 176.1867 & 0.2296   & 0.59 &      &      & G    & HQ LinKS, SuGOHI (C)        \\
34 & KIDS\_176.5\_-2.5  & J114654.718-021707.62 & 176.728  & -2.2855  & 0.48 & 0.14 & 0.4  & G    &                         \\
35 & KIDS\_177.0\_-1.5  & J114815.363-010907.79 & 177.064  & -1.1522  & 0.46 & 0.8  & 0.34 & G    &                         \\
36 & KIDS\_178.0\_-1.5  & J115015.646-015752.32 & 177.5652 & -1.9645  & 0.42 & 0.68 & 0.29 & G    &                         \\
37 & KIDS\_178.0\_-1.5  & J115008.624-010948.48 & 177.5359 & -1.1635  & 0.66 &      &      & G    &                         \\
38 & KIDS\_178.0\_0.5  & J115252.263+004733.11 & 178.2178 & 0.7925  & 0.49 & 0.14 & 0.47 & G & HQ LinKS, SuGOHI (A), HSC \\
39 & KIDS\_179.0\_1.5   & J115729.592+012844.07 & 179.3733 & 1.4789   & 0.43 & 0.75 & 0.27 & G    &                         \\
40 & KIDS\_180.0\_-0.5  & J120139.878-001226.26 & 180.4162 & -0.2073  & 0.32 &      &      & G    & HQ KiDS, SuGOHI (B)         \\
41 & KIDS\_182.0\_-0.5  & J120917.360-003415.91 & 182.3223 & -0.5711  & 0.41 & 0.02 & 0.39 & G    &                         \\
42 & KIDS\_182.0\_-1.5  & J120743.259-010355.44 & 181.9302 & -1.0654  & 0.27 &      &      & G    & SuGOHI (B)                  \\
43 & KIDS\_182.0\_1.5   & J120650.154+011816.95 & 181.709  & 1.3047   & 0.57 &      &      & G    &                         \\
44 & KIDS\_183.0\_-0.5  & J121105.273-004905.30 & 182.772  & -0.8181  & 0.62 & 0.71 & 0.75 & G    & SuGOHI                  \\
45 & KIDS\_183.0\_1.5   & J121215.544+015448.72 & 183.0648 & 1.9135   & 0.46 & 0.12 & 0.4  & G    & HQ LinKS, SuGOHI (C)        \\
46 & KIDS\_183.0\_1.5   & J121152.582+011119.99 & 182.9691 & 1.1889   & 0.5  & 0.25 & 0.36 & G    &                         \\
47 & KIDS\_183.0\_1.5   & J121317.558+015311.22 & 183.3232 & 1.8865   & 0.48 & 0.22 & 0.4  & G    & LinKS, SuGOHI (C)          \\
48 & KIDS\_184.0\_-0.5  & J121424.037-003206.32 & 183.6002 & -0.5351  & 0.35 &      &      & G    & HQ LinKS                \\
49 & KIDS\_185.0\_-0.5  & J122015.955-001401.72 & 185.0665 & -0.2338  & 0.36 & 0.3  & 0.33 & G    &                         \\
50 & KIDS\_212.0\_-1.5  & J140929.697-011410.82 & 212.3737 & -1.2363  & 0.76 &      &      & G    & SuGOHI (A)                 \\
51 & KIDS\_214.0\_-0.5  & J141503.117-003105.93 & 213.763  & -0.5183  & 0.56 & 0.01 & 0.51 & G    & SuGOHI (C)                  \\
52 & KIDS\_214.0\_1.5   & J141649.819+013822.23 & 214.2076 & 1.6395   & 0.58 & 0.77 & 0.43 & G    & SuGOHI (A)                 \\
53 & KIDS\_215.6\_2.5   & J142116.381+023741.67 & 215.3183 & 2.6282   & 0.41 &      &      & G    &                         \\
54 & KIDS\_216.0\_-0.5  & J142332.130-003115.73 & 215.8839 & -0.521   & 0.49 & 0.02 & 0.36 & G    &                         \\
55 & KIDS\_216.0\_0.5   & J142541.213+002216.85 & 216.4217 & 0.3713   & 0.31 & 0.66 & 0.23 & G    &                         \\
56 & KIDS\_216.0\_0.5   & J142557.674+005618.75 & 216.4903 & 0.9385   & 0.61 &      &      & G    & SuGOHI (C)                  \\
57 & KIDS\_216.0\_1.5   & J142353.070+013446.95 & 215.9711 & 1.5797   & 0.52 &      &      & G    & SuGOHI (B)                  \\
58 & KIDS\_218.6\_2.5   & J143519.027+023214.40 & 218.8293 & 2.5373   & 0.17 &      &      & G    &                         \\
59 & KIDS\_220.0\_-0.5  & J144133.015-005404.08 & 220.3876 & -0.9011  & 0.75 &      &      & G    & SuGOHI (A), SGAGS           \\
60 & KIDS\_221.0\_0.5   & J144515.550+004133.42 & 221.3148 & 0.6926   & 0.45 &      &      & G    &                         \\
61 & KIDS\_222.0\_-0.5  & J144723.506-001207.74 & 221.8479 & -0.2022  & 0.66 &      &      & G    &                         \\
62 & KIDS\_222.0\_0.5   & J144950.700+005536.65 & 222.4613 & 0.9268   & 0.5  & 0.23 & 0.42 & G    & HQ KiDS, SuGOHI (C)        \\
63 & KIDS\_222.0\_0.5   & J144930.567+000324.77 & 222.3774 & 0.0569   & 0.34 & 0.81 & 0.39 & G    &                         \\
64 & KIDS\_345.6\_-34.1 & J230226.294-335637.56 & 345.6096 & -33.9438 & 0.62 & 0.18 & 0.32 & G    &                         \\
65 & KIDS\_346.2\_-30.2 & J230621.790-303040.05 & 346.5908 & -30.5111 & 0.78 &      &      & G    &                         \\
66 & KIDS\_348.0\_-34.1 & J231411.445-334054.60 & 348.5477 & -33.6818 & 0.64 &      &      & G    &                         \\
67 & KIDS\_348.1\_-33.1 & J231242.301-332318.44 & 348.1763 & -33.3885 & 0.69 & 0.81 & 0.55 & G    & HQ KiDS                 \\
68 & KIDS\_349.6\_-30.2 & J232041.173-301353.02 & 350.1716 & -30.2314 & 0.73 &      &      & G    &                         \\
69 & KIDS\_349.6\_-30.2 & J232011.140-294158.79 & 350.0464 & -29.6997 & 0.51 &      &      & G    & LinKS                   \\
70 & KIDS\_350.4\_-34.1 & J232007.419-340513.81 & 350.0309 & -34.0872 & 0.47 & 0.77 & 0.36 & G    & LinKS                    \\
\hline
\end{tabular}
\end{adjustbox}
\end{table*}

\section{Application on 221 deg$^2$ of KiDS DR4 }
\label{sec:220_Deg}

The overlapping region between the GAMA and KiDS surveys spans 221 deg$^2$, with the GAMA survey regions' coordinates provided in Tab. \ref{tab:ganma_survey}. We employ the RFT model to identify SGL in this portion of the sky. The preselection yields a target sample of 5\,538\,525 elements, from which RFT identifies ML{\footnotesize cand}=51\,626, the full list of these objects with relative KiDS ID, coordinates, label and prediction probability is in Tab. \ref{tab:all_checked}.
Throughout this section and the following, we use the term "TEGLIE" candidates or (TP$_{TEGLIE}$) to refer to the comprehensive list of all lenses graded 1 or 2. The subset of "HQ TEGLIE" candidates, denoted by the label 1, encompasses high-confidence SGL candidates. 

Upon visual examination, the final list of SGL candidates within the GAMA footprint includes 175 candidates, 56 of which are categorized as HQ SGLs. 
These HQ candidates, along with those discovered during model testing outside the GAMA region, are combined and presented in Tab. \ref{tab:all_candidates_1} and displayed in Fig. \ref{fig:all_candidates}. Henceforth, we will use the indices in the tables to reference specific grade 1 candidates.
The merged catalog consists of 264 lens candidates, including 71 HQ candidates. The less certain grade 2 candidates discovered within the GAMA footprint are depicted in the Appendix, illustrated in Fig. \ref{fig:grade_2_0} and listed in Tab. \ref{tab:all_candidates_2}.  

To ascertain which of our TEGLIE candidates represent novel discoveries, we cross-check them against catalogs from other survey searches: KiDS, HSC (SuGOHI), Pan-STARRS, and HST (SL2S). It is noteworthy that most SGL catalogs published in the literature comprise candidate lenses, with only a small proportion of these having been spectroscopically confirmed so far. Due to the size of our cutouts  (20 $\times$ 20 $arcsec^2$), we cross-match with other catalogs using a 10 $arcsec$ tolerance. If multiple objects are found, the closest one is considered.

In Tab. \ref{tab:all_candidates_1}, the ``crosscheck'' column indicates the surveys or searches in which these high-confidence candidates were identified. Table \ref{tab:crosscheck} presents the number of coinciding elements with the SGL searches that exhibit the highest overlap. The top row lists the names of the various searches, including TEGLIE. HQ KiDS refers to the high-quality SGL candidates proposed by the KiDS collaboration in \cite{links, Li_2020}, LiNKS encompasses all objects flagged as strong lens candidates by at least one visual inspector \cite{links}, and SuGOHI from \cite{Jaelani_2021_sugohi}. Within the SuGOHI survey, we also subdivide the candidates based on their grades: A indicates "definitely a lens,", B indicates "likely a lens", while C signifies "possibly a lens." The leftmost column defines the different grade subsets, with 1+2 constituting the sum of candidates with grades 1 and 2.

\begin{table*}[]
\centering
\caption{Summary of the cross-check results of this work candidates, with every number indicating the common candidates between our (Tab. \ref{tab:all_checked}) and other SGL candidates catalogs. The leftmost column represents the grade assigned by visual inspectors in the 221 tiles we evaluated, as described in Section \ref{sec:220_Deg}. With ``1+2'' we indicate the number of images with grade 1 or grade 2.
The top columns, instead, represent the other surveys that were examined. Only the surveys with the most frequent detections are included while TEGLIE is incorporated to present the number of elements with each grade. SuGOHI SGL candidates are divided into their grades from A to C, with A being the most confident one.  }
\label{tab:crosscheck}
\begin{tabular}{llllllll}
\cline{3-8}
\noalign{\vspace{2.5pt}}
\cline{3-8}
\noalign{\smallskip}
&& \multirow{2}{*}{This work}   & \multirow{2}{*}{HQ KiDS} & \multirow{2}{*}{LinKS}   & \multicolumn{3}{l}{SuGOHI} \\
\multicolumn{2}{l}{}  & &     &    & A       & B      & C       \\ \hline
\noalign{\smallskip}
\multirow{5}{*}{This work} & \multicolumn{1}{l|}{1} &56 & 13 & 21 & 8 & 5 & 8 \\  
&\multicolumn{1}{l|}{2} & 173 & 10  & 16  & 2 & 4 & 10 \\
&\multicolumn{1}{l|}{1+2}& 231  & 23 & 37 & 10 & 9 & 18  \\
&\multicolumn{1}{l|}{3} & 121 & 0& 1& 0   & 1 & 1    \\
&\multicolumn{1}{l|}{0} & 51275 & 5  & 316       & 11      & 33     & 111    \\ \hline
\end{tabular}
\end{table*}

Of the HQ KiDS SGL candidates, which were incorporated into our fine-tuning set, HQ$_{KiDS}$=38 fall within this portion of the sky. Of these HQ$_{KiDS}$ candidates, 10 were not identified by our model: 3 failed to meet our redshift cutoff, and 7  (shown in Fig. \ref{fig:check_kids}, image IDs from 5 to 11) did not exceed the prediction probability threshold, even though the model has been trained on these images.
Out of the 28 candidates that cleared the prediction probability threshold, 23 have been labeled as TP, while 5 were deemed FP.
The FP candidates are presented in Fig. \ref{fig:check_kids} (images from 0 to 4).
Images 0, 1, and 4 all exhibit an arc-like structure even though the arc in image 1 requires very close inspection to be noticed. These candidates have been overlooked by the visual inspector during the validation process. This demonstrates the challenges of labeling lenses, as both distraction and fatigue can lead to missed candidates.
Finally, candidate 3 displays diffuse emission but lacks distinct arcs.   

The LinKS dataset comprises TP$_{LinKS}$=1\,983, of which TP$_{LinKS, GAMA}$=503 fall within the GAMA survey area.  Out of these LinKS candidates, 10 are HQ$_{KiDS}$ and of the TP$_{LinKS, GAMA}$ candidates, 354 were also identified by our model (ML{\footnotesize cand}$_{TEGLIE}$), and 37 were graded as true positives, including 21 in high-confidence (HQ$_{TEGLIE}$) candidates. Of TP$_{LinKS}$ in HQ$_{KiDS}$ 5 have been tagged as FP.

The Hyper Suprime-Cam Subaru Strategic Program \citep[HSC SSP;][]{Miyazaki_2012} has also surveyed this portion of the sky, generating the SuGOHI catalog of 3\,057 SGL candidates \citep{Sonnenfeld_2017_sugohi, Chan_2020_sugohi, Jaelani_2020_sugohi, Sonnenfeld_2020_sugohi, Wong_2022_sugohi}. 
Among the around 50,000  ML{\footnotesize cand}$_{TEGLIE}$ candidates identified by our ML model, 194 were found in the SuGOHI catalog. Of these, 37 were confirmed as true positives, and among them, 21 were classified as high-confidence candidates. Notably, 111 of the grade 0 SGLs held a ``possibly a lens'' (grade C) classification in SuGOHI. This discrepancy suggests that our grading standards might have been stricter than those used for the HSC lenses. This could be due to factors like different grader expertise or specific selection criteria employed by each project. While most of the higher-grade SuGOHI lenses have been missed by our graders due to the superior resolution and depth of the HSC data.

Furthermore, two candidates have been located and suggested as SGL candidates by the Canada-France-Hawaii Telescope Legacy Survey-Strong legacy survey \citep[SL2S;][]{More_2012}. However, only object 21 (SDSSJ1143-0144) has been labeled as HQ true positive, while SDSSJ0912+0029, not showing clear lensing structures, has been labeled as a false positive.

Three additional lenses (Pan-STARRS ID: PS1J0846-0149, PS1J0921+0214, PS1J1422+0209) in ML{\footnotesize cand}$_{TEGLIE}$ have been found in the Pan-STARRS candidates sample \citep{Canameras_2020} . However, only PS1J0921+0214 has been labeled as SGL candidate. Upon visual inspection, no clear double images or arc-like structures were found in the images of the missed candidates.

Finally, each candidate in Tab. \ref{tab:all_candidates_1} underwent inspection on the SIMBAD astronomical database \citep{Wenger_2000_simbad}. This check led to the exclusion of one candidate (object 8) due to the lensed object classification as a white dwarf. Candidate 59,  is a confirmed strong lens and part of the SDSS Giant Arcs Sample \citep[SGAGS; presented in][]{Hennawi_2008,Bayliss_2011} with ID SGAS J144133.2-005401 \citep{Rigby_2014}. Candidate 38, is a confirmed quadruple-lensed source, initially discovered with the Hyper Suprime-Cam (HSC) Survey \citep{More_2017}. 

Of the grade 2 objects, 23 have been already located either in SuGOHI or in the KiDS searches.

In summary, our model identified a total of 264 TP$_{TEGLIE}$, among which 71 HQ candidates with 27 re-discoveries, yielding 44 new HQ candidates discovered exclusively by our model.

\begin{figure*}[!htbp]
\includegraphics[width=\textwidth]{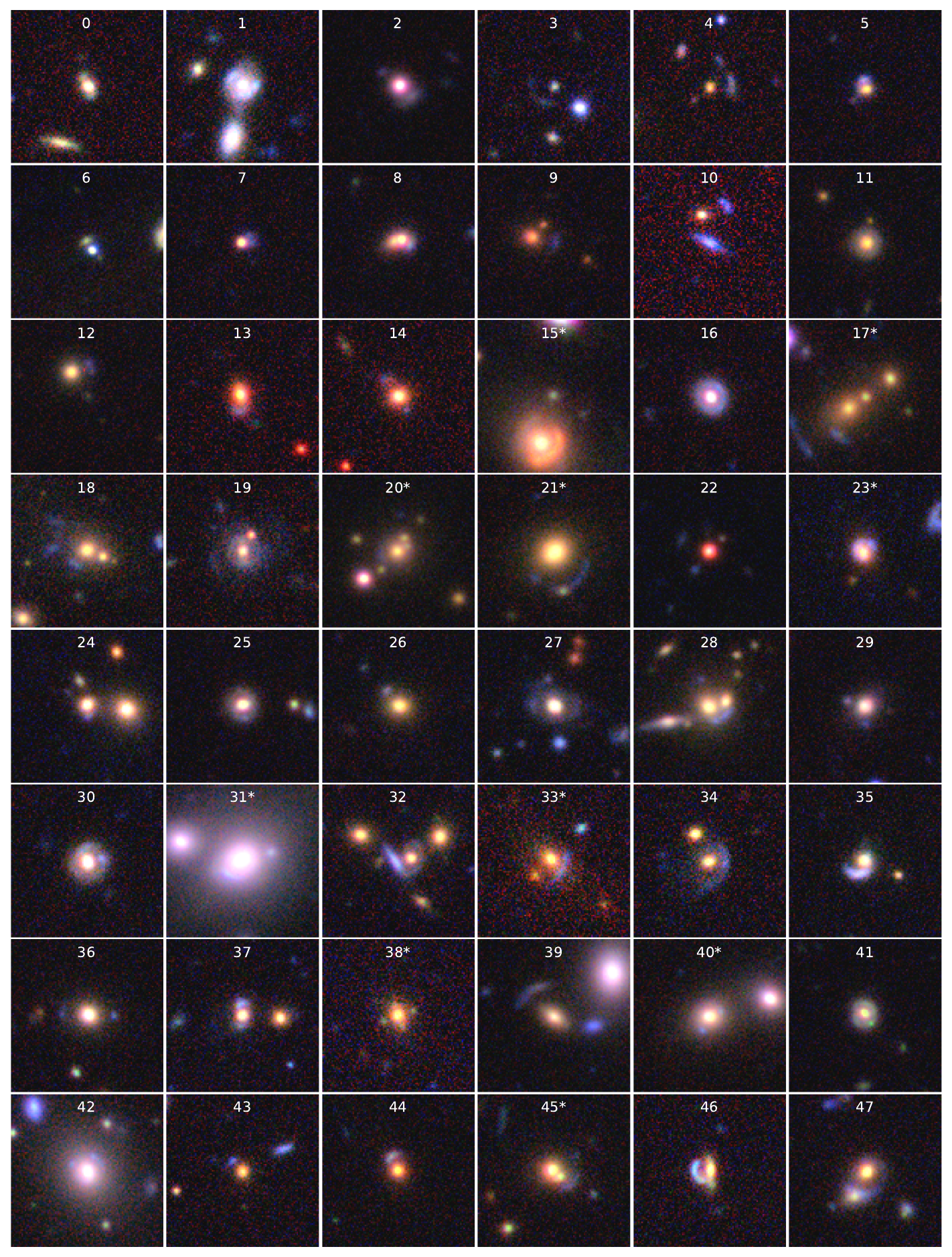}
\centering
\caption{First 48 high-confidence SGL candidates found in this work. The numbers on the top of every cutout correspond to the ID of the object in Table \ref{tab:all_candidates_1}. Images marked with an asterisk are high-quality SGL candidates already identified by the KiDS collaboration.}
\label{fig:all_candidates}
\end{figure*}
\addtocounter{figure}{-1}
\begin{figure*}[!htbp]
\includegraphics[width=\textwidth]{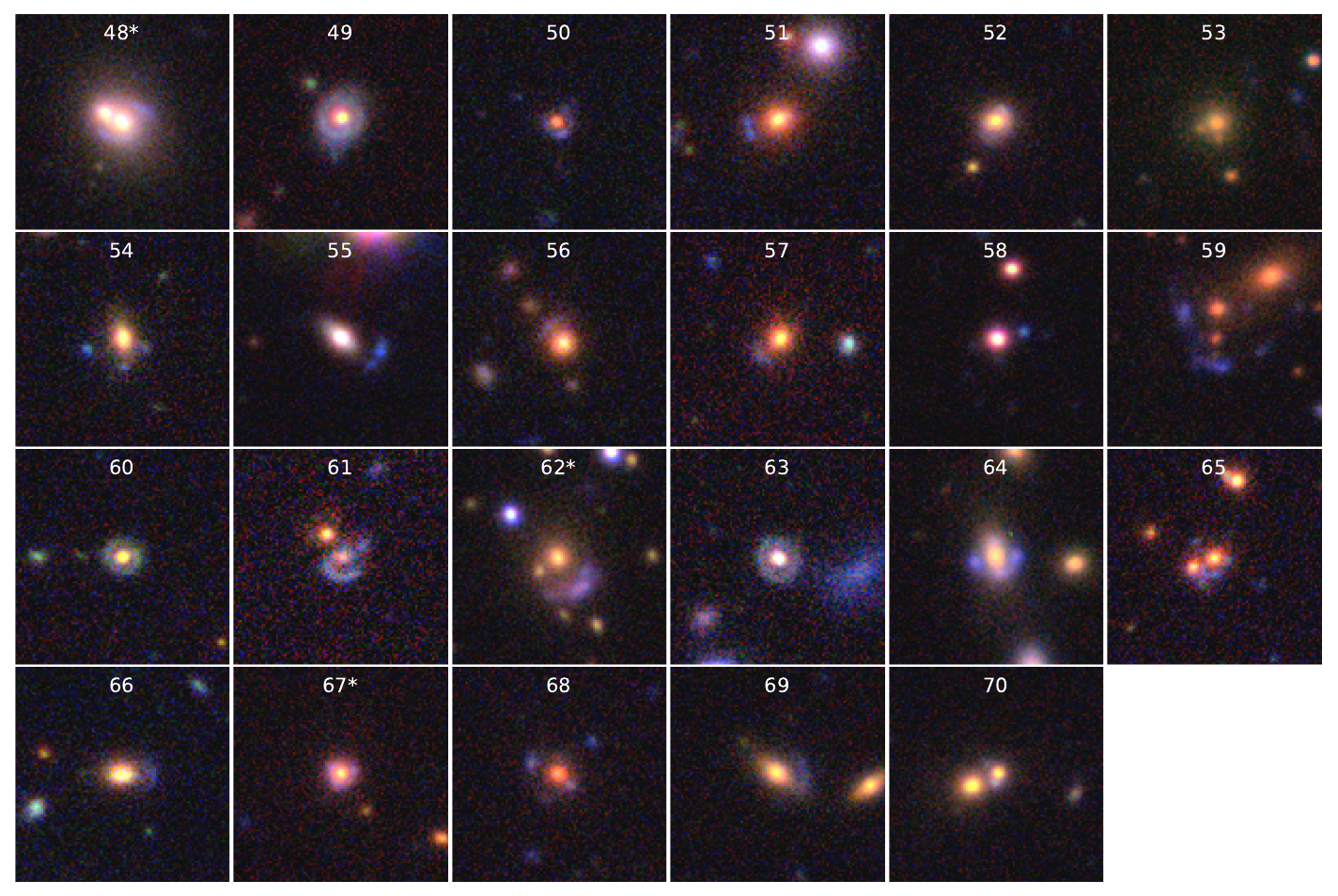}
\centering
\caption{(Continued) - remaining 20 high-confidence SGL candidates found in this work.}
\label{fig:all_candidates_1}
\end{figure*}

\begin{figure*}[!htbp]
\includegraphics[width=\textwidth]{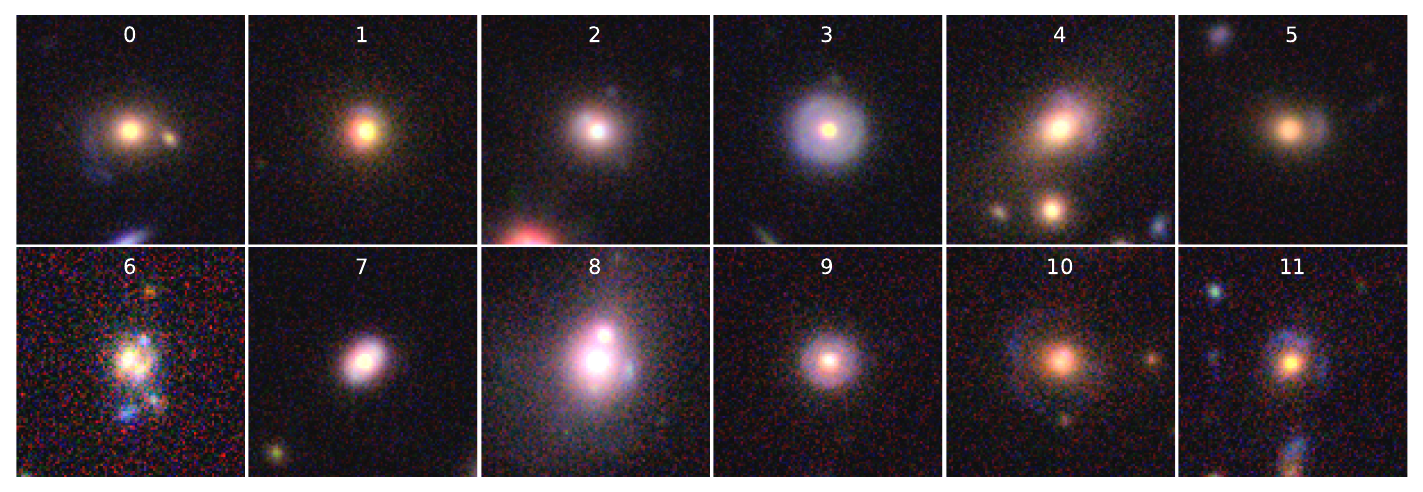}
\centering
\caption{
\textit{From 0 to 4:}
Strong gravitational lens candidates found in previous KiDS automated searches \citep{links,Li_2020} proposed as SGL candidates by our model but rejected by the visual inspectors.
\textit{From 5 to 11:} HQ SGL found in previous KiDS automated searches not proposed as SGL candidates by our model even though they were in the training set.}
\label{fig:check_kids}
\end{figure*}

\begin{figure}[!htbp]
\includegraphics[width=8cm]{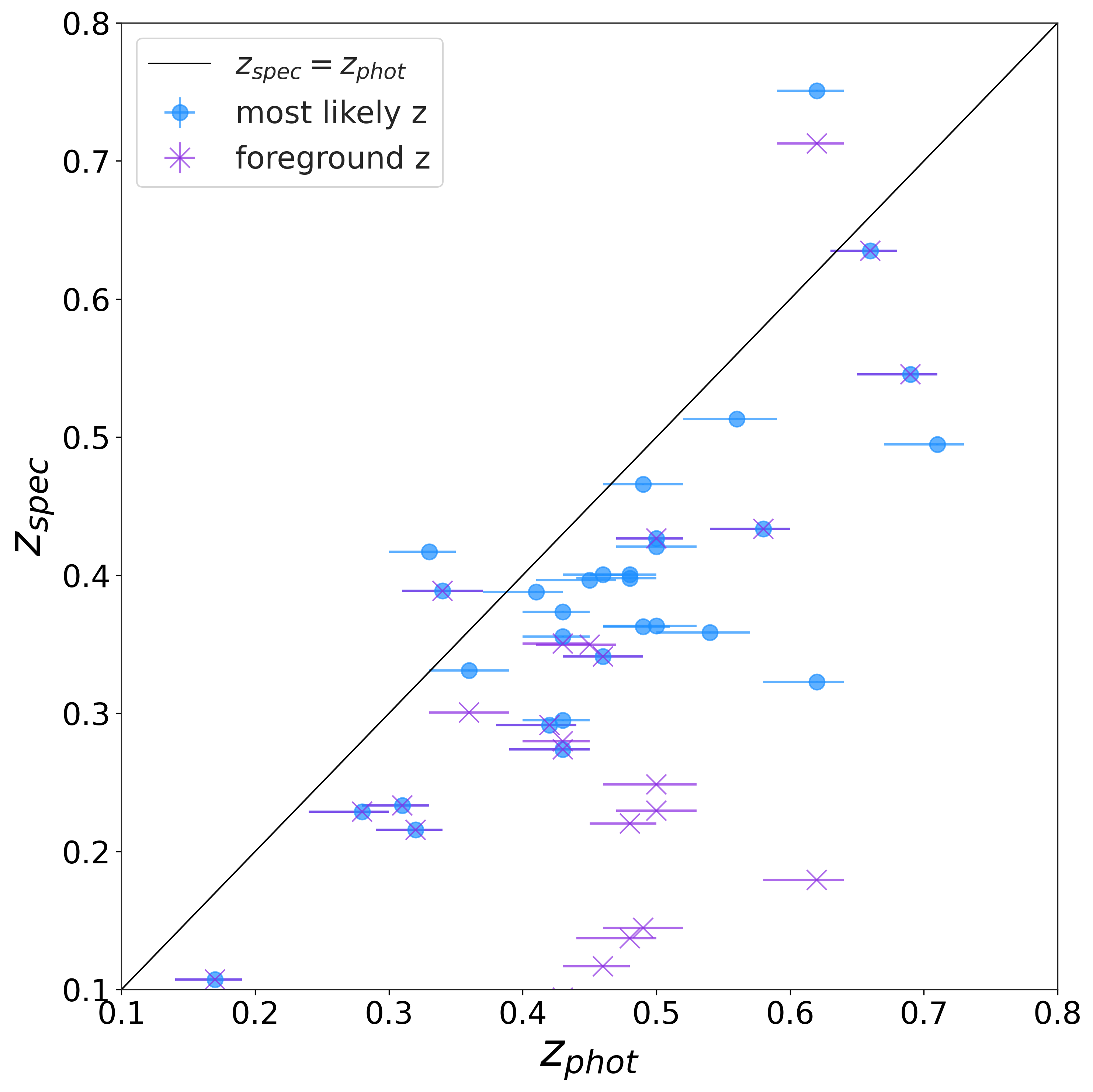}
\centering
\caption{Comparison of the spectroscopic redshifts ($z_{spec}$) from the GAMA survey with the photometric redshifts ($z_{phot}$) derived from the nearest neighbor search in the KiDS survey of the 31 grade 1 galaxies having $R\geq1.2$ (Eq.~\ref{eq:R}). For objects in GAMA, we utilize the redshift from the \code{AATSpecAutozAllv27\_DR4} catalog. 
The purple crosses represent the KiDS $z_{phot}$ vs the lower $z_{spec}$ between the two redshift estimates coming from the two highest correlation peaks from the \code{AUTOZ} output.
The blue circles illustrate the  KiDS $z_{phot}$ vs the $z_{spec}$ 'best-fitting' redshift (highest cross-correlation) from \code{AUTOZ}. 
The black solid line represents a perfect one-to-one correspondence between the two redshift measurements.}
\label{fig:zspec_sphot} 
\end{figure}

\begin{figure}[!htbp]
\includegraphics[width=\columnwidth]{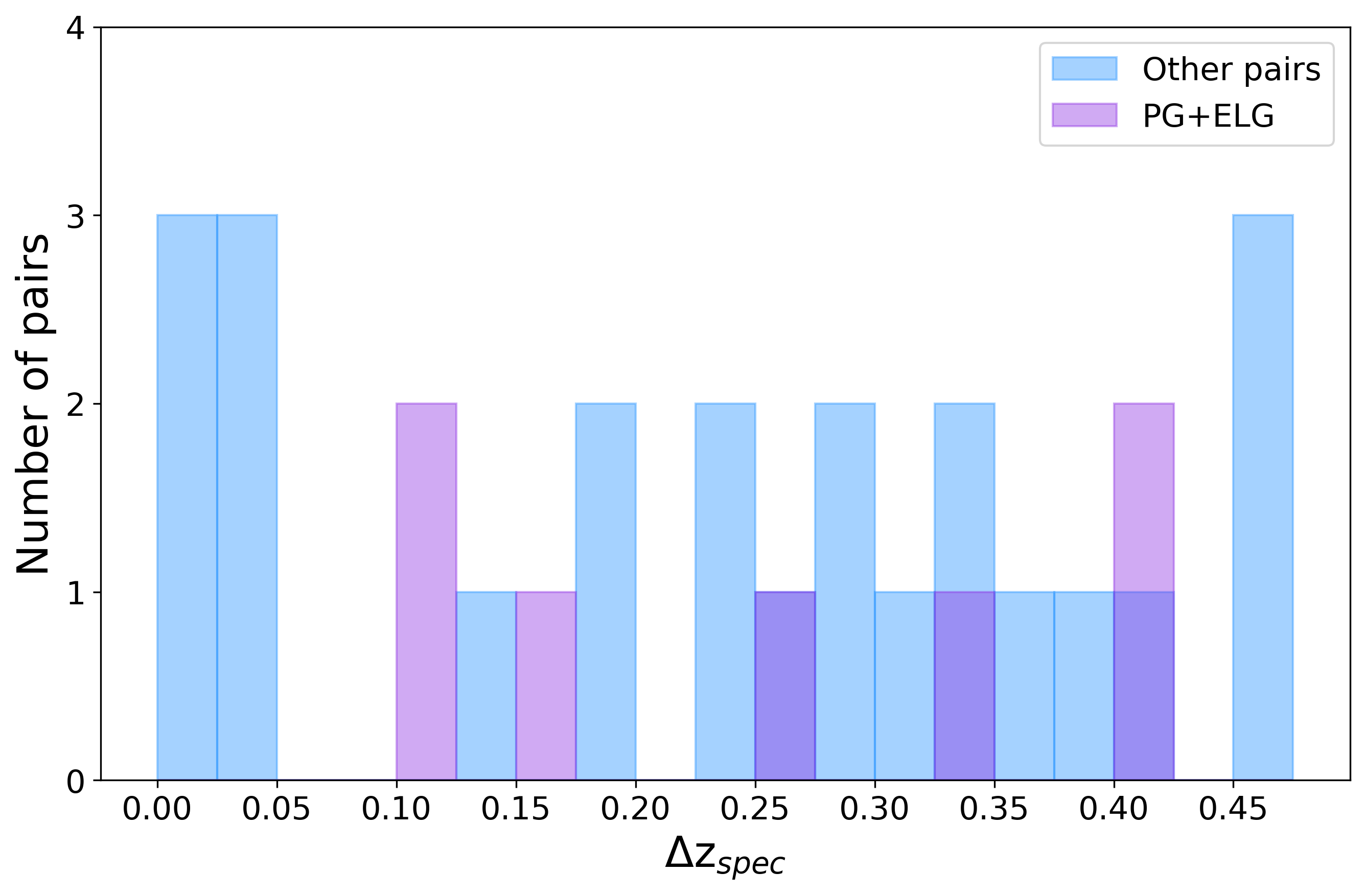}
\centering
\caption{Absolute difference in redshift between the foreground and background galaxies of the 31 grade 1 candidates with a ratio R (Eq. \ref{eq:R}) greater than 1.2.
Purple blocks represent the passive galaxy (PG) + emitting line galaxies (ELG) (background + foreground) and blue blocks include all the other galaxy pairs (PG+PG, ELG+ELG, ELG+PG). }
\label{fig:delta_zspec_zphot}
\end{figure}

\section{Cross-Matching with GAMA}

The sky region investigated in this study completely overlaps with the GAMA fields, allowing us to search for spectroscopic counterparts.
We perform the nearest neighbor matching by right ascension and declination within a 3-arcsec positional tolerance between the GAMA DR4 catalog and the candidates found in this work. Out of the 264 TEGLIE candidates, we find 144 matches, with 35 of these being high-confidence candidates.  
Considering that a 3 arcsec positional tolerance might be too broad for some objects, we implemented an additional verification step. We compare the \code{FLUX\_RADIUS} values from the SExtractor run of each object in the KiDS catalog. If the separation between the KiDS and GAMA positions is less than the \code{FLUX\_RADIUS} converted in arcseconds, we consider the crossmatch as reliable. Twelve sources do not pass this test (8 grade 2 and 4 grade 1).

Using the \code{AATSpecAutozAllv27\_DR4} catalog, as detailed in Sect. \ref{sec:gama}, we investigate the TEGLIE candidates spectra. Among the 101 grade 2 candidates with reliable GAMA counterparts, 7 (ID in Tab. \ref{tab:all_candidates_2}: 9,60,111,112,150,153,186) have either the first or second best-fitting template of a star, while all grade 1 images have galaxy templates.
To identify double redshift candidates, we apply the R (Eq. \ref{eq:R}) selection criteria outlined by \cite{Holwerda_2015} and \citep{Knabel2023} to the grade 1 candidates. We designate candidates with R$\geq$1.85, resulting in the identification of 26 grade 1 candidates. Relaxing the threshold to R$\geq$1.2 increases the number of selected candidates to 31.  Among the 31 selected candidates, 13 are ELG+PG, 8 ELG+ELG, 7 PG+ELG, and 3 PG+PG pairs. 

We include the spectroscopic redshifts of the first ($z_{spec}$) and second  ($z_{spec,2}$) correlation peaks of the grade 1 candidates in Tab. \ref{tab:all_candidates_1}. 
Figure \ref{fig:zspec_sphot} compares the lens KiDS photometric redshifts with the GAMA spectroscopic redshifts for the "high-confidence" double redshift candidates (selected with R $>$ 1.2). The error bars represent the minimum and maximum redshift ranges provided in the KiDS catalog. The blue circles represent the "most likely" redshift derived from spectral fitting for each source.
The purple crosses depict an alternative spec-$z$ value, which corresponds to the lower redshift among the two obtained from the two highest correlation peaks identified in the spectra. 
The figure suggests a possible overestimation of the photo-z values compared to the spec-z, particularly for some sources. This discrepancy could be attributed to the close proximity of the lensing galaxy (central source) to other objects in the field, potentially corresponding to multiple images. Such complex environments can pose challenges for accurate photometric redshift estimation algorithms. {Whether this is a systematic effect or specific to our sample requires further investigation.

\citet{Holwerda_2015} considered pairs with passive galaxies as the foreground object and emission-line galaxies as the background object as potential SGL candidates. Our sample includes seven such PG+ELG configurations, identified with IDs 13, 29, 30, 46, 52, 55 and 67 in Tab. \ref{tab:all_candidates_1}.

A cross-match (within a 3 arcsec radius) between the \citet{Holwerda_2015} sample and the ML candidates (Tab. \ref{tab:all_checked}) revealed 64 overlapping. Among these, 22 have PG+ELG galaxy templates and hence are considered as SGL candidates by \citet{Holwerda_2015}.  Of these 22 double-z SGL candidates, 21 have been labeled as non-lenses during our visual inspection (shown in Fig. \ref{fig:pg_elg_0}) and 1 as grade 3 (ID 20 in Tab. \ref{tab:grade_3}). Inspecting Fig. \ref{fig:pg_elg_0}, there is no object showing clear lensing features, the objects appear to be serendipitous overlaps of two galaxies rather than gravitationally lensed systems.
This highlights the limitations of the blended spectra method for definitive SGL identification. While effective for selecting candidate galaxy pairs, follow-up observations with either high-resolution imaging or further spectroscopical data are crucial for confirmation.   
This has already been noted by \citet{Chan_2016}. 
In their work, only 6 out of 14 PG+ELG candidates from the \citet{Holwerda_2015} sample within the Hyper Suprime-Cam field were confirmed as probable lenses, suggesting a success rate of approximately 50\% for the blended spectra approach.  However,  confirmation for the \citet{Chan_2016} candidates themselves is still required to solidify this success rate estimate.

\citet{Knabel_2020} further refined the selection criteria established by \citet{Holwerda_2015} by requiring an absolute difference in the spectroscopic redshifts of the first and second correlation peaks ($\Delta z$) to be greater than 0.1. Figure \ref{fig:delta_zspec_zphot} shows the distribution of $\Delta z$ for PG+ELG pairs (purple) and all other source combinations (blue). As expected, all PG+ELG pairs exhibit $\Delta z>$0.1, making them stronger SGL candidates. The source combinations with $\Delta z$ < 0.1 include 3 ELG+PG, 3 ELG+ELG, and 1 PG+PG pair.

\citet{Knabel2023} pre-selected sources with foreground redshifts exceeding 0.05. All configurations in our sample satisfy this criterion except for 3 ELG+ELG and 4 ELG+PG pairs, leaving us with 25 potential SGL candidates.

Due to the limited number of source matches between GAMA and KiDS surveys, subsequent sections will rely on photometric redshifts for characterizing the properties of the candidates. Additionally, spectroscopic redshifts are unavailable for some of the candidate double redshifts identified in previous studies by \citep{links, Li_2020, Li_2021}. To ensure consistency when comparing our results with these studies, we will employ the KiDS photometric redshifts for further analysis.

\section{Properties of the candidates}
\label{sec:candidates_propriety}

In this section, we delve into the characteristics of the 71 HQ$_{TEGLIE}$ included in Tab. \ref{tab:all_candidates_1}, excluding the one classified as a star in the SIMBAD database.

Figure \ref{fig:r_auto_z} presents photometric redshift vs. $r$-band magnitude and the distribution of two quantities, as in \cite{Li_2020}, for the HQ SGL candidates identified by KiDS collaboration and in this work. We denote $\widetilde{x}$ as the median of variable x.
The TEGLIE candidates have an $r$-band magnitude that is slightly fainter than that of the SGL candidates found by the KiDS collaboration, with $\widetilde{r}_{TEGLIE} = 19.33$ mag compared to $\widetilde{r}_{KiDS} = 18.66$ mag. Whereas the redshift median is similar for both datasets, $\widetilde{z}_{TEGLIE} = 0.46$ in this work and $\widetilde{z}_{KiDS} = 0.45$ for KiDS.
This suggests that the TEGLIE candidates are slightly less luminous but share a similar redshift distribution, albeit with a less pronounced peak, to the KiDS sample. Moreover, the redshift distribution of KiDS confirms that selecting a redshift range from 0 to 0.8 was a safe decision, as most of the lenses fall within this range.
The lenses identified by KiDS exhibit a correlation between redshift and magnitude, likely attributable to the preselection of bright galaxies. In contrast, the TEGLIE candidates appear more dispersed across the redshift-magnitude space, possibly due to the absence of such a preselection.

In Figure \ref{fig:color_z}, we present the observer-frame $g-r$ color in relation to the redshift of TEGLIE and KiDS HQ candidates. As mentioned in Sect. \ref{sec:prev_ml}, the KiDS collaboration preselected a sample of BG and a subsample of LRG. 
This color cut resulted in the selection of KiDS lenses with red colors, ${g-r}_{KiDS}$ $\sim$ 0.8 mag at $z$ $\sim$ 0 and $g-r _{KiDS}$ $\sim$ 1.6 at $z$ $\sim$ 0.5. In our case, without any color cut, we still obtain similar results to the KiDS sample. The influence of this color cut is further investigated in a color-color plot in Fig. \ref{fig:color_color}.  
All sources passing the BG preselection are plotted as red crosses. The LRG selection is determined by an additional magnitude cut and two inequalities as per Eq. \ref{eq:color_cut}.
The solid pink line represents the boundaries of the lower inequality in Eq. \ref{eq:color_cut}, while the pink area indicates the region where the inequality holds. The top inequality in Eq. \ref{eq:color_cut} depends on the $r$-band magnitude of each object - all candidates satisfying this inequality are plotted as red squares in Fig. \ref{fig:color_color}. As a result, all squares within the pink area of the plot are considered LRGs according to the KiDS formulation. On the other hand, all objects (crosses and squares) outside the pink stripe are thus BGs. Most of the TEGLIE candidates (blue-filled circles) fall within the color cut $c_{perp}$ (pink block) but not all meet the BG preselection criteria. 
We highlight these ``not BG" candidates with a wider blue circumference. Of the HQ TEGLIE candidates, 57 are BGs based on the criteria outlined in Sect. \ref{sec:prev_ml}. Specifically, 7 candidates have a magnitude greater than 21, and 4 have \code{SG2DPHOT} values equal to 4. These latter candidates (ID 2, 7, 22, 58) exhibit a small separation between the two sources and are therefore classified as potential stars, although none of them are tagged as such in SIMBAD.

\begin{figure}[!htbp]
\includegraphics[width=\columnwidth]{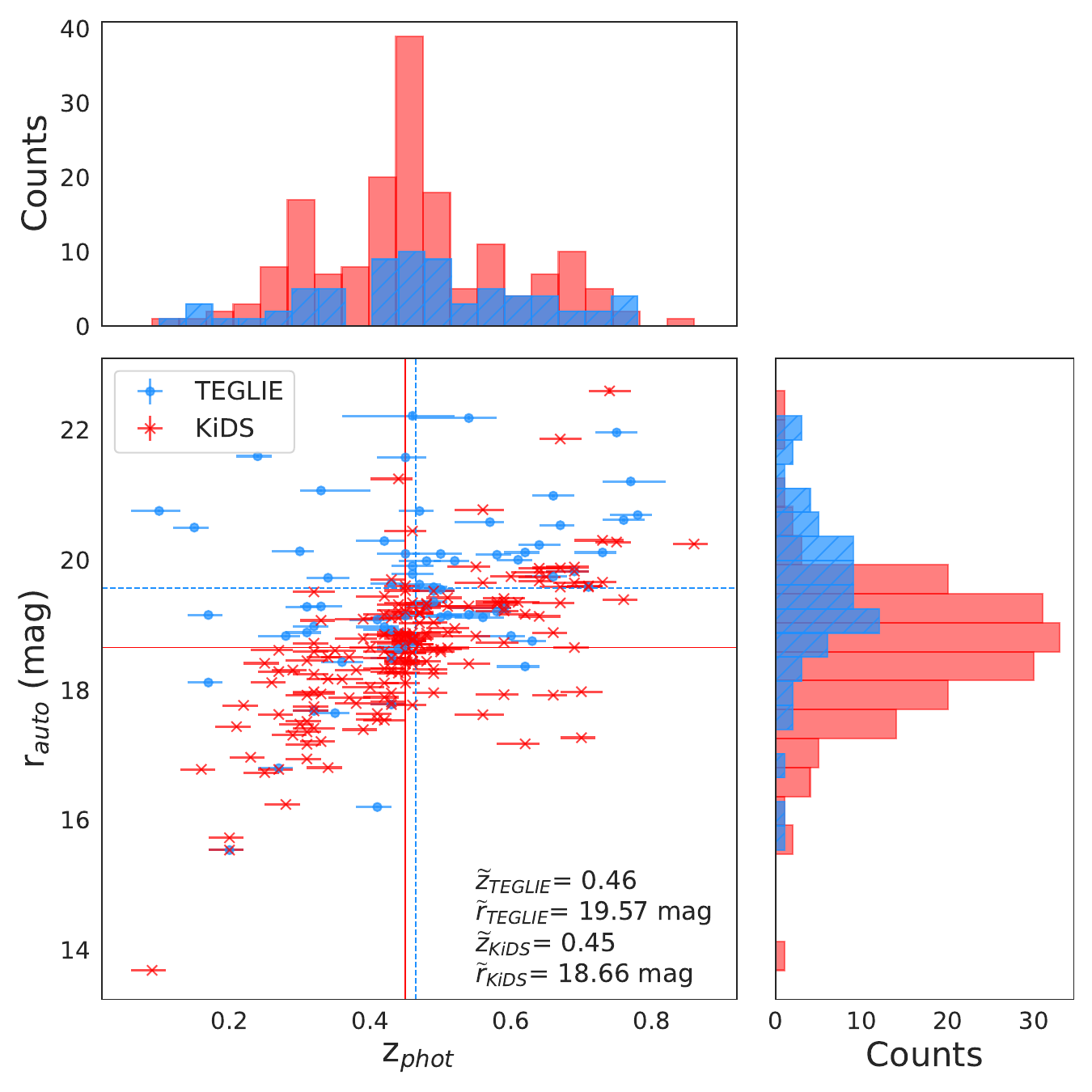}
\centering
\caption{Photometric redshift vs $r$-band magnitude space.  Red crosses represent lenses found by the KiDS collaboration, with error bars indicating the range between the predicted maximal and minimal estimated redshift given by the \code{BPZ} code. Histograms of redshift and magnitude, for both populations, are plotted along each axis to better visualize the distribution of values. Values of the median magnitude and redshift for both samples are represented by dashed lines (TEGLIE), and solid lines (KiDS) and reported in the plot. }
\label{fig:r_auto_z}
\end{figure}

\begin{figure}[!htbp]
\includegraphics[width=\columnwidth]{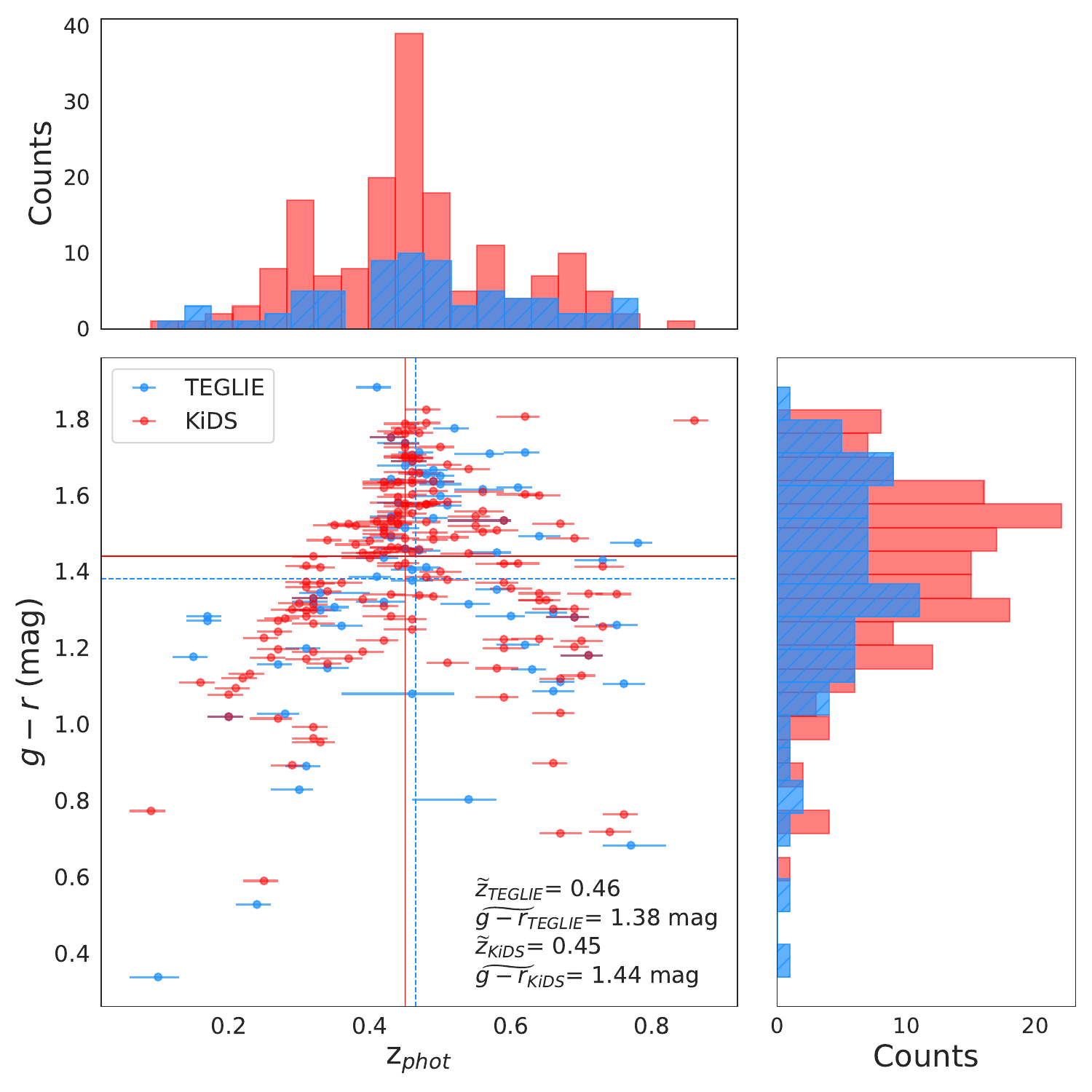}
\centering
\caption{The $g-r$ observer-frame color corrected for Galactic extinction vs photometric redshift. Red crosses represent lenses found by the KiDS survey, with error bars indicating the range between the predicted maximum and minimum estimated redshift. Blue dots represent newly found high-quality candidates.
Histograms of redshift and color, for both populations, are plotted along each axis to better visualize the distribution of values. Values of the median color and estimated redshift for both populations are represented by dashed lines (TEGLIE), solid lines (KiDS) and reported in the plot. }
\label{fig:color_z}
\end{figure}

\begin{figure}[!htbp]
\includegraphics[width=\columnwidth]{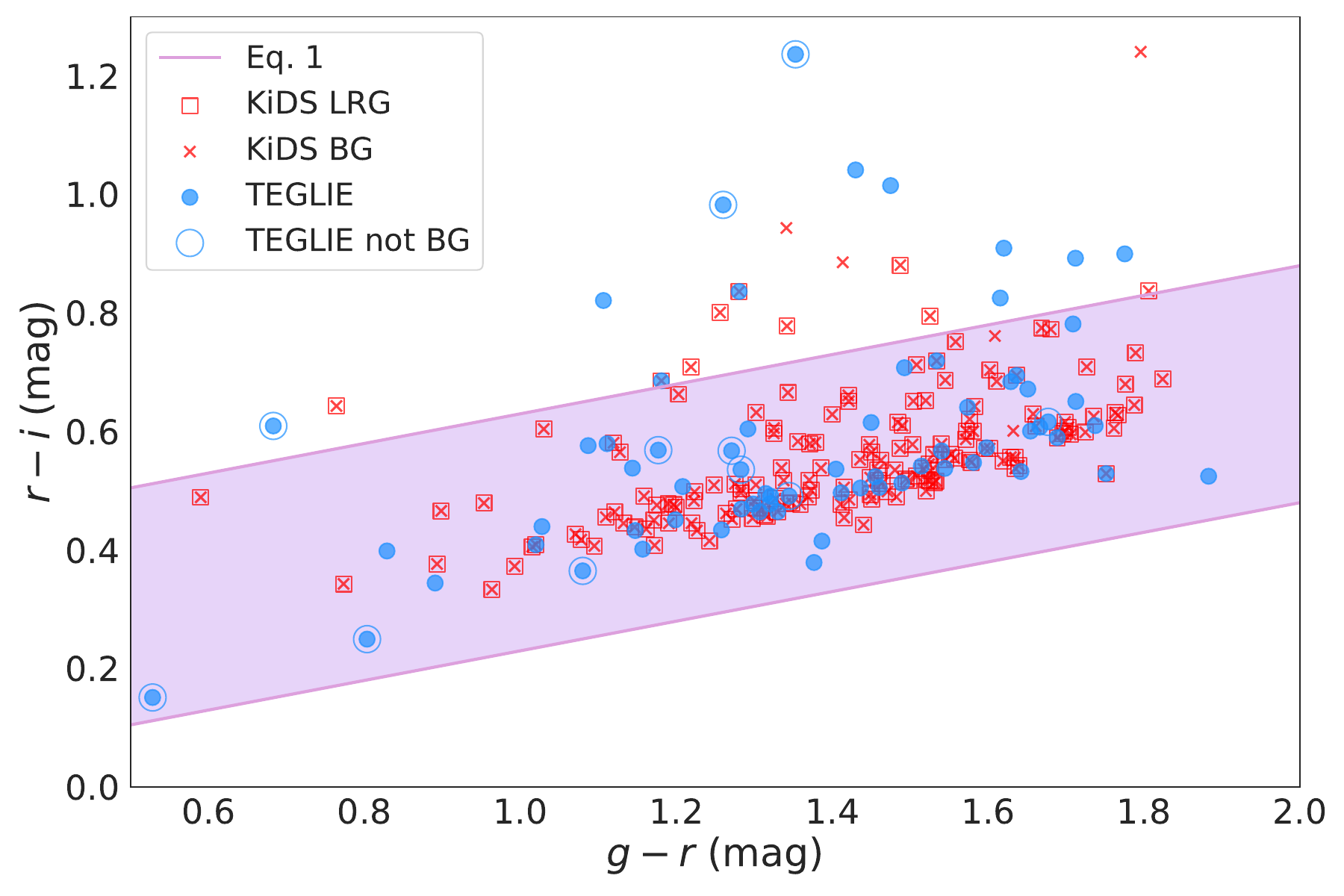}
\centering 
\caption{Visualization of the KiDS selection criteria for identifying LRGs (Luminous Red Galaxies) and BGs (Bright Galaxies) through a color cut, as described in Sect. \ref{sec:prev_ml}. Our model was applied to a sample without this specific color selection, allowing it to detect candidates that would have otherwise been excluded by the KiDS LRG/BG cut.
The blue dots with a circle around them are all the TEGLIE HQ (grade 1) candidates that would have been missed with the KiDS preselection.
A solid pink line limits the boundaries of the bottom inequality in Eq. \ref{eq:color_cut}. The red crosses are the HQ KiDS lenses that satisfy the top inequality in Eq. \ref{eq:color_cut}; the red crosses represent all the objects passing the BG selection, as described in Sect. \ref{sec:prev_ml}. Blue dots are the TEGLIE candidate while the blue circles are all the HQ TEGLIE that would have not been passed the KiDS color cut.   }
\label{fig:color_color}
\end{figure}

\section{Discussion}
\label{sec:discussion}

\subsection{Comparison of models' performance}

\begin{table*}[]
\centering
\caption{ML SGL Searches in KiDS Survey (details in Sect. \ref{sec:prev_ml}). Column 2 specifies the bands given to the ML model, column 3 describes the type of sample of galaxies (BG = bright galaxies, LRG = luminous red galaxies, $z<$0.8 is the redshift cut at 0.8) to create the column 5 given to the ML model.
Column 4 is the prediction probability threshold used to divide the lens candidates class from the non-lens class.  Column 6 is the number of SGL candidates identified by each model, with column 7 the percentage of candidates in relation to the sample size. 
TP (Column 8) is the number of confirmed candidates within parenthesis the number of high-confidence ones, and FP (Column 9) is the number of false positives with $\%$ FP (Column 10) the percentage of false positives in the ML candidates sample.
Finally, column 11 displays the precision ($\%$ P, Eq. \ref{eq:P}) achieved by each approach. Note that all data for this work is based on the 221 deg$^2$ search described in Sect. \ref{sec:220_Deg}.}
\label{tab:tp_fp}
\begin{tabular}{lllllllllll}
\hline\hline  \noalign{\smallskip}
1   & 2 & 3    & 4 & 5  & 6 & 7 & 8  & 9   & 10  & 11 \\
Lens Search   & band(s) & Preselection    & $p$& Sample  & ML$_{cand.}$ & \% ML$_{cand.}$ & TP (HQ)  & FP    & \% FP  & \% P  \\ \hline  \noalign{\smallskip}
Petrillo+2017 & $r$       & LRG             & 0.8         & 21\,789   & 761      & 3.49        & 56       & 705   & 92.64 & 7.36  \\
Petrillo+2018 & $r$ $i$ $g$   & LRG             & 0.8         & 21\,789   & 390      & 1.79        & 60       & 330   & 84.61 & 15.38 \\
Petrillo+2019 &$r$ $i$ $g$   & LRG             & 0.8         & 88\,327   & 1\,689     & 1.91        & 89       & 1\,600  & 94.73 & 5.27  \\
Li+2020       & $r$       & BG              & 0.98        & 3\,808\,963 & 3\,552     & 0.09        & 133      & 3\,419  & 96.26 & 3.74  \\
Li+2020       & $r$ $i$ $g$   & LRG             & 0.75        & 126\,884  & 2\,848     & 2.24        & 153      & 2\,695  & 94.63 & 5.37  \\
Li+2021       &$r$ $i$ $g$   & BG              & 0.9         & 1\,432\,348 & 3\,740     & 0.26        & 295      & 3\,548  & 94.87 & 7.89  \\
Li+2021       &$r$ $i$ $g$   & LRG             & 0.8         & 72\,010   & 1\,299     & 1.80        & 192      & 1\,004  & 77.29 & 14.79 \\
This work        & $r$ $i$ $g$ $u$ & $z<0.8$      & 0.8         & 5\,538\,525 & 51\,638    & 0.93        & 231 (56) & 51\,396 & 99.55 &0.45  \\
This work        & $r$ $i$ $g$ $u$ & LRG             & 0.8         & 32\,709   & 13\,560    & 41.46       & 62 (17)  & 13\,498 & 99.88 & 0.46  \\
This work        & $r$ $i$ $g$ $u$ & BG              & 0.8         & 865\,022  & 40\,738    & 4.71        & 192 (52) & 40\,546 & 99.88 & 0.47  \\
This work        & $r$ $i$ $g$ $u$ & BG              & 0.90        & 865\,022  & 32\,858    & 3.80        & 158 (47) & 32\,700 & 99.86 & 0.48  \\
This work        & $r$ $i$ $g$ $u$ & BG              & 0.98        & 865\,022  & 19\,985    & 2.31        & 122 (38) & 19\,863 & 99.82 & 0.62  \\ \hline
\end{tabular}
\end{table*}

Table \ref{tab:tp_fp} summarizes the results of various SGL searches on the KiDS survey, including this work. For this comparison, we consider only the final search on the 221 deg$^2$ overlapping with GAMA.
For studies employing multiple CNN architectures \citep{links, Li_2021}, we report only the 3-bands ($r,i,g$) model results for a more accurate comparison with our 4-bands ($r,i,g,u$) TE. Both models leverage color information in addition to morphology, in contrast with the $r$-band one which considers only morphology. 

The differences in sample sizes between distinct searches - per same preprocessing (Preselection) - are due to differences in sky coverage. The quantity of SGL candidates found by the ML model (ML candidates) instead, is influenced by the prediction probability threshold ($p$) used to divide the SGL candidates class from the non-lens ones and by the sample size. In the same way, the FPs increase together with the ML candidates and sample size:  the more data is provided to the ML model the more chances of finding false positives.
To standardize performance comparison, we calculate the percentage of ML candidates relative to the entire sample (\% ML cand), the percentage of false positives within the ML candidates (\% FP)  long with the precision (P). These metrics provide an overall assessment of model performance.

While expert visual confirmation remains essential for identifying TPs, across studies, significant variations exist in the number of inspectors involved and the grading system employed. This inconsistency can considerably impact the reported percentage of ML candidates.
In our work, the majority of visual inspections were conducted by a single author, potentially introducing labeling bias and increasing the risk of overlooking high-quality candidates. However, the high false positive rate necessitated this approach, as assigning the same lenses to multiple inspectors for confirmation proved impractical and too time-consuming. This highlights the crucial importance of considering these methodological variations when comparing results across different studies.

Table \ref{tab:tp_fp} highlights the difference in our model's precision compared to other SGL searches. While the lower precision in our study could be partially attributed to the sample size investigated, similar-sized samples like in \cite{Li_2020} report a significantly higher precision (one order of magnitude greater). This suggests that our model's performance currently falls short of the competitive level.
The primary factor contributing to this discrepancy likely lies in the contrasting training strategies employed. Our model was trained entirely on simulated data before being fine-tuned for the KiDS survey. In contrast, \cite{Li_2020} trained their model on real KiDS galaxies with superimposed simulated lensing features. This inherent difference in the training data could explain the observed variation in precision.
The next sections will delve deeper into this aspect, exploring the potential influence of training data as well as galaxy population preselection on the model performance.

\subsection{Data preselection cuts}

Another key difference lies in preprocessing strategies. Early SGL searches \citep{Petrillo_2017, Petrillo_2018, links} targeted only the LRGs subsample,  capitalizing on their higher lensing probability ($\sim$ 80\%) compared to the 20\% probability of spiral galaxies \citep[see, for instance,][]{Turner_1984}. Later studies expanded to include both LRGs and BGs, aiming to generalize the sample beyond a simple color cut.
Due to the aforementioned reason, the CNN models in \cite{Petrillo_2017,Petrillo_2018, links, Li_2020, Li_2021} were trained on simulated arcs superimposed on LRGs. This potential bias towards LRGs leads to higher FPs contamination in the BG sample, justifying the higher prediction probability threshold used for this sample in the studies considering this galaxy population. 

We consider the impact of different $p$ thresholds and sample selection for the TEGLIE 
candidates on the GAMA footprint by using preprocessing strategies used in other KiDS SGL searches. Table \ref{tab:tp_fp} also shows the outcomes of these selections.
This work exhibits a significantly lower precision (around one order of magnitude) than other SGL searches.  The BG cut with $p > 0.98$  yielded improved precision compared to the original redshift cut ($z<$ 0.8), but still remained lower than other studies. 
Interestingly, the percentage of ML candidates in our work closely aligns with other studies, except for the LRG cut which shows the lowest performance.
Furthermore, the TEGLIE candidates with $z<$ 0.8 would have reduced both TP and FPs by $\sim$ 20\% for BGs and $\sim$ 70\% for LRGs, making the LRG cut overly restrictive.
Initially, the ML sample contained 231 candidates with 51\,396 FPs (99.55\% of all the candidates).
Increasing the prediction probability threshold from 0.8 to 0.9 shrunk the ML sample by roughly 37\%. Further raising the threshold to 0.98 filtered out an additional $\sim$ 60\%. This stricter threshold also led to a significant reduction in FPs, with a decrease of 40\% and 53\% for $p$ = 0.9 and $p$ = 0.98, respectively.
These findings underline the impact of preselection in the framework of SGL searches.

Table \ref{tab:tp_fp} provides insights into both total and high-confidence (grade 1) TPs, with the latter displayed in parentheses within the TP column. Analyzing all TEGLIE candidates reveals a proportional decrease in total TPs as the number of ML candidates shrinks. The LRG cut stands out as an exception, being overly restrictive and leading to a significant reduction in candidates.
However, the picture changes when focusing on high-confidence TPs. The LRG cut remains overly restrictive, hindering potential discoveries.
The BG cut, instead, maintains most of the confirmed candidates. With $p$=0.8 we retain a 94\% completeness of the original candidates while achieving a $\sim$ 20\% reduction in FPs.
Even more stringent cuts, like $p$ = 0.98, demonstrate significant potential. This cut offers a 61\% reduction in FPs while still preserving 68\% of the original high-confidence TPs identified with the less restrictive z $<$ 0.8 cut.

\subsection{Selecting training data}

Another fundamental aspect to consider when comparing ML models is the training set. The transformer encoder was trained on the Bologna Lens Challenge dataset outperforming, with a 99\% accuracy, all the models that participated in the challenge \citep{Metcalf_2019}. The simulated dataset contains 80\% fully simulated images and 15\% KiDS BG images with added simulated arcs. On the simulated images, KiDS PSF maps of each band were applied. 
This training set is significantly different from the one used in other SGL searches where the lensed arcs were all superimposed on LRGs images in KiDS.
Fine-tuning with real SGL candidates from KiDS improved performance, but not to the level of direct training on KiDS observed galaxies. This suggests that fine-tuning helps mitigate false positives, but it cannot fully compensate for training with images resembling the actual observed data with its specific PSF.
This sensibility is known for CNN but it is valid also for transformer encoders even if they should be able to extract and learn the relation between pixels and understand better the morphology of SGL.
Another difference, as evidenced already by \cite{Davies_2019} is that the simulated lenses on the Bologna lens challenge had too wide Einstein rings with respect to the ones found in observations.

The presence of False positives is a constant issue in SGL searches. In Tab. \ref{tab:tp_fp}, we included the percentage of FP in respect to the ML candidates and for most of the searches above 90\%, with an average of  $\approx$ 90\%.
In this work, we explored a way to mitigate the prevalence of false positives generated by ML model when transitioning to a new target set, particularly when moving from simulated data to real observations. SGLs are extremely rare phenomena, making it challenging to assemble a substantial collection of real observations to effectively train ML algorithms for their identification. Due to the scarcity of real SGL observations, simulated images are employed for training ML algorithms. 
Another reason for the presence of FP is that real-world observations are inherently more intricate than simulated data, and SGLs may not always exhibit the classic bright arcs paired with a redder central source resulting in more FP. 
For instance, training an algorithm to identify arcs in an image does not guarantee a low false positive rate. Numerous celestial objects possess elongated features that can mimic lensed galaxies, such as the tidal tails of merging galaxies. Additionally, some edge-on galaxies can be misclassified as lensed arcs if they lie in close proximity to another source. Similarly, ring galaxies, which are rare galaxies with a bright ring of stars surrounding a central core, can be confused with Einstein rings \citep{Timmis_2017}.
In this work, we identified 121 objects during visual inspection that exhibited such characteristics and were assigned a grade of 3. These objects are illustrated in Fig. \ref{fig:grade_3_0}.

In Fig \ref{fig:grade_0_0}, we presented some of the non-lenses used in the fine-tuning, these cutouts do not exactly show dubious cases in which a human would have difficulties in the labeling. The discrepancy between human and machine perception of relevant features emphasizes the importance of having a diverse dataset, including images similar to lenses. 
While incorporating genuine SGL images is essential for training models, it is equally important to include misleading objects for both human experts and ML models. Figure \ref{fig:grade_3_0} exemplifies such cases. 
Certain features like elongation and the presence of brighter arcs or redder central sources are not conclusive indicators of SGLs, as the presence of multiple objects with similar colors can lead to ambiguity, making visual inspection challenging even for experienced inspectors, as discussed in \citet{Rojas_2023}.

As previously discussed in \cite{rojas2021strong}, a dataset that includes ring galaxies and other potential false positives will be valuable for training more accurate models in the future. 
Collecting datasets with misleading objects is important for training ML algorithms, however, as discussed in \cite{Wilde_2022_ML}, this approach has a fundamental interpretability problem. It is not possible to directly understand how the model works, which features of SGL the model learns, or if there is a substantial bias towards a class of lenses. The ``black-box" issue is a critical problem in ML studies and understanding the biases of these algorithms is vital \citep[e.g.,][]{Obermeyer_2019,yip2021peeking,YuLiang_2022, Chen_2023, ferrara_2023}.  Machine learning methods excel at modeling high-dimensional data by capturing intricate non-linear relationships. Consequently, handling models with billions of parameters is a challenging endeavor.

\begin{figure*}[!htbp]
\includegraphics[width=\textwidth]{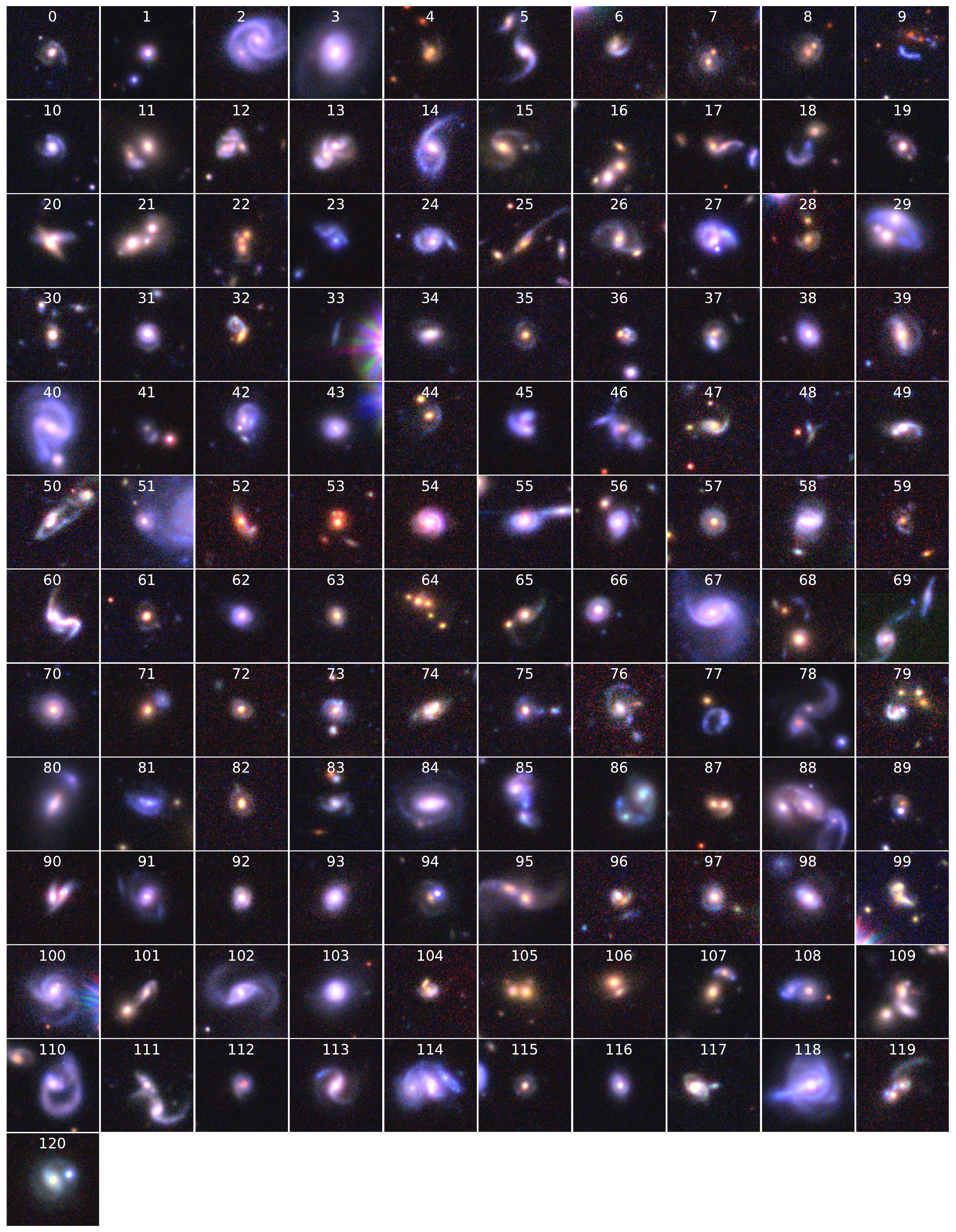}
\centering
\caption{Grade 3 objects, mostly mergers and ring galaxies. }
\label{fig:grade_3_0}
\end{figure*}

 \subsection{Model interpretability and biases}
 
Several techniques, such as GradCAM \citep{Selvaraju_2019} and LIME \citep{ribeiro_2016}, can be employed to pinpoint the features that play a crucial role in a model's decision-making process. However, these methods are not suitable for the present scenario due to their constraints. GradCAM is only applicable to CNNs, while LIME is incapable of handling images with more than three dimensions, making it unsuitable for the four-band image data used to train the model.
Techniques like GradCAM and other input-altering approaches, such as masking specific regions of the image, have been investigated in \cite{Wilde_2022_ML} to enhance the interpretability of CNNs for SGL detection. These methods suggest that models tend to prioritize lenses exhibiting blue ring-like features. However, observations of Einstein rings and lensed dusty forming galaxies have been reported in the literature \citep{Negrello_2010, Geach_2015, Negrello_2016, Spilker_2016, Rivera_2019, Neri_2020}, highlighting a potential limitation of ML models for SGL detection if they rely heavily on the lensing systems' colors.

One approach to mitigate bias in ML classifications could involve ensembling different architectures with varying hyperparameters, trained on diverse datasets. The final label could then be determined by averaging the prediction probabilities of these individual networks, as demonstrated by \cite{canameras2023holismokes} and \cite{Schaefer_2018}. Alternatively, a more structured approach could involve incorporating the calibrated probabilities generated within a Bayesian framework, as presented in \cite{holloway_2023}.
Averaging the predictions of multiple models can lead to a more robust detection algorithm. This, combined with enhanced training data and the contributions of citizen science initiatives, holds the potential to drive significant scientific advancements in the years to come.

For next-generation surveys like the LSST, a combination of methods and machine learning models will be essential. This process could begin with a robust preselection step to identify galaxies with a statistically high likelihood of being strong lenses. Subsequently, a variety of machine learning models and algorithms could be employed for precise feature detection and false positive reduction. Finally, by combining machine learning with citizen science initiatives like the Space Warps project \citep{Marshall_2016}, the accurate labeling of all SGL candidates could be achieved.

\section{Conclusions }
\label{sec:conclusions}

This paper presents a methodology to mitigate the challenge of false positives arising from machine learning models employed for SGL detection. Our approach incorporates images of real SGL candidates and non-SGL examples alongside data augmentation techniques. Utilizing this augmented dataset, we fine-tune a pre-trained model, originally developed using simulated data that approximates the KiDS survey observations, to enhance its performance on real KiDS survey data.
Through an evaluation of six different data augmentation techniques, we identify the combination of rotation, flipping, and transposed image augmentation as the most effective, achieving a 70\% reduction in false positives.

Applying our method to a 221 deg$^2$ region co-observed by KiDS and GAMA surveys yielded 56 high-confidence and 175 less-certain SGL candidate detections. These candidates originated from a relaxed galaxy selection compared to previous searches, incorporating all galaxies up to redshift 0.8. Incorporating the SGL candidates identified during the testing phase of the fine-tuned model, the total number of HQ SGL candidates increases to 71. Of these 71, 44 candidates are entirely new detections, not previously identified by other methods.

Our analysis revealed that stricter selection criteria would have significantly reduced the number of candidates. Focusing solely on bright galaxies would have decreased both true positives (grade 1 and 2) and false positives by 20\%. Further restricting to the luminous red galaxies population would have reduced both categories by 70\%, though hindering discoveries. However, with the bright galaxies cut and considering only high-confidence candidates, we would have retrieved 94\% of them while still achieving a 20\% reduction in false positives.

Among the high-confidence SGLs, 31 pass the double-redshift selection using the GAMA survey data. Of these, 7 show blended spectra of passive galaxy in the foreground as emission line galaxy in the background, confirming their chance of being strong lenses and enabling further analysis for lens modeling and property investigation in future work. 

Additionally, we provide a catalog of 121 false positives exhibiting features resembling SGLs, such as bright arcs, bent structures, and ring galaxies. This catalog provides valuable training data for future machine learning models, aiding in their ability to differentiate true SGLs and further minimize false positive detections.

Our work highlights the efficacy of fine-tuning with data augmentation in enhancing the performance of machine learning models for SGL detection. The fine-tuned model achieved a significant reduction in false positives, detecting 70\% fewer compared to the non-fine-tuned model. We also evidence the importance of the selection cuts in galaxy populations which enables the identification of novel SGL candidates while simultaneously reducing the number of false positives.
However, our model has still not reached competitive performance. While the model architecture demonstrates strong performance on simulated data (as reported by \cite{Hareesh}), the current limitation suggests the need for more realistic simulations of SGL systems during the training phase. Incorporating such data has the potential to significantly enhance the model's ability to detect real-world SGLs. Furthermore, a multifaceted approach - including fine-tuning, citizen science efforts, and model ensembles - is likely the most effective strategy for minimizing false positives. Citizen science can play a crucial role in scrutinizing all of the SGL candidates identified by machine learning models.
Looking ahead, the adoption of novel architectures like transformer encoders and widely used models like CNNs holds promise for furthering the field of strong lens identification in wide-area surveys such as LSST. 

\begin{acknowledgements}
The authors wish to express their gratitude to Alessandro Sonnenfeld for the enlightening discussion, to Junais for the useful comments, and to Shuaibo Geng for helping with the visual inspection.

This research was supported by the Polish National Science Centre grant UMO-2018/30/M/ST9/00757 and by Polish Ministry of Science and Higher Education grant DIR/WK/2018/12. This article/publication is based upon work from COST Action CA21136 – “Addressing observational tensions in cosmology with systematics and fundamental physics (CosmoVerse)”, supported by COST (European Cooperation in Science and Technology).
ML and VE acknowledge support from the South African Radio Astronomy Observatory and the National Research Foundation (NRF) towards this research. Opinions expressed and conclusions arrived at, are those of the authors and are not necessarily to be attributed to the NRF.

Based on observations made with ESO Telescopes at the La Silla Paranal Observatory under programme IDs 177.A-3016, 177.A-3017, 177.A-3018 and 179.A-2004, and on data products produced by the KiDS consortium. The KiDS production team acknowledges support from: Deutsche Forschungsgemeinschaft, ERC, NOVA and NWO-M grants; Target; the University of Padova, and the University Federico II (Naples).

GAMA is a joint European-Australasian project based around a spectroscopic campaign using the Anglo-Australian Telescope. The GAMA input catalogue is based on data taken from the Sloan Digital Sky Survey and the UKIRT Infrared Deep Sky Survey. Complementary imaging of the GAMA regions is being obtained by a number of independent survey programmes including GALEX MIS, VST KiDS, VISTA VIKING, WISE, Herschel-ATLAS, GMRT and ASKAP providing UV to radio coverage. GAMA is funded by the STFC (UK), the ARC (Australia), the AAO, and the participating institutions. The GAMA website is https://www.gama-survey.org/ .

\end{acknowledgements}

%
\bibliographystyle{aa} 
%
\bibliography{mybibliography.bib}

\begin{thebibliography}{133}
\expandafter\ifx\csname natexlab\endcsname\relax\def\natexlab#1{#1}\fi

\bibitem[{Alard(2006)}]{alard2006automated}
Alard, C. 2006, Automated detection of gravitational arcs

\bibitem[{Allam \& McEwen(2023)}]{allam2023paying}
Allam, Tarek, J. \& McEwen, J.~D. 2023, {Paying attention to astronomical transients: Introducing the time-series transformer for photometric classification}

\bibitem[{Aslahishahri {et~al.}(2023)Aslahishahri, Ubbens, \& Stavness}]{aslahishahri2023darts}
Aslahishahri, M., Ubbens, J., \& Stavness, I. 2023, DARTS: Double Attention Reference-based Transformer for Super-resolution

\bibitem[{Baldry {et~al.}(2014)Baldry, Alpaslan, Bauer, Bland-Hawthorn, Brough, Cluver, Croom, Davies, Driver, Gunawardhana, Holwerda, Hopkins, Kelvin, Liske, Lopez-Sanchez, Loveday, Norberg, Peacock, Robotham, \& Taylor}]{Baldry_2014}
Baldry, I.~K., Alpaslan, M., Bauer, A.~E., {et~al.} 2014, \mnras, 441, 2440–2451

\bibitem[{Barnabè {et~al.}(2012)Barnabè, Dutton, Marshall, Auger, Brewer, Treu, Bolton, Koo, \& Koopmans}]{Barnabe_2012}
Barnabè, M., Dutton, A.~A., Marshall, P.~J., {et~al.} 2012, \mnras, 423, 1073

\bibitem[{Bayliss {et~al.}(2011)Bayliss, Gladders, Oguri, Hennawi, Sharon, Koester, \& Dahle}]{Bayliss_2011}
Bayliss, M.~B., Gladders, M.~D., Oguri, M., {et~al.} 2011, The Astrophysical Journal Letters, 727, L26

\bibitem[{{Ben{\'\i}tez}(2011)}]{Benitez_2011}
{Ben{\'\i}tez}, N. 2011, {BPZ: Bayesian Photometric Redshift Code}, Astrophysics Source Code Library, record ascl:1108.011

\bibitem[{{Bertin} \& {Arnouts}(1996)}]{Bertin_1996}
{Bertin}, E. \& {Arnouts}, S. 1996, \aaps, 117, 393

\bibitem[{Bolton {et~al.}(2008)Bolton, Burles, Koopmans, Treu, Gavazzi, Moustakas, Wayth, \& Schlegel}]{Bolton_2008}
Bolton, A.~S., Burles, S., Koopmans, L. V.~E., {et~al.} 2008, The Astrophysical Journal, 682, 964

\bibitem[{{Boylan-Kolchin} {et~al.}(2009){Boylan-Kolchin}, {Springel}, {White}, {Jenkins}, \& {Lemson}}]{Boylan_2009}
{Boylan-Kolchin}, M., {Springel}, V., {White}, S. D.~M., {Jenkins}, A., \& {Lemson}, G. 2009, \mnras, 398, 1150

\bibitem[{Brownstein {et~al.}(2011)Brownstein, Bolton, Schlegel, Eisenstein, Kochanek, Connolly, Maraston, Pandey, Seitz, Wake, Wood-Vasey, Brinkmann, Schneider, \& Weaver}]{Brownstein_2012}
Brownstein, J.~R., Bolton, A.~S., Schlegel, D.~J., {et~al.} 2011, The Astrophysical Journal, 744, 41

\bibitem[{Canameras {et~al.}(2023)Canameras, Schuldt, Shu, Suyu, Taubenberger, Andika, Bag, Inoue, Jaelani, Leal-Taixe, Meinhardt, Melo, \& More}]{canameras2023holismokes}
Canameras, R., Schuldt, S., Shu, Y., {et~al.} 2023, HOLISMOKES -- XI. Evaluation of supervised neural networks for strong-lens searches in ground-based imaging surveys

\bibitem[{Ca{\~{n}}ameras {et~al.}(2020)Ca{\~{n}}ameras, Schuldt, Suyu, Taubenberger, Meinhardt, Leal-Taix{\'{e}}, Lemon, Rojas, \& Savary}]{Canameras_2020}
Ca{\~{n}}ameras, R., Schuldt, S., Suyu, S.~H., {et~al.} 2020, \aap, 644, A163

\bibitem[{Canameras {et~al.}(2021)}]{Canameras_2021}
Canameras, R. {et~al.} 2021, Astron. Astrophys., 653, L6

\bibitem[{Cao {et~al.}(2015)Cao, Biesiada, Gavazzi, Pi\'{o}rkowska, \& Zhu}]{Cao_2015}
Cao, S., Biesiada, M., Gavazzi, R., Pi\'{o}rkowska, A., \& Zhu, Z.-H. 2015, \apj, 806, 185

\bibitem[{{Cao} {et~al.}(2017){Cao}, {Li}, {Biesiada}, {Xu}, {Cai}, \& {Zhu}}]{Cao2017}
{Cao}, S., {Li}, X., {Biesiada}, M., {et~al.} 2017, \apj, 835, 92

\bibitem[{Capaccioli \& Schipani(2011)}]{capaccioli2011vlt}
Capaccioli, M. \& Schipani, P. 2011, The Messenger, 146, 27

\bibitem[{Carion {et~al.}(2020)Carion, Massa, Synnaeve, Usunier, Kirillov, \& Zagoruyko}]{carion2020endtoend}
Carion, N., Massa, F., Synnaeve, G., {et~al.} 2020, in Computer Vision -- ECCV 2020, ed. A.~Vedaldi, H.~Bischof, T.~Brox, \& J.-M. Frahm (Cham: Springer International Publishing), 213--229

\bibitem[{{Chan} {et~al.}(2016){Chan}, {Suyu}, {More}, {Oguri}, {Chiueh}, {Coupon}, {Hsieh}, {Komiyama}, {Miyazaki}, {Murayama}, {Nishizawa}, {Price}, {Tait}, {Terai}, {Utsumi}, \& {Wang}}]{Chan_2016}
{Chan}, J. H.~H., {Suyu}, S.~H., {More}, A., {et~al.} 2016, \apj, 832, 135

\bibitem[{Chan {et~al.}(2020)Chan, Suyu, Sonnenfeld, Jaelani, More, Yonehara, Kubota, Coupon, Lee, Oguri, Rusu, \& Wong}]{Chan_2020_sugohi}
Chan, J. H.~H., Suyu, S.~H., Sonnenfeld, A., {et~al.} 2020, \aap, 636, A87

\bibitem[{Chen {et~al.}(2023)Chen, Wu, \& Wang}]{Chen_2023}
Chen, P., Wu, L., \& Wang, L. 2023, Applied Sciences, 13

\bibitem[{Chou {et~al.}(2022)Chou, Moreira, Bruza, Ouyang, \& Jorge}]{YuLiang_2022}
Chou, Y.-L., Moreira, C., Bruza, P., Ouyang, C., \& Jorge, J. 2022, Inf. Fusion, 81, 59–83

\bibitem[{{Collett}(2015)}]{Collett_2015}
{Collett}, T.~E. 2015, \apj, 811, 20

\bibitem[{Collett \& Auger(2014)}]{Collett_2014}
Collett, T.~E. \& Auger, M.~W. 2014, MNRAS, 443, 969

\bibitem[{{Davies} {et~al.}(2019){Davies}, {Serjeant}, \& {Bromley}}]{Davies_2019}
{Davies}, A., {Serjeant}, S., \& {Bromley}, J.~M. 2019, \mnras, 487, 5263

\bibitem[{{de Jong} {et~al.}(2013){de Jong}, {Verdoes Kleijn}, {Kuijken}, \& {Valentijn}}]{dejong2013}
{de Jong}, J. T.~A., {Verdoes Kleijn}, G.~A., {Kuijken}, K.~H., \& {Valentijn}, E.~A. 2013, Experimental Astronomy, 35, 25

\bibitem[{{de Jong, Jelte T. A.} {et~al.}(2017){de Jong, Jelte T. A.}, {Kleijn, Gijs A. Verdoes}, {Erben, Thomas}, {Hildebrandt, Hendrik}, {Kuijken, Konrad}, {Sikkema, Gert}, {Brescia, Massimo}, {Bilicki, Maciej}, {Napolitano, Nicola R.}, {Amaro, Valeria}, {Begeman, Kor G.}, {Boxhoorn, Danny R.}, {Buddelmeijer, Hugo}, {Cavuoti, Stefano}, {Getman, Fedor}, {Grado, Aniello}, {Helmich, Ewout}, {Huang, Zhuoyi}, {Irisarri, Nancy}, {La Barbera, Francesco}, {Longo, Giuseppe}, {McFarland, John P.}, {Nakajima, Reiko}, {Paolillo, Maurizio}, {Puddu, Emanuella}, {Radovich, Mario}, {Rifatto, Agatino}, {Tortora, Crescenzo}, {Valentijn, Edwin A.}, {Vellucci, Civita}, {Vriend, Willem-Jan}, {Amon, Alexandra}, {Blake, Chris}, {Choi, Ami}, {Conti, Ian Fenech}, {Gwyn, Stephen D. J.}, {Herbonnet, Ricardo}, {Heymans, Catherine}, {Hoekstra, Henk}, {Klaes, Dominik}, {Merten, Julian}, {Miller, Lance}, {Schneider, Peter}, \& {Viola, Massimo}}]{deJong2017}
{de Jong, Jelte T. A.}, {Kleijn, Gijs A. Verdoes}, {Erben, Thomas}, {et~al.} 2017, \aap, 604, A134

\bibitem[{{de Jong, Jelte T. A.} {et~al.}(2015){de Jong, Jelte T. A.}, {Verdoes Kleijn, Gijs A.}, {Boxhoorn, Danny R.}, {Buddelmeijer, Hugo}, {Capaccioli, Massimo}, {Getman, Fedor}, {Grado, Aniello}, {Helmich, Ewout}, {Huang, Zhuoyi}, {Irisarri, Nancy}, {Kuijken, Konrad}, {La Barbera, Francesco}, {McFarland, John P.}, {Napolitano, Nicola R.}, {Radovich, Mario}, {Sikkema, Gert}, {Valentijn, Edwin A.}, {Begeman, Kor G.}, {Brescia, Massimo}, {Cavuoti, Stefano}, {Choi, Ami}, {Cordes, Oliver-Mark}, {Covone, Giovanni}, {Dall\'{}Ora, Massimo}, {Hildebrandt, Hendrik}, {Longo, Giuseppe}, {Nakajima, Reiko}, {Paolillo, Maurizio}, {Puddu, Emanuella}, {Rifatto, Agatino}, {Tortora, Crescenzo}, {van Uitert, Edo}, {Buddendiek, Axel}, {Harnois-D\'eraps, Joachim}, {Erben, Thomas}, {Eriksen, Martin B.}, {Heymans, Catherine}, {Hoekstra, Henk}, {Joachimi, Benjamin}, {Kitching, Thomas D.}, {Klaes, Dominik}, {Koopmans, L\'eon V. E.}, {K\"ohlinger, Fabian}, {Roy, Nivya}, {Sif\'on, Crist\'obal}, {Schneider, Peter}, {Sutherland, Will
  J.}, {Viola, Massimo}, \& {Vriend, Willem-Jan}}]{deJong2015}
{de Jong, Jelte T. A.}, {Verdoes Kleijn, Gijs A.}, {Boxhoorn, Danny R.}, {et~al.} 2015, \aap, 582, A62

\bibitem[{Devlin {et~al.}(2019)Devlin, Chang, Lee, \& Toutanova}]{devlin_2019_bert}
Devlin, J., Chang, M.-W., Lee, K., \& Toutanova, K. 2019, BERT: Pre-training of Deep Bidirectional Transformers for Language Understanding

\bibitem[{Diehl {et~al.}(2017)Diehl, Buckley-Geer, Lindgren, Nord, Gaitsch, Gaitsch, Lin, Allam, Collett, Furlanetto, Gill, More, Nightingale, Odden, Pellico, Tucker, da~Costa, Neto, Kuropatkin, Soares-Santos, Welch, Zhang, Frieman, Abdalla, Annis, Benoit-Lévy, Bertin, Brooks, Burke, Rosell, Kind, Carretero, Cunha, D’Andrea, Desai, Dietrich, Drlica-Wagner, Evrard, Finley, Flaugher, García-Bellido, Gerdes, Goldstein, Gruen, Gruendl, Gschwend, Gutierrez, James, Kuehn, Kuhlmann, Lahav, Li, Lima, Maia, Marshall, Menanteau, Miquel, Nichol, Nugent, Ogando, Plazas, Reil, Romer, Sako, Sanchez, Santiago, Scarpine, Schindler, Schubnell, Sevilla-Noarbe, Sheldon, Smith, Sobreira, Suchyta, Swanson, Tarle, Thomas, Walker, \& Collaboration)}]{Diehl_2017}
Diehl, H.~T., Buckley-Geer, E.~J., Lindgren, K.~A., {et~al.} 2017, The Astrophysical Journal Supplement Series, 232, 15

\bibitem[{{Donoso-Oliva, C.} {et~al.}(2023){Donoso-Oliva, C.}, {Becker, I.}, {Protopapas, P.}, {Cabrera-Vives, G.}, {Vishnu, M.}, \& {Vardhan, H.}}]{Donoso_2023_trans}
{Donoso-Oliva, C.}, {Becker, I.}, {Protopapas, P.}, {et~al.} 2023, \aap, 670, A54

\bibitem[{Dosovitskiy {et~al.}(2021)Dosovitskiy, Beyer, Kolesnikov, Weissenborn, Zhai, Unterthiner, Dehghani, Minderer, Heigold, Gelly, Uszkoreit, \& Houlsby}]{dosovitskiy2020image}
Dosovitskiy, A., Beyer, L., Kolesnikov, A., {et~al.} 2021, in 9th International Conference on Learning Representations, {ICLR} 2021, Virtual Event, Austria, May 3-7, 2021 (OpenReview.net)

\bibitem[{{Driver} {et~al.}(2022){Driver}, {Bellstedt}, {Robotham}, {Baldry}, {Davies}, {Liske}, {Obreschkow}, {Taylor}, {Wright}, {Alpaslan}, {Bamford}, {Bauer}, {Bland-Hawthorn}, {Bilicki}, {Bravo}, {Brough}, {Casura}, {Cluver}, {Colless}, {Conselice}, {Croom}, {de Jong}, {D'Eugenio}, {De Propris}, {Dogruel}, {Drinkwater}, {Dvornik}, {Farrow}, {Frenk}, {Giblin}, {Graham}, {Grootes}, {Gunawardhana}, {Hashemizadeh}, {H{\"a}u{\ss}ler}, {Heymans}, {Hildebrandt}, {Holwerda}, {Hopkins}, {Jarrett}, {Heath Jones}, {Kelvin}, {Koushan}, {Kuijken}, {Lara-L{\'o}pez}, {Lange}, {L{\'o}pez-S{\'a}nchez}, {Loveday}, {Mahajan}, {Meyer}, {Moffett}, {Napolitano}, {Norberg}, {Owers}, {Radovich}, {Raouf}, {Peacock}, {Phillipps}, {Pimbblet}, {Popescu}, {Said}, {Sansom}, {Seibert}, {Sutherland}, {Thorne}, {Tuffs}, {Turner}, {van der Wel}, {van Kampen}, \& {Wilkins}}]{Driver_2022}
{Driver}, S.~P., {Bellstedt}, S., {Robotham}, A. S.~G., {et~al.} 2022, \mnras, 513, 439

\bibitem[{Driver {et~al.}(2011)Driver, Hill, Kelvin, Robotham, Liske, Norberg, Baldry, Bamford, Hopkins, Loveday, Peacock, Andrae, Bland-Hawthorn, Brough, Brown, Cameron, Ching, Colless, Conselice, Croom, Cross, De~Propris, Dye, Drinkwater, Ellis, Graham, Grootes, Gunawardhana, Jones, van Kampen, Maraston, Nichol, Parkinson, Phillipps, Pimbblet, Popescu, Prescott, Roseboom, Sadler, Sansom, Sharp, Smith, Taylor, Thomas, Tuffs, Wijesinghe, Dunne, Frenk, Jarvis, Madore, Meyer, Seibert, Staveley-Smith, Sutherland, \& Warren}]{Driver_2011}
Driver, S.~P., Hill, D.~T., Kelvin, L.~S., {et~al.} 2011, \mnras, 413, 971

\bibitem[{{Driver} {et~al.}(2009){Driver}, {Norberg}, {Baldry}, {Bamford}, {Hopkins}, {Liske}, {Loveday}, {Peacock}, {Hill}, {Kelvin}, {Robotham}, {Cross}, {Parkinson}, {Prescott}, {Conselice}, {Dunne}, {Brough}, {Jones}, {Sharp}, {van Kampen}, {Oliver}, {Roseboom}, {Bland-Hawthorn}, {Croom}, {Ellis}, {Cameron}, {Cole}, {Frenk}, {Couch}, {Graham}, {Proctor}, {De Propris}, {Doyle}, {Edmondson}, {Nichol}, {Thomas}, {Eales}, {Jarvis}, {Kuijken}, {Lahav}, {Madore}, {Seibert}, {Meyer}, {Staveley-Smith}, {Phillipps}, {Popescu}, {Sansom}, {Sutherland}, {Tuffs}, \& {Warren}}]{Driver_2009}
{Driver}, S.~P., {Norberg}, P., {Baldry}, I.~K., {et~al.} 2009, Astronomy and Geophysics, 50, 5.12

\bibitem[{Dye \& Warren(2005)}]{Dye_2005}
Dye, S. \& Warren, S.~J. 2005, The Astrophysical Journal, 623, 31

\bibitem[{Eisenstein {et~al.}(2001)Eisenstein, Annis, Gunn, Szalay, Connolly, Nichol, Bahcall, Bernardi, Burles, Castander, Fukugita, Hogg, Ivezi{\'{c}}, Knapp, Lupton, Narayanan, Postman, Reichart, Richmond, Schneider, Schlegel, Strauss, SubbaRao, Tucker, Berk, Vogeley, Weinberg, \& Yanny}]{Eisenstein_2001}
Eisenstein, D.~J., Annis, J., Gunn, J.~E., {et~al.} 2001, The Astronomical Journal, 122, 2267

\bibitem[{Faure {et~al.}(2008)Faure, Kneib, Covone, Tasca, Leauthaud, Capak, Jahnke, Smolcic, de~la Torre, Ellis, Finoguenov, Koekemoer, Fevre, Massey, Mellier, Refregier, Rhodes, Scoville, Schinnerer, Taylor, Waerbeke, \& Walcher}]{Faure_2008}
Faure, C., Kneib, J.-P., Covone, G., {et~al.} 2008, The Astrophysical Journal Supplement Series, 176, 19

\bibitem[{Ferrara(2023)}]{ferrara_2023}
Ferrara, E. 2023, Fairness And Bias in Artificial Intelligence: A Brief Survey of Sources, Impacts, And Mitigation Strategies

\bibitem[{{Garvin, Emily O.} {et~al.}(2022){Garvin, Emily O.}, {Kruk, Sandor}, {Cornen, Claude}, {Bhatawdekar, Rachana}, {Cañameras, Raoul}, \& {Merín, Bruno}}]{Garvin_2022}
{Garvin, Emily O.}, {Kruk, Sandor}, {Cornen, Claude}, {et~al.} 2022, \aap, 667, A141

\bibitem[{Geach {et~al.}(2015)Geach, More, Verma, Marshall, Jackson, Belles, Beswick, Baeten, Chavez, Cornen, Cox, Erben, Erickson, Garrington, Harrison, Harrington, Hughes, Ivison, Jordan, Lin, Leauthaud, Lintott, Lynn, Kapadia, Kneib, Macmillan, Makler, Miller, Montaña, Mujica, Muxlow, Narayanan, Briain, O'Brien, Oguri, Paget, Parrish, Ross, Rozo, Rusu, Rykoff, Sanchez-Argüelles, Simpson, Snyder, Schloerb, Tecza, Wang, Van~Waerbeke, Wilcox, Viero, Wilson, Yun, \& Zeballos}]{Geach_2015}
Geach, J.~E., More, A., Verma, A., {et~al.} 2015, \mnras, 452, 502

\bibitem[{Gentile {et~al.}(2021)Gentile, Tortora, Covone, Koopmans, Spiniello, Fan, Li, Liu, Napolitano, Vaccari, \& Fu}]{Gentile_2021}
Gentile, F., Tortora, C., Covone, G., {et~al.} 2021, MNRAS, 510, 500

\bibitem[{He {et~al.}(2016)He, Zhang, Ren, \& Sun}]{he_2015_resnet}
He, K., Zhang, X., Ren, S., \& Sun, J. 2016, in 2016 IEEE Conference on Computer Vision and Pattern Recognition (CVPR), 770--778

\bibitem[{{He} {et~al.}(2020){He}, {Er}, {Long}, {Liu}, {Liu}, {Li}, {Liu}, {Deng}, \& {Fan}}]{He_2020}
{He}, Z., {Er}, X., {Long}, Q., {et~al.} 2020, \mnras, 497, 556

\bibitem[{Hennawi {et~al.}(2008)Hennawi, Gladders, Oguri, Dalal, Koester, Natarajan, Strauss, Inada, Kayo, Lin, Lampeitl, Annis, Bahcall, \& Schneider}]{Hennawi_2008}
Hennawi, J.~F., Gladders, M.~D., Oguri, M., {et~al.} 2008, The Astronomical Journal, 135, 664

\bibitem[{Hezaveh {et~al.}(2016)Hezaveh, Dalal, Holder, Kisner, Kuhlen, \& Levasseur}]{Hezaveh_2016}
Hezaveh, Y., Dalal, N., Holder, G., {et~al.} 2016, Journal of Cosmology and Astroparticle Physics, 2016, 048

\bibitem[{Holloway {et~al.}(2023)Holloway, Marshall, Verma, More, Cañameras, Jaelani, Ishida, \& Wong}]{holloway_2023}
Holloway, P., Marshall, P.~J., Verma, A., {et~al.} 2023, A Bayesian Approach to Strong Lens Finding in the Era of Wide-area Surveys

\bibitem[{Holwerda {et~al.}(2015)Holwerda, Baldry, Alpaslan, Bauer, Bland-Hawthorn, Brough, Brown, Cluver, Conselice, Driver, Hopkins, Jones, López-Sánchez, Loveday, Meyer, \& Moffett}]{Holwerda_2015}
Holwerda, B.~W., Baldry, I.~K., Alpaslan, M., {et~al.} 2015, \mnras, 449, 4277

\bibitem[{Huang {et~al.}(2018)Huang, Liu, van~der Maaten, \& Weinberger}]{huang_2018}
Huang, G., Liu, Z., van~der Maaten, L., \& Weinberger, K.~Q. 2018, Densely Connected Convolutional Networks

\bibitem[{{Huang} {et~al.}(2022){Huang}, {Chih-Fan Chen}, {Chang}, {Lin}, {Hsu}, {Thengane}, \& {Yao-Yu Lin}}]{Huang_transforemr}
{Huang}, K.-W., {Chih-Fan Chen}, G., {Chang}, P.-W., {et~al.} 2022, arXiv e-prints, arXiv:2210.04143

\bibitem[{Huang {et~al.}(2021)Huang, Storfer, Gu, Ravi, Pilon, Sheu, Venguswamy, Banka, Dey, Landriau, \& et~al.}]{Huang_2021}
Huang, X., Storfer, C., Gu, A., {et~al.} 2021, \apj, 909, 27

\bibitem[{Huang {et~al.}(2020)Huang, Storfer, Ravi, Pilon, Domingo, Schlegel, Bailey, Dey, Gupta, Herrera, \& et~al.}]{Huang_2020}
Huang, X., Storfer, C., Ravi, V., {et~al.} 2020, \apj, 894, 78

\bibitem[{Hwang {et~al.}(2023)Hwang, Sabiu, Park, \& Hong}]{hwang2023universe}
Hwang, S.~Y., Sabiu, C.~G., Park, I., \& Hong, S.~E. 2023, The Universe is worth $64^3$ pixels: Convolution Neural Network and Vision Transformers for Cosmology

\bibitem[{{Jacobs} {et~al.}(2019){Jacobs}, {Collett}, {Glazebrook}, {Buckley-Geer}, {Diehl}, {Lin}, {McCarthy}, {Qin}, {Odden}, {Caso Escudero}, {Dial}, {Yung}, {Gaitsch}, {Pellico}, {Lindgren}, {Abbott}, {Annis}, {Avila}, {Brooks}, {Burke}, {Carnero Rosell}, {Carrasco Kind}, {Carretero}, {da Costa}, {De Vicente}, {Fosalba}, {Frieman}, {Garc{\'\i}a-Bellido}, {Gaztanaga}, {Goldstein}, {Gruen}, {Gruendl}, {Gschwend}, {Hollowood}, {Honscheid}, {Hoyle}, {James}, {Krause}, {Kuropatkin}, {Lahav}, {Lima}, {Maia}, {Marshall}, {Miquel}, {Plazas}, {Roodman}, {Sanchez}, {Scarpine}, {Serrano}, {Sevilla-Noarbe}, {Smith}, {Sobreira}, {Suchyta}, {Swanson}, {Tarle}, {Vikram}, {Walker}, {Zhang}, \& {DES Collaboration}}]{Jacobs_2019}
{Jacobs}, C., {Collett}, T., {Glazebrook}, K., {et~al.} 2019, \apjs, 243, 17

\bibitem[{Jaelani {et~al.}(2020)Jaelani, More, Oguri, Sonnenfeld, Suyu, Rusu, Wong, Chan, Kayo, Lee, Chao, Coupon, Inoue, \& Futamase}]{Jaelani_2020_sugohi}
Jaelani, A.~T., More, A., Oguri, M., {et~al.} 2020, \mnras, 495, 1291

\bibitem[{Jaelani {et~al.}(2021)Jaelani, Rusu, Kayo, More, Sonnenfeld, Silverman, Schramm, Anguita, Inada, Kondo, Schechter, Lee, Oguri, Chan, Wong, \& Inoue}]{Jaelani_2021_sugohi}
Jaelani, A.~T., Rusu, C.~E., Kayo, I., {et~al.} 2021, \mnras, 502, 1487

\bibitem[{{Jia} {et~al.}(2023){Jia}, {Sun}, {Li}, {Song}, {Ning}, {Wei}, \& {Luo}}]{Jia_Transformers}
{Jia}, P., {Sun}, R., {Li}, N., {et~al.} 2023, \aj, 165, 26

\bibitem[{Khan {et~al.}(2022)Khan, Naseer, Hayat, Zamir, Khan, \& Shah}]{Khan_2022}
Khan, S., Naseer, M., Hayat, M., {et~al.} 2022, {ACM} Computing Surveys, 54, 1

\bibitem[{{Knabel} {et~al.}(2023){Knabel}, {Holwerda}, {Nightingale}, {Treu}, {Bilicki}, {Brough}, {Driver}, {Finnerty}, {Haberzettl}, {Hegde}, {Hopkins}, {Kuijken}, {Liske}, {Pimblett}, {Steele}, \& {Wright}}]{Knabel2023}
{Knabel}, S., {Holwerda}, B.~W., {Nightingale}, J., {et~al.} 2023, \mnras, 520, 804

\bibitem[{Knabel {et~al.}(2020)Knabel, Steele, Holwerda, Bridge, Jacques, Hopkins, Bamford, Brown, Brough, Kelvin, Bilicki, \& Kielkopf}]{Knabel_2020}
Knabel, S., Steele, R.~L., Holwerda, B.~W., {et~al.} 2020, The Astronomical Journal, 160, 223

\bibitem[{Krizhevsky {et~al.}(2012)Krizhevsky, Sutskever, \& Hinton}]{Alexnet_2012}
Krizhevsky, A., Sutskever, I., \& Hinton, G.~E. 2012, in Advances in Neural Information Processing Systems, ed. F.~Pereira, C.~Burges, L.~Bottou, \& K.~Weinberger, Vol.~25 (Curran Associates, Inc.)

\bibitem[{{Kuijken}(2011)}]{Kuijken2011}
{Kuijken}, K. 2011, The Messenger, 146, 8

\bibitem[{Kuijken {et~al.}(2019)Kuijken, Heymans, Dvornik, Hildebrandt, de~Jong, Wright, Erben, Bilicki, Giblin, Shan, Getman, Grado, Hoekstra, Miller, Napolitano, Paolilo, Radovich, Schneider, Sutherland, Tewes, Tortora, Valentijn, \& Kleijn}]{Kuijken_2019}
Kuijken, K., Heymans, C., Dvornik, A., {et~al.} 2019, \aap, 625, A2

\bibitem[{{Kuijken, K.}(2008)}]{gaap}
{Kuijken, K.} 2008, \aap, 482, 1053

\bibitem[{{La Barbera} {et~al.}(2008){La Barbera}, {de Carvalho}, {Kohl-Moreira}, {Gal}, {Soares-Santos}, {Capaccioli}, {Santos}, \& {Sant'Anna}}]{La_Barbera_2008}
{La Barbera}, F., {de Carvalho}, R.~R., {Kohl-Moreira}, J.~L., {et~al.} 2008, \pasp, 120, 681

\bibitem[{{Laureijs} {et~al.}(2011){Laureijs}, {Amiaux}, {Arduini}, {Augu{\`e}res}, {Brinchmann}, {Cole}, {Cropper}, {Dabin}, {Duvet}, {Ealet}, {Garilli}, {Gondoin}, {Guzzo}, {Hoar}, {Hoekstra}, {Holmes}, {Kitching}, {Maciaszek}, {Mellier}, {Pasian}, {Percival}, {Rhodes}, {Saavedra Criado}, {Sauvage}, {Scaramella}, {Valenziano}, {Warren}, {Bender}, {Castander}, {Cimatti}, {Le F{\`e}vre}, {Kurki-Suonio}, {Levi}, {Lilje}, {Meylan}, {Nichol}, {Pedersen}, {Popa}, {Rebolo Lopez}, {Rix}, {Rottgering}, {Zeilinger}, {Grupp}, {Hudelot}, {Massey}, {Meneghetti}, {Miller}, {Paltani}, {Paulin-Henriksson}, {Pires}, {Saxton}, {Schrabback}, {Seidel}, {Walsh}, {Aghanim}, {Amendola}, {Bartlett}, {Baccigalupi}, {Beaulieu}, {Benabed}, {Cuby}, {Elbaz}, {Fosalba}, {Gavazzi}, {Helmi}, {Hook}, {Irwin}, {Kneib}, {Kunz}, {Mannucci}, {Moscardini}, {Tao}, {Teyssier}, {Weller}, {Zamorani}, {Zapatero Osorio}, {Boulade}, {Foumond}, {Di Giorgio}, {Guttridge}, {James}, {Kemp}, {Martignac}, {Spencer}, {Walton}, {Bl{\"u}mchen}, {Bonoli},
  {Bortoletto}, {Cerna}, {Corcione}, {Fabron}, {Jahnke}, {Ligori}, {Madrid}, {Martin}, {Morgante}, {Pamplona}, {Prieto}, {Riva}, {Toledo}, {Trifoglio}, {Zerbi}, {Abdalla}, {Douspis}, {Grenet}, {Borgani}, {Bouwens}, {Courbin}, {Delouis}, {Dubath}, {Fontana}, {Frailis}, {Grazian}, {Koppenh{\"o}fer}, {Mansutti}, {Melchior}, {Mignoli}, {Mohr}, {Neissner}, {Noddle}, {Poncet}, {Scodeggio}, {Serrano}, {Shane}, {Starck}, {Surace}, {Taylor}, {Verdoes-Kleijn}, {Vuerli}, {Williams}, {Zacchei}, {Altieri}, {Escudero Sanz}, {Kohley}, {Oosterbroek}, {Astier}, {Bacon}, {Bardelli}, {Baugh}, {Bellagamba}, {Benoist}, {Bianchi}, {Biviano}, {Branchini}, {Carbone}, {Cardone}, {Clements}, {Colombi}, {Conselice}, {Cresci}, {Deacon}, {Dunlop}, {Fedeli}, {Fontanot}, {Franzetti}, {Giocoli}, {Garcia-Bellido}, {Gow}, {Heavens}, {Hewett}, {Heymans}, {Holland}, {Huang}, {Ilbert}, {Joachimi}, {Jennins}, {Kerins}, {Kiessling}, {Kirk}, {Kotak}, {Krause}, {Lahav}, {van Leeuwen}, {Lesgourgues}, {Lombardi}, {Magliocchetti}, {Maguire},
  {Majerotto}, {Maoli}, {Marulli}, {Maurogordato}, {McCracken}, {McLure}, {Melchiorri}, {Merson}, {Moresco}, {Nonino}, {Norberg}, {Peacock}, {Pello}, {Penny}, {Pettorino}, {Di Porto}, {Pozzetti}, {Quercellini}, {Radovich}, {Rassat}, {Roche}, {Ronayette}, {Rossetti}, {Sartoris}, {Schneider}, {Semboloni}, {Serjeant}, {Simpson}, {Skordis}, {Smadja}, {Smartt}, {Spano}, {Spiro}, {Sullivan}, {Tilquin}, {Trotta}, {Verde}, {Wang}, {Williger}, {Zhao}, {Zoubian}, \& {Zucca}}]{laureijs2011euclid}
{Laureijs}, R., {Amiaux}, J., {Arduini}, S., {et~al.} 2011, arXiv e-prints, arXiv:1110.3193

\bibitem[{LeCun {et~al.}(1998)LeCun, Bottou, Bengio, \& Haffner}]{Lecun}
LeCun, Y., Bottou, L., Bengio, Y., \& Haffner, P. 1998, Proceedings of the IEEE, 86, 2278

\bibitem[{Lewis \& Gale(1994)}]{lewis_1994_uncert_sampl}
Lewis, D.~D. \& Gale, W.~A. 1994, A Sequential Algorithm for Training Text Classifiers

\bibitem[{Li {et~al.}(2021)Li, Napolitano, Spiniello, Tortora, Kuijken, Koopmans, Schneider, Getman, Xie, Long, Shu, Vernardos, Huang, Covone, Dvornik, Heymans, Hildebrandt, Radovich, \& Wright}]{Li_2021}
Li, R., Napolitano, N.~R., Spiniello, C., {et~al.} 2021, The Astrophysical Journal, 923, 16

\bibitem[{Li {et~al.}(2020)Li, Napolitano, Tortora, Spiniello, Koopmans, Huang, Roy, Vernardos, Chatterjee, Giblin, \& et~al.}]{Li_2020}
Li, R., Napolitano, N.~R., Tortora, C., {et~al.} 2020, \apj, 899, 30

\bibitem[{Li {et~al.}(2023)Li, Ding, Zhang, Yuan, Pang, Cheng, Chen, Liu, \& Loy}]{li2023transformerbased}
Li, X., Ding, H., Zhang, W., {et~al.} 2023, Transformer-Based Visual Segmentation: A Survey

\bibitem[{{Liske} {et~al.}(2015){Liske}, {Baldry}, {Driver}, {Tuffs}, {Alpaslan}, {Andrae}, {Brough}, {Cluver}, {Grootes}, {Gunawardhana}, {Kelvin}, {Loveday}, {Robotham}, {Taylor}, {Bamford}, {Bland-Hawthorn}, {Brown}, {Drinkwater}, {Hopkins}, {Meyer}, {Norberg}, {Peacock}, {Agius}, {Andrews}, {Bauer}, {Ching}, {Colless}, {Conselice}, {Croom}, {Davies}, {De Propris}, {Dunne}, {Eardley}, {Ellis}, {Foster}, {Frenk}, {H{\"a}u{\ss}ler}, {Holwerda}, {Howlett}, {Ibarra}, {Jarvis}, {Jones}, {Kafle}, {Lacey}, {Lange}, {Lara-L{\'o}pez}, {L{\'o}pez-S{\'a}nchez}, {Maddox}, {Madore}, {McNaught-Roberts}, {Moffett}, {Nichol}, {Owers}, {Palamara}, {Penny}, {Phillipps}, {Pimbblet}, {Popescu}, {Prescott}, {Proctor}, {Sadler}, {Sansom}, {Seibert}, {Sharp}, {Sutherland}, {V{\'a}zquez-Mata}, {van Kampen}, {Wilkins}, {Williams}, \& {Wright}}]{Liske_2015}
{Liske}, J., {Baldry}, I.~K., {Driver}, S.~P., {et~al.} 2015, \mnras, 452, 2087

\bibitem[{{LSST Science Collaboration} {et~al.}(2009){LSST Science Collaboration}, {Abell}, {Allison}, {Anderson}, {Andrew}, {Angel}, {Armus}, {Arnett}, {Asztalos}, {Axelrod}, {Bailey}, {Ballantyne}, {Bankert}, {Barkhouse}, {Barr}, {Barrientos}, {Barth}, {Bartlett}, {Becker}, {Becla}, {Beers}, {Bernstein}, {Biswas}, {Blanton}, {Bloom}, {Bochanski}, {Boeshaar}, {Borne}, {Bradac}, {Brandt}, {Bridge}, {Brown}, {Brunner}, {Bullock}, {Burgasser}, {Burge}, {Burke}, {Cargile}, {Chandrasekharan}, {Chartas}, {Chesley}, {Chu}, {Cinabro}, {Claire}, {Claver}, {Clowe}, {Connolly}, {Cook}, {Cooke}, {Cooray}, {Covey}, {Culliton}, {de Jong}, {de Vries}, {Debattista}, {Delgado}, {Dell'Antonio}, {Dhital}, {Di Stefano}, {Dickinson}, {Dilday}, {Djorgovski}, {Dobler}, {Donalek}, {Dubois-Felsmann}, {Durech}, {Eliasdottir}, {Eracleous}, {Eyer}, {Falco}, {Fan}, {Fassnacht}, {Ferguson}, {Fernandez}, {Fields}, {Finkbeiner}, {Figueroa}, {Fox}, {Francke}, {Frank}, {Frieman}, {Fromenteau}, {Furqan}, {Galaz}, {Gal-Yam}, {Garnavich},
  {Gawiser}, {Geary}, {Gee}, {Gibson}, {Gilmore}, {Grace}, {Green}, {Gressler}, {Grillmair}, {Habib}, {Haggerty}, {Hamuy}, {Harris}, {Hawley}, {Heavens}, {Hebb}, {Henry}, {Hileman}, {Hilton}, {Hoadley}, {Holberg}, {Holman}, {Howell}, {Infante}, {Ivezic}, {Jacoby}, {Jain}, {R}, {Jedicke}, {Jee}, {Garrett Jernigan}, {Jha}, {Johnston}, {Jones}, {Juric}, {Kaasalainen}, {Styliani}, {Kafka}, {Kahn}, {Kaib}, {Kalirai}, {Kantor}, {Kasliwal}, {Keeton}, {Kessler}, {Knezevic}, {Kowalski}, {Krabbendam}, {Krughoff}, {Kulkarni}, {Kuhlman}, {Lacy}, {Lepine}, {Liang}, {Lien}, {Lira}, {Long}, {Lorenz}, {Lotz}, {Lupton}, {Lutz}, {Macri}, {Mahabal}, {Mandelbaum}, {Marshall}, {May}, {McGehee}, {Meadows}, {Meert}, {Milani}, {Miller}, {Miller}, {Mills}, {Minniti}, {Monet}, {Mukadam}, {Nakar}, {Neill}, {Newman}, {Nikolaev}, {Nordby}, {O'Connor}, {Oguri}, {Oliver}, {Olivier}, {Olsen}, {Olsen}, {Olszewski}, {Oluseyi}, {Padilla}, {Parker}, {Pepper}, {Peterson}, {Petry}, {Pinto}, {Pizagno}, {Popescu}, {Prsa}, {Radcka}, {Raddick},
  {Rasmussen}, {Rau}, {Rho}, {Rhoads}, {Richards}, {Ridgway}, {Robertson}, {Roskar}, {Saha}, {Sarajedini}, {Scannapieco}, {Schalk}, {Schindler}, {Schmidt}, {Schmidt}, {Schneider}, {Schumacher}, {Scranton}, {Sebag}, {Seppala}, {Shemmer}, {Simon}, {Sivertz}, {Smith}, {Allyn Smith}, {Smith}, {Spitz}, {Stanford}, {Stassun}, {Strader}, {Strauss}, {Stubbs}, {Sweeney}, {Szalay}, {Szkody}, {Takada}, {Thorman}, {Trilling}, {Trimble}, {Tyson}, {Van Berg}, {Vanden Berk}, {VanderPlas}, {Verde}, {Vrsnak}, {Walkowicz}, {Wandelt}, {Wang}, {Wang}, {Warner}, {Wechsler}, {West}, {Wiecha}, {Williams}, {Willman}, {Wittman}, {Wolff}, {Wood-Vasey}, {Wozniak}, {Young}, {Zentner}, \& {Zhan}}]{lsstsciencecollaboration2009lsst}
{LSST Science Collaboration}, {Abell}, P.~A., {Allison}, J., {et~al.} 2009, arXiv e-prints, arXiv:0912.0201

\bibitem[{{Lupton} {et~al.}(2004){Lupton}, {Blanton}, {Fekete}, {Hogg}, {O'Mullane}, {Szalay}, \& {Wherry}}]{Lupton_2004}
{Lupton}, R., {Blanton}, M.~R., {Fekete}, G., {et~al.} 2004, \pasp, 116, 133

\bibitem[{{Marshall} {et~al.}(2016){Marshall}, {Verma}, {More}, {Davis}, {More}, {Kapadia}, {Parrish}, {Snyder}, {Wilcox}, {Baeten}, {Macmillan}, {Cornen}, {Baumer}, {Simpson}, {Lintott}, {Miller}, {Paget}, {Simpson}, {Smith}, {K{\"u}ng}, {Saha}, \& {Collett}}]{Marshall_2016}
{Marshall}, P.~J., {Verma}, A., {More}, A., {et~al.} 2016, \mnras, 455, 1171

\bibitem[{Merz {et~al.}(2023)Merz, Liu, Burke, Aleo, Liu, Carrasco Kind, Kindratenko, \& Liu}]{Merz_2023_trans}
Merz, G., Liu, Y., Burke, C.~J., {et~al.} 2023, \mnras, 526, 1122

\bibitem[{Metcalf {et~al.}(2019)Metcalf, Meneghetti, Avestruz, Bellagamba, Bom, Bertin, Cabanac, Courbin, Davies, Decencière, \& et~al.}]{Metcalf_2019}
Metcalf, R.~B., Meneghetti, M., Avestruz, C., {et~al.} 2019, \aap, 625, A119

\bibitem[{{Metcalf} \& {Petkova}(2014)}]{Metcalf_2014_GLAMER}
{Metcalf}, R.~B. \& {Petkova}, M. 2014, \mnras, 445, 1942

\bibitem[{{Miyazaki} {et~al.}(2012){Miyazaki}, {Komiyama}, {Nakaya}, {Kamata}, {Doi}, {Hamana}, {Karoji}, {Furusawa}, {Kawanomoto}, {Morokuma}, {Ishizuka}, {Nariai}, {Tanaka}, {Uraguchi}, {Utsumi}, {Obuchi}, {Okura}, {Oguri}, {Takata}, {Tomono}, {Kurakami}, {Namikawa}, {Usuda}, {Yamanoi}, {Terai}, {Uekiyo}, {Yamada}, {Koike}, {Aihara}, {Fujimori}, {Mineo}, {Miyatake}, {Yasuda}, {Nishizawa}, {Saito}, {Tanaka}, {Uchida}, {Katayama}, {Wang}, {Chen}, {Lupton}, {Loomis}, {Bickerton}, {Price}, {Gunn}, {Suzuki}, {Miyazaki}, {Muramatsu}, {Yamamoto}, {Endo}, {Ezaki}, {Itoh}, {Miwa}, {Yokota}, {Matsuda}, {Ebinuma}, \& {Takeshi}}]{Miyazaki_2012}
{Miyazaki}, S., {Komiyama}, Y., {Nakaya}, H., {et~al.} 2012, in Society of Photo-Optical Instrumentation Engineers (SPIE) Conference Series, Vol. 8446, Ground-based and Airborne Instrumentation for Astronomy IV, ed. I.~S. {McLean}, S.~K. {Ramsay}, \& H.~{Takami}, 84460Z

\bibitem[{More {et~al.}(2012)More, Cabanac, More, Alard, Limousin, Kneib, Gavazzi, \& Motta}]{More_2012}
More, A., Cabanac, R., More, S., {et~al.} 2012, The Astrophysical Journal, 749, 38

\bibitem[{{More} {et~al.}(2017){More}, {Lee}, {Oguri}, {Ono}, {Suyu}, {Chan}, {Silverman}, {More}, {Schulze}, {Komiyama}, {Matsuoka}, {Miyazaki}, {Nagao}, {Ouchi}, {Tait}, {Tanaka}, {Tanaka}, {Usuda}, \& {Yasuda}}]{More_2017}
{More}, A., {Lee}, C.-H., {Oguri}, M., {et~al.} 2017, \mnras, 465, 2411

\bibitem[{More {et~al.}(2015)More, Verma, Marshall, More, Baeten, Wilcox, Macmillan, Cornen, Kapadia, Parrish, Snyder, Davis, Gavazzi, Lintott, Simpson, Miller, Smith, Paget, Saha, Küng, \& Collett}]{More_2015}
More, A., Verma, A., Marshall, P.~J., {et~al.} 2015, \mnras, 455, 1191

\bibitem[{Mumuni \& Mumuni(2022)}]{Mumuni_2022}
Mumuni, A. \& Mumuni, F. 2022, Array, 16, 100258

\bibitem[{Negrello {et~al.}(2016)Negrello, Amber, Amvrosiadis, Cai, Lapi, Gonzalez-Nuevo, De~Zotti, Furlanetto, Maddox, Allen, Bakx, Bussmann, Cooray, Covone, Danese, Dannerbauer, Fu, Greenslade, Gurwell, Hopwood, Koopmans, Napolitano, Nayyeri, Omont, Petrillo, Riechers, Serjeant, Tortora, Valiante, Verdoes~Kleijn, Vernardos, Wardlow, Baes, Baker, Bourne, Clements, Crawford, Dye, Dunne, Eales, Ivison, Marchetti, Michałowski, Smith, Vaccari, \& van~der Werf}]{Negrello_2016}
Negrello, M., Amber, S., Amvrosiadis, A., {et~al.} 2016, \mnras, 465, 3558

\bibitem[{Negrello {et~al.}(2010)Negrello, Hopwood, Zotti, Cooray, Verma, Bock, Frayer, Gurwell, Omont, Neri, Dannerbauer, Leeuw, Barton, Cooke, Kim, da~Cunha, Rodighiero, Cox, Bonfield, Jarvis, Serjeant, Ivison, Dye, Aretxaga, Hughes, Ibar, Bertoldi, Valtchanov, Eales, Dunne, Driver, Auld, Buttiglione, Cava, Grady, Clements, Dariush, Fritz, Hill, Hornbeck, Kelvin, Lagache, Lopez-Caniego, Gonzalez-Nuevo, Maddox, Pascale, Pohlen, Rigby, Robotham, Simpson, Smith, Temi, Thompson, Woodgate, York, Aguirre, Beelen, Blain, Baker, Birkinshaw, Blundell, Bradford, Burgarella, Danese, Dunlop, Fleuren, Glenn, Harris, Kamenetzky, Lupu, Maddalena, Madore, Maloney, Matsuhara, Michaowski, Murphy, Naylor, Nguyen, Popescu, Rawlings, Rigopoulou, Scott, Scott, Seibert, Smail, Tuffs, Vieira, van~der Werf, \& Zmuidzinas}]{Negrello_2010}
Negrello, M., Hopwood, R., Zotti, G.~D., {et~al.} 2010, Science, 330, 800

\bibitem[{{Neri, R.} {et~al.}(2020){Neri, R.}, {Cox, P.}, {Omont, A.}, {Beelen, A.}, {Berta, S.}, {Bakx, T.}, {Lehnert, M.}, {Baker, A. J.}, {Buat, V.}, {Cooray, A.}, {Dannerbauer, H.}, {Dunne, L.}, {Dye, S.}, {Eales, S.}, {Gavazzi, R.}, {Harris, A. I.}, {Herrera, C. N.}, {Hughes, D.}, {Ivison, R.}, {Jin, S.}, {Krips, M.}, {Lagache, G.}, {Marchetti, L.}, {Messias, H.}, {Negrello, M.}, {Perez-Fournon, I.}, {Riechers, D. A.}, {Serjeant, S.}, {Urquhart, S.}, {Vlahakis, C.}, {Wei\ss{}, A.}, {van der Werf, P.}, {Yang, C.}, \& {Young, A. J.}}]{Neri_2020}
{Neri, R.}, {Cox, P.}, {Omont, A.}, {et~al.} 2020, \aap, 635, A7

\bibitem[{{Nightingale} {et~al.}(2019){Nightingale}, {Massey}, {Harvey}, {Cooper}, {Etherington}, {Tam}, \& {Hayes}}]{Nightingale_2019}
{Nightingale}, J.~W., {Massey}, R.~J., {Harvey}, D.~R., {et~al.} 2019, \mnras, 489, 2049

\bibitem[{Obermeyer {et~al.}(2019)Obermeyer, Powers, Vogeli, \& Mullainathan}]{Obermeyer_2019}
Obermeyer, Z., Powers, B., Vogeli, C., \& Mullainathan, S. 2019, Science, 366, 447

\bibitem[{O'Donnell {et~al.}(2022)O'Donnell, Wilkinson, Diehl, Aros-Bunster, Bechtol, Birrer, Buckley-Geer, Rosell, Kind, da~Costa, Lozano, Gruendl, Hilton, Lin, Lindgren, Martin, Pieres, Rykoff, Sevilla-Noarbe, Sheldon, Sif{\'{o} }n, Tucker, Yanny, Abbott, Aguena, Allam, Andrade-Oliveira, Annis, Bertin, Brooks, Burke, Carretero, Costanzi, Vicente, Desai, Dietrich, Eckert, Everett, Ferrero, Flaugher, Fosalba, Frieman, Garc{\'{\i}}a-Bellido, Gaztanaga, Gerdes, Gruen, Gschwend, Gill, Gutierrez, Hinton, Hollowood, Honscheid, James, Jeltema, Kuehn, Lahav, Lima, Maia, Marshall, Melchior, Menanteau, Miquel, Morgan, Nord, Ogando, Paz-Chinch{\'{o}}n, Pereira, Malag{\'{o}}n, Rodriguez-Monroy, Romer, Roodman, Sanchez, Scarpine, Schubnell, Serrano, Smith, Suchyta, Swanson, Tarle, Thomas, To, \& Varga}]{O_Donnell_2022}
O'Donnell, J.~H., Wilkinson, R.~D., Diehl, H.~T., {et~al.} 2022, The Astrophysical Journal Supplement Series, 259, 27

\bibitem[{Oguri \& Marshall(2010)}]{Oguri_2010}
Oguri, M. \& Marshall, P.~J. 2010, \mnras, 405, 2579

\bibitem[{Pan \& Yang(2010)}]{Pan_2010}
Pan, S.~J. \& Yang, Q. 2010, IEEE Transactions on Knowledge and Data Engineering, 22, 1345

\bibitem[{Paul \& Chen(2022)}]{Paul_Chen_2022}
Paul, S. \& Chen, P.-Y. 2022, Proceedings of the AAAI Conference on Artificial Intelligence, 36, 2071

\bibitem[{Petkova {et~al.}(2014)Petkova, Metcalf, \& Giocoli}]{Petkova_2014_glamer}
Petkova, M., Metcalf, R.~B., \& Giocoli, C. 2014, \mnras, 445, 1954

\bibitem[{Petrillo {et~al.}(2018)Petrillo, Tortora, Chatterjee, Vernardos, Koopmans, Kleijn, Napolitano, Covone, Kelvin, \& Hopkins}]{Petrillo_2018}
Petrillo, C.~E., Tortora, C., Chatterjee, S., {et~al.} 2018, MNRAS

\bibitem[{{Petrillo} {et~al.}(2017){Petrillo}, {Tortora}, {Chatterjee}, {Vernardos}, {Koopmans}, {Verdoes Kleijn}, {Napolitano}, {Covone}, {Schneider}, {Grado}, \& {McFarland}}]{Petrillo_2017}
{Petrillo}, C.~E., {Tortora}, C., {Chatterjee}, S., {et~al.} 2017, MNRAS, 472, 1129

\bibitem[{Petrillo {et~al.}(2019)Petrillo, Tortora, Vernardos, Koopmans, Verdoes Kleijn, Bilicki, Napolitano, Chatterjee, Covone, Dvornik, Erben, Getman, Giblin, Heymans, de Jong, Kuijken, Schneider, Shan, Spiniello, \& Wright}]{links}
Petrillo, C.~E., Tortora, C., Vernardos, G., {et~al.} 2019, \mnras, 484, 3879

\bibitem[{Rezaei {et~al.}(2022)Rezaei, McKean, Biehl, de Roo, \& Lafontaine}]{Rezaei_2022_ML}
Rezaei, S., McKean, J.~P., Biehl, M., de Roo, W., \& Lafontaine, A. 2022, \mnras, 517, 1156

\bibitem[{Ribeiro {et~al.}(2016)Ribeiro, Singh, \& Guestrin}]{ribeiro_2016}
Ribeiro, M.~T., Singh, S., \& Guestrin, C. 2016, "Why Should I Trust You?": Explaining the Predictions of Any Classifier

\bibitem[{Rigby {et~al.}(2014)Rigby, Bayliss, Gladders, Sharon, Wuyts, \& Dahle}]{Rigby_2014}
Rigby, J.~R., Bayliss, M.~B., Gladders, M.~D., {et~al.} 2014, The Astrophysical Journal, 790, 44

\bibitem[{Rivera {et~al.}(2019)Rivera, Baker, Gallardo, Gralla, Harris, Huffenberger, Hughes, Keeton, López-Caraballo, Marriage, Partridge, Sievers, Tagore, Walter, Weiß, \& Wollack}]{Rivera_2019}
Rivera, J., Baker, A.~J., Gallardo, P.~A., {et~al.} 2019, The Astrophysical Journal, 879, 95

\bibitem[{Rojas {et~al.}(2023)Rojas, Collett, Ballard, Magee, Birrer, Buckley-Geer, Chan, Clément, Diego, Gentile, González, Joseph, Mastache, Schuldt, Tortora, Verdugo, Verma, Daylan, Millon, Jackson, Dye, Melo, Mahler, Ogando, Courbin, Fritz, Herle, Acevedo Barroso, Cañameras, Cornen, Dhanasingham, Glazebrook, Martinez, Ryczanowski, Savary, Góis-Silva, Arturo Ureña-López, Wiesner, Wilde, Valim Calçada, Cabanac, Pan, Sierra, Despali, Cavalcante-Gomes, Macmillan, Maresca, Grudskaia, O’Donnell, Paic, Niemiec, de la Bella, Bromley, Williams, More, \& Levine.}]{Rojas_2023}
Rojas, K., Collett, T.~E., Ballard, D., {et~al.} 2023, \mnras, 523, 4413

\bibitem[{{Rojas} {et~al.}(2021){Rojas}, {Savary}, {Cl{\'e}ment}, {Maus}, {Courbin}, {Lemon}, {Chan}, {Vernardos}, {Joseph}, {Ca{\~n}ameras}, \& {Galan}}]{rojas2021strong}
{Rojas}, K., {Savary}, E., {Cl{\'e}ment}, B., {et~al.} 2021, arXiv e-prints, arXiv:2109.00014

\bibitem[{Schaefer {et~al.}(2018)Schaefer, Geiger, Kuntzer, \& Kneib}]{Schaefer_2018}
Schaefer, C., Geiger, M., Kuntzer, T., \& Kneib, J.-P. 2018, \aap, 611, A2

\bibitem[{{Schneider} {et~al.}(1992){Schneider}, {Ehlers}, \& {Falco}}]{Schneider_1992}
{Schneider}, P., {Ehlers}, J., \& {Falco}, E.~E. 1992, {Gravitational Lenses}

\bibitem[{Selvaraju {et~al.}(2019)Selvaraju, Cogswell, Das, Vedantam, Parikh, \& Batra}]{Selvaraju_2019}
Selvaraju, R.~R., Cogswell, M., Das, A., {et~al.} 2019, International Journal of Computer Vision, 128, 336–359

\bibitem[{{Shu, Yiping} {et~al.}(2022){Shu, Yiping}, {Ca\~nameras, Raoul}, {Schuldt, Stefan}, {Suyu, Sherry H.}, {Taubenberger, Stefan}, {Inoue, Kaiki Taro}, \& {Jaelani, Anton T.}}]{Shu_2022}
{Shu, Yiping}, {Ca\~nameras, Raoul}, {Schuldt, Stefan}, {et~al.} 2022, \aap, 662, A4

\bibitem[{Simonyan \& Zisserman(2015)}]{simonyan_2015}
Simonyan, K. \& Zisserman, A. 2015, Very Deep Convolutional Networks for Large-Scale Image Recognition

\bibitem[{Sonnenfeld {et~al.}(2017)Sonnenfeld, Chan, Shu, More, Oguri, Suyu, Wong, Lee, Coupon, Yonehara, Bolton, Jaelani, Tanaka, Miyazaki, \& Komiyama}]{Sonnenfeld_2017_sugohi}
Sonnenfeld, A., Chan, J. H.~H., Shu, Y., {et~al.} 2017, Publications of the Astronomical Society of Japan, 70

\bibitem[{Sonnenfeld {et~al.}(2020)Sonnenfeld, Verma, More, Baeten, Macmillan, Wong, Chan, Jaelani, Lee, Oguri, Rusu, Veldthuis, Trouille, Marshall, Hutchings, Allen, Donnell, Cornen, Davis, McMaster, Lintott, \& Miller}]{Sonnenfeld_2020_sugohi}
Sonnenfeld, A., Verma, A., More, A., {et~al.} 2020, \aap, 642, A148

\bibitem[{Spilker {et~al.}(2016)Spilker, Marrone, Aravena, Béthermin, Bothwell, Carlstrom, Chapman, Crawford, de~Breuck, Fassnacht, Gonzalez, Greve, Hezaveh, Litke, Ma, Malkan, Rotermund, Strandet, Vieira, Weiss, \& Welikala}]{Spilker_2016}
Spilker, J.~S., Marrone, D.~P., Aravena, M., {et~al.} 2016, The Astrophysical Journal, 826, 112

\bibitem[{Stein {et~al.}(2022)Stein, Blaum, Harrington, Medan, \& Lukić}]{Stein_2022}
Stein, G., Blaum, J., Harrington, P., Medan, T., \& Lukić, Z. 2022, The Astrophysical Journal, 932, 107

\bibitem[{Storfer {et~al.}(2023)Storfer, Huang, Gu, Sheu, Banka, Dey, Jain, Kwon, Lang, Lee, Meisner, Moustakas, Myers, Tabares-Tarquinio, Schlafly, \& Schlegel}]{storfer_2023}
Storfer, C., Huang, X., Gu, A., {et~al.} 2023, New Strong Gravitational Lenses from the DESI Legacy Imaging Surveys Data Release 9

\bibitem[{Szegedy {et~al.}(2015)Szegedy, Liu, Jia, Sermanet, Reed, Anguelov, Erhan, Vanhoucke, \& Rabinovich}]{Szegedy_2015}
Szegedy, C., Liu, W., Jia, Y., {et~al.} 2015, in 2015 IEEE Conference on Computer Vision and Pattern Recognition (CVPR), 1--9

\bibitem[{Tan \& Le(2020)}]{tan_2020_efficientnet}
Tan, M. \& Le, Q.~V. 2020, EfficientNet: Rethinking Model Scaling for Convolutional Neural Networks

\bibitem[{{Thuruthipilly} {et~al.}(2022{\natexlab{a}}){Thuruthipilly}, {Grespan}, \& {Zadro{\.z}ny}}]{Hareesh2}
{Thuruthipilly}, H., {Grespan}, M., \& {Zadro{\.z}ny}, A. 2022{\natexlab{a}}, arXiv e-prints, arXiv:2212.12915

\bibitem[{{Thuruthipilly} {et~al.}(2024){Thuruthipilly}, {Junais}, {Pollo}, {Sureshkumar}, {Grespan}, {Sawant}, {Ma{\l}ek}, \& {Zadrozny}}]{haressh_lsbs}
{Thuruthipilly}, H., {Junais}, {Pollo}, A., {et~al.} 2024, \aap, 682, A4

\bibitem[{{Thuruthipilly} {et~al.}(2022{\natexlab{b}}){Thuruthipilly}, {Zadrozny}, {Pollo}, \& {Biesiada}}]{Hareesh}
{Thuruthipilly}, H., {Zadrozny}, A., {Pollo}, A., \& {Biesiada}, M. 2022{\natexlab{b}}, A\&A, 664, A4

\bibitem[{Timmis \& Shamir(2017)}]{Timmis_2017}
Timmis, I. \& Shamir, L. 2017, The Astrophysical Journal Supplement Series, 231, 2

\bibitem[{Tran {et~al.}(2022)Tran, Harshan, Glazebrook, Vasan, Jones, Jacobs, Kacprzak, Barone, Collett, Gupta, Henderson, Kewley, Lopez, Nanayakkara, Sanders, \& Sweet}]{Tran_2022}
Tran, K.-V.~H., Harshan, A., Glazebrook, K., {et~al.} 2022, The Astronomical Journal, 164, 148

\bibitem[{{Turner} {et~al.}(1984){Turner}, {Ostriker}, \& {Gott}}]{Turner_1984}
{Turner}, E.~L., {Ostriker}, J.~P., \& {Gott}, J.~R., I. 1984, \apj, 284, 1

\bibitem[{Vaswani {et~al.}(2017)Vaswani, Shazeer, Parmar, Uszkoreit, Jones, Gomez, Kaiser, \& Polosukhin}]{vaswani2017attention}
Vaswani, A., Shazeer, N., Parmar, N., {et~al.} 2017, in Advances in Neural Information Processing Systems 30: Annual Conference on Neural Information Processing Systems 2017, December 4-9, 2017, Long Beach, CA, {USA}, 5998--6008

\bibitem[{{Verma} {et~al.}(2019){Verma}, {Collett}, {Smith}, {Strong Lensing Science Collaboration}, \& {the DESC Strong Lensing Science Working Group}}]{verma2019strong}
{Verma}, A., {Collett}, T., {Smith}, G.~P., {Strong Lensing Science Collaboration}, \& {the DESC Strong Lensing Science Working Group}. 2019, arXiv e-prints, arXiv:1902.05141

\bibitem[{Wang {et~al.}(2019)Wang, Ng, Ma, Nallapati, \& Xiang}]{wang2019multipassage}
Wang, Z., Ng, P., Ma, X., Nallapati, R., \& Xiang, B. 2019, Multi-passage BERT: A Globally Normalized BERT Model for Open-domain Question Answering

\bibitem[{{Wei} {et~al.}(2022){Wei}, {Chen}, {Cao}, \& {Wu}}]{Wei2022}
{Wei}, J.-J., {Chen}, Y., {Cao}, S., \& {Wu}, X.-F. 2022, \apjl, 927, L1

\bibitem[{Wenger {et~al.}(2000)Wenger, Ochsenbein, Egret, Dubois, Bonnarel, Borde, Genova, Jasniewicz, Laloë, Lesteven, \& Monier}]{Wenger_2000_simbad}
Wenger, M., Ochsenbein, F., Egret, D., {et~al.} 2000, Astronomy and Astrophysics Supplement Series, 143, 9

\bibitem[{Wilde {et~al.}(2022)Wilde, Serjeant, Bromley, Dickinson, Koopmans, \& Metcalf}]{Wilde_2022_ML}
Wilde, J., Serjeant, S., Bromley, J.~M., {et~al.} 2022, \mnras, 512, 3464

\bibitem[{Wong {et~al.}(2022)Wong, Chan, Chao, Jaelani, Kayo, Lee, More, \& Oguri}]{Wong_2022_sugohi}
Wong, K.~C., Chan, J. H.~H., Chao, D. C.-Y., {et~al.} 2022, Publications of the Astronomical Society of Japan, 74, 1209

\bibitem[{Wortsman {et~al.}(2022)Wortsman, Ilharco, Gadre, Roelofs, Gontijo-Lopes, Morcos, Namkoong, Farhadi, Carmon, Kornblith, \& Schmidt}]{Mitchell}
Wortsman, M., Ilharco, G., Gadre, S.~Y., {et~al.} 2022, in International Conference on Machine Learning

\bibitem[{Xu {et~al.}(2023)Xu, Yoon, Fuentes, \& Park}]{Xu_2023}
Xu, M., Yoon, S., Fuentes, A., \& Park, D.~S. 2023, Pattern Recognition, 137, 109347

\bibitem[{Yip {et~al.}(2021)Yip, Changeat, Nikolaou, Morvan, Edwards, Waldmann, \& Tinetti}]{yip2021peeking}
Yip, K.~H., Changeat, Q., Nikolaou, N., {et~al.} 2021, Peeking inside the Black Box: Interpreting Deep Learning Models for Exoplanet Atmospheric Retrievals

\bibitem[{Yosinski {et~al.}(2014)Yosinski, Clune, Bengio, \& Lipson}]{Yosinski_2014}
Yosinski, J., Clune, J., Bengio, Y., \& Lipson, H. 2014, in Advances in Neural Information Processing Systems, ed. Z.~Ghahramani, M.~Welling, C.~Cortes, N.~Lawrence, \& K.~Weinberger, Vol.~27 (Curran Associates, Inc.)

\bibitem[{Yu {et~al.}(2022)Yu, Wang, Vasudevan, Yeung, Seyedhosseini, \& Wu}]{Jiahui}
Yu, J., Wang, Z., Vasudevan, V., {et~al.} 2022, Trans. Mach. Learn. Res., 2022

\bibitem[{Zaborowski {et~al.}(2023)Zaborowski, Drlica-Wagner, Ashmead, Wu, Morgan, Bom, Shajib, Birrer, Cerny, Buckley-Geer, Mutlu-Pakdil, Ferguson, Glazebrook, Lozano, Gordon, Martinez, Manwadkar, O'Donnell, Poh, Riley, Sakowska, Santana-Silva, Santiago, Sluse, Tan, Tollerud, Verma, Carballo-Bello, Choi, James, Kuropatkin, Mart{\'{\i} }nez-V{\'{a}}zquez, Nidever, Castellon, Noël, Olsen, Pace, Mau, Yanny, Zenteno, Abbott, Aguena, Alves, Andrade-Oliveira, Bocquet, Brooks, Burke, Rosell, Kind, Carretero, Castander, Conselice, Costanzi, Pereira, Vicente, Desai, Dietrich, Doel, Everett, Ferrero, Flaugher, Friedel, Frieman, Garc{\'{\i}}a-Bellido, Gruen, Gruendl, Gutierrez, Hinton, Hollowood, Honscheid, Kuehn, Lin, Marshall, Melchior, Mena-Fern{\'{a}}ndez, Menanteau, Miquel, Palmese, Paz-Chinch{\'{o}}n, Pieres, Malag{\'{o}}n, Prat, Rodriguez-Monroy, Romer, Sanchez, Scarpine, Sevilla-Noarbe, Smith, Suchyta, To, \& Weaverdyck}]{Zaborowski_2023}
Zaborowski, E.~A., Drlica-Wagner, A., Ashmead, F., {et~al.} 2023, The Astrophysical Journal, 954, 68

\end{thebibliography}
\onecolumn
\begin{appendix}

\section{Metrics for labeled dataset} 
\label{sec:metric_labeled}

Metrics are necessary to evaluate the effectiveness of machine learning models. In our case, metrics provide a quantitative measure of a model's ability to distinguish between lenses and non-lenses.
The number of true positives (TP) and true negatives (TN), representing correctly classified images, and the number of false positives (FP) and negatives (FN), representing misclassified images, are ideal metrics but often unavailable for real-world datasets due to the difficulty of obtaining labeled data. In the Bologna Lens Challenge, where labels were readily available, the models were evaluated using the area under the receiver operating characteristic curve (AUROC) and true positive rate (TPR). 

The receiver operating characteristic (ROC) curve is constructed by plotting the TPR against the false positive rate (FPR) at different thresholds. The area enclosed by this curve represents the AUROC value. A perfect model would have an AUROC of 1.0, while a completely random model would have an AUROC of 0.5.

TPR, also called sensitivity, is defined as: 
\begin{equation}
    TPR = \frac{TP}{TP + FN} \;,
\end{equation}
and the FPR as:
\begin{equation}
    FPR = \frac{FP} {FP + TN}.
\end{equation} 

\section{Grade 2 candidates}

\begin{table*}[!htbp]
 \footnotesize
 \centering
   \caption{Candidates identified through visual inspection with grade 2. 
    `ID' is the index of the object in Fig. \ref{fig:grade_2_0} and \ref{fig:grade_2_1}. 
    `KiDS ID' is the object's ID in the KiDS catalog where we omitted 'KiDSDR4 ' present in every ID before the ICRS coordinates. 
    `R.A.' and `Dec.' are the object's coordinates in degrees and  z$_{phot}$ is the photometric redshift as reported in the KiDS catalog. `Test' specifies whether the object was found before (BF) or after (AF) fine-tuning of the model or whether it was found during the search within the GAMA (G) footprint. 
 \label{tab:all_candidates_2}}
 \begin{adjustbox}{width=1.05\textwidth}
\begin{tabular}{llllll|llllll}
\hline\hline\noalign{\smallskip}
ID & KiDS ID & R.A. J2000 & Dec. J2000 & $z_{phot}$ & test & ID & KiDS ID & R.A. J2000 & Dec. J2000 & $z_{phot}$ & test \\
\hline\noalign{\smallskip}
0  & KIDS\_0.0\_-28.2   & 0.104   & -28.091 & 0.28 & BF   & 97  & KIDS\_183.0\_-0.5  & 182.829 & -0.547  & 0.45 & G    \\
1  & KIDS\_0.0\_-28.2   & 0.071   & -28.072 & 0.6  & BF   & 98  & KIDS\_183.0\_-1.5  & 183.442 & -1.799  & 0.62 & G    \\
2  & KIDS\_156.0\_1.5   & 156.397 & 1.362   & 0.4  & BF   & 99  & KIDS\_183.0\_-1.5  & 182.601 & -1.61   & 0.52 & G    \\
3  & KIDS\_168.0\_1.5   & 167.691 & 1.977   & 0.32 & BF   & 100 & KIDS\_183.0\_-1.5  & 182.778 & -1.334  & 0.43 & G    \\
4  & KIDS\_212.6\_2.5   & 212.181 & 2.14    & 0.58 & BF   & 101 & KIDS\_183.0\_-1.5  & 183.365 & -1.34   & 0.38 & G    \\
5  & KIDS\_212.6\_2.5   & 212.573 & 2.22    & 0.07 & BF   & 102 & KIDS\_183.0\_-1.5  & 182.6   & -1.564  & 0.42 & G    \\
6  & KIDS\_53.5\_-35.1  & 53.1    & -35.491 & 0.4  & BF   & 103 & KIDS\_183.0\_-1.5  & 182.881 & -1.189  & 0.73 & G    \\
7  & KIDS\_138.0\_0.5   & 137.658 & 0.401   & 0.6  & BF   & 104 & KIDS\_183.0\_1.5   & 182.918 & 1.915   & 0.29 & G    \\
8  & KIDS\_40.6\_-28.2  & 40.968  & -28.52  & 0.04 & AF   & 105 & KIDS\_184.0\_-0.5  & 183.89  & -0.979  & 0.58 & G    \\
9  & KIDS\_131.0\_1.5   & 130.532 & 1.703   & 0.16 & AF   & 106 & KIDS\_184.0\_0.5   & 184.208 & 0.047   & 0.44 & G    \\
10 & KIDS\_179.0\_-0.5  & 178.639 & -0.701  & 0.29 & AF   & 107 & KIDS\_184.5\_-2.5  & 184.054 & -2.461  & 0.47 & G    \\
11 & KIDS\_167.0\_-0.5  & 166.581 & -0.932  & 0.71 & AF   & 108 & KIDS\_185.0\_-1.5  & 184.767 & -1.807  & 0.7  & G    \\
12 & KIDS\_32.7\_-28.2  & 32.669  & -27.753 & 0.46 & AF   & 109 & KIDS\_185.0\_-1.5  & 184.641 & -1.053  & 0.42 & G    \\
13 & KIDS\_167.0\_-0.5  & 167.07  & -0.943  & 0.47 & AF   & 110 & KIDS\_185.0\_0.5   & 185.304 & 0.188   & 0.5  & G    \\
14 & KIDS\_131.0\_1.5   & 131.145 & 1.185   & 0.6  & AF   & 111 & KIDS\_185.0\_0.5   & 185.201 & 0.363   & 0.47 & G    \\
15 & KIDS\_167.0\_-0.5  & 167.42  & -0.309  & 0.27 & AF   & 112 & KIDS\_185.0\_1.5   & 185.444 & 1.563   & 0.37 & G    \\
16 & KIDS\_131.0\_1.5   & 130.546 & 1.643   & 0.51 & AF   & 113 & KIDS\_212.0\_-1.5  & 211.772 & -1.216  & 0.72 & G    \\
17 & KIDS\_340.8\_-28.2 & 340.287 & -28.284 & 0.25 & AF   & 114 & KIDS\_212.0\_-1.5  & 211.89  & -1.823  & 0.42 & G    \\
18 & KIDS\_179.0\_-0.5  & 179.286 & -0.613  & 0.2  & AF   & 115 & KIDS\_212.6\_2.5   & 212.638 & 2.376   & 0.51 & G    \\
19 & KIDS\_179.0\_-0.5  & 179.489 & -0.957  & 0.48 & AF   & 116 & KIDS\_213.0\_0.5   & 213.46  & 0.869   & 0.34 & G    \\
20 & KIDS\_129.4\_2.5   & 129.509 & 2.881   & 0.44 & G    & 117 & KIDS\_214.0\_-0.5  & 213.645 & -0.987  & 0.36 & G    \\
21 & KIDS\_129.4\_2.5   & 129.566 & 2.684   & 0.33 & G    & 118 & KIDS\_214.0\_-0.5  & 213.683 & -0.56   & 0.45 & G    \\
22 & KIDS\_130.0\_-0.5  & 130.011 & -0.459  & 0.42 & G    & 119 & KIDS\_214.0\_-1.5  & 214.167 & -1.262  & 0.41 & G    \\
23 & KIDS\_130.0\_-1.5  & 129.909 & -1.695  & 0.42 & G    & 120 & KIDS\_214.0\_1.5   & 214.115 & 1.626   & 0.42 & G    \\
24 & KIDS\_130.0\_-1.5  & 129.783 & -1.388  & 0.48 & G    & 121 & KIDS\_214.0\_1.5   & 213.927 & 1.539   & 0.23 & G    \\
25 & KIDS\_130.4\_2.5   & 130.82  & 2.488   & 0.24 & G    & 122 & KIDS\_214.6\_2.5   & 214.367 & 1.993   & 0.49 & G    \\
26 & KIDS\_132.0\_-1.5  & 131.842 & -1.578  & 0.49 & G    & 123 & KIDS\_214.6\_2.5   & 214.624 & 2.518   & 0.65 & G    \\
27 & KIDS\_132.0\_0.5   & 131.846 & 0.824   & 0.48 & G    & 124 & KIDS\_214.6\_2.5   & 214.808 & 2.943   & 0.42 & G    \\
28 & KIDS\_133.4\_2.5   & 133.417 & 2.576   & 0.61 & G    & 125 & KIDS\_215.0\_-0.5  & 214.827 & -0.42   & 0.35 & G    \\
29 & KIDS\_134.0\_-1.5  & 133.69  & -1.568  & 0.27 & G    & 126 & KIDS\_215.0\_0.5   & 214.972 & 0.029   & 0.27 & G    \\
30 & KIDS\_134.0\_-1.5  & 133.537 & -1.629  & 0.41 & G    & 127 & KIDS\_215.0\_1.5   & 214.775 & 1.83    & 0.44 & G    \\
31 & KIDS\_134.0\_-1.5  & 134.316 & -1.178  & 0.2  & G    & 128 & KIDS\_215.0\_1.5   & 214.861 & 1.178   & 0.42 & G    \\
32 & KIDS\_134.0\_-1.5  & 133.913 & -1.297  & 0.2  & G    & 129 & KIDS\_215.0\_1.5   & 214.851 & 1.678   & 0.45 & G    \\

\hline
\end{tabular}
\end{adjustbox}
\end{table*}

\addtocounter{table}{-1}
\begin{table*}[!htbp]
\footnotesize
\centering 
\caption{ (Continued)}
 \begin{adjustbox}{width=1.05\textwidth}
\begin{tabular}{llllll|llllll}
\hline\hline\noalign{\smallskip}
ID & KiDS ID & R.A. J2000 & Dec. J2000 & $z_{phot}$ & test & ID & KiDS ID & R.A. J2000 & Dec. J2000 & $z_{phot}$ & test \\
\hline\noalign{\smallskip}

33 & KIDS\_134.0\_0.5   & 133.824 & 0.35    & 0.56 & G    & 130 & KIDS\_216.0\_-0.5  & 215.707 & -0.553  & 0.48 & G    \\
34 & KIDS\_135.0\_-0.5  & 135.151 & -0.116  & 0.43 & G    & 131 & KIDS\_216.0\_-0.5  & 215.64  & -0.608  & 0.51 & G    \\
35 & KIDS\_135.0\_-1.5  & 135.459 & -1.355  & 0.43 & G    & 132 & KIDS\_216.0\_-1.5  & 216.367 & -1.251  & 0.6  & G    \\
36 & KIDS\_135.0\_-1.5  & 135.218 & -1.042  & 0.46 & G    & 133 & KIDS\_216.0\_0.5   & 215.913 & 0.011   & 0.22 & G    \\
37 & KIDS\_136.0\_-0.5  & 135.889 & -0.975  & 0.49 & G    & 134 & KIDS\_216.0\_0.5   & 215.709 & 0.106   & 0.35 & G    \\
38 & KIDS\_136.0\_-0.5  & 135.728 & -0.963  & 0.61 & G    & 135 & KIDS\_216.0\_0.5   & 216.094 & 0.061   & 0.56 & G    \\
39 & KIDS\_136.0\_-0.5  & 136.393 & -0.471  & 0.23 & G    & 136 & KIDS\_216.0\_0.5   & 216.303 & 0.491   & 0.6  & G    \\
40 & KIDS\_136.0\_-0.5  & 136.035 & -0.136  & 0.64 & G    & 137 & KIDS\_216.0\_1.5   & 215.551 & 1.522   & 0.38 & G    \\
41 & KIDS\_136.0\_-0.5  & 136.32  & -0.123  & 0.56 & G    & 138 & KIDS\_217.0\_-0.5  & 217.224 & -0.829  & 0.49 & G    \\
42 & KIDS\_136.0\_-0.5  & 136.197 & -0.89   & 0.45 & G    & 139 & KIDS\_217.0\_0.5   & 217.04  & 0.202   & 0.44 & G    \\
43 & KIDS\_136.0\_-0.5  & 135.942 & -0.815  & 0.46 & G    & 140 & KIDS\_217.0\_0.5   & 216.778 & 0.721   & 0.42 & G    \\
44 & KIDS\_136.0\_-0.5  & 135.776 & -0.818  & 0.66 & G    & 141 & KIDS\_217.0\_0.5   & 217.066 & 0.229   & 0.43 & G    \\
45 & KIDS\_137.0\_0.5   & 137.017 & 0.358   & 0.3  & G    & 142 & KIDS\_217.0\_0.5   & 217.167 & 0.982   & 0.74 & G    \\
46 & KIDS\_138.0\_-0.5  & 138.231 & -0.705  & 0.27 & G    & 143 & KIDS\_217.0\_0.5   & 216.838 & 0.593   & 0.44 & G    \\
47 & KIDS\_138.0\_-0.5  & 137.982 & -0.732  & 0.56 & G    & 144 & KIDS\_218.0\_-1.5  & 217.633 & -1.075  & 0.14 & G    \\
48 & KIDS\_138.0\_1.5   & 137.703 & 1.683   & 0.61 & G    & 145 & KIDS\_218.0\_-1.5  & 218.321 & -1.18   & 0.32 & G    \\
49 & KIDS\_138.4\_2.5   & 138.818 & 2.623   & 0.49 & G    & 146 & KIDS\_218.0\_-1.5  & 218.31  & -1.951  & 0.58 & G    \\
50 & KIDS\_139.0\_-0.5  & 138.848 & -0.004  & 0.72 & G    & 147 & KIDS\_218.0\_0.5   & 218.315 & 0.503   & 0.44 & G    \\
51 & KIDS\_139.0\_-1.5  & 138.86  & -1.489  & 0.5  & G    & 148 & KIDS\_218.6\_2.5   & 218.455 & 2.857   & 0.42 & G    \\
52 & KIDS\_139.0\_-1.5  & 138.894 & -1.895  & 0.41 & G    & 149 & KIDS\_220.0\_-1.5  & 220.433 & -1.297  & 0.41 & G    \\
53 & KIDS\_139.0\_-1.5  & 138.719 & -1.073  & 0.56 & G    & 150 & KIDS\_220.0\_-1.5  & 219.773 & -1.783  & 0.45 & G    \\
54 & KIDS\_139.0\_-1.5  & 138.675 & -1.508  & 0.19 & G    & 151 & KIDS\_221.0\_-1.5  & 221.267 & -1.836  & 0.48 & G    \\
55 & KIDS\_139.0\_-1.5  & 138.977 & -1.278  & 0.6  & G    & 152 & KIDS\_221.0\_-1.5  & 220.681 & -1.554  & 0.33 & G    \\
56 & KIDS\_139.0\_1.5   & 138.649 & 1.858   & 0.46 & G    & 153 & KIDS\_222.0\_-0.5  & 221.947 & -0.443  & 0.34 & G    \\
57 & KIDS\_140.0\_-1.5  & 139.543 & -1.594  & 0.32 & G    & 154 & KIDS\_222.0\_-0.5  & 222.072 & -0.088  & 0.42 & G    \\
58 & KIDS\_140.0\_-1.5  & 139.8   & -1.682  & 0.59 & G    & 155 & KIDS\_339.0\_-31.2 & 339.414 & -31.247 & 0.61 & G    \\
59 & KIDS\_140.0\_-1.5  & 139.565 & -1.455  & 0.33 & G    & 156 & KIDS\_339.6\_-34.1 & 339.884 & -34.525 & 0.39 & G    \\
60 & KIDS\_140.0\_-1.5  & 139.708 & -1.924  & 0.44 & G    & 157 & KIDS\_340.4\_-30.2 & 340.17  & -29.902 & 0.43 & G    \\
61 & KIDS\_140.0\_-1.5  & 139.917 & -1.275  & 0.44 & G    & 158 & KIDS\_341.6\_-30.2 & 341.442 & -29.933 & 0.77 & G    \\
62 & KIDS\_140.0\_1.5   & 139.936 & 1.623   & 0.48 & G    & 159 & KIDS\_341.6\_-30.2 & 342.106 & -30.251 & 0.14 & G    \\
63 & KIDS\_175.0\_-1.5  & 175.203 & -1.691  & 0.31 & G    & 160 & KIDS\_342.0\_-34.1 & 341.802 & -33.641 & 0.33 & G    \\
64 & KIDS\_175.0\_0.5   & 175.335 & 0.424   & 0.51 & G    & 161 & KIDS\_342.2\_-33.1 & 341.604 & -33.225 & 0.32 & G    \\
65 & KIDS\_175.0\_0.5   & 175.341 & 0.644   & 0.41 & G    & 162 & KIDS\_342.5\_-31.2 & 342.498 & -31.313 & 0.48 & G    \\
66 & KIDS\_175.0\_0.5   & 174.567 & 0.683   & 0.59 & G    & 163 & KIDS\_343.4\_-33.1 & 343.102 & -33.579 & 0.5  & G    \\
67 & KIDS\_175.0\_0.5   & 174.636 & 0.33    & 0.6  & G    & 164 & KIDS\_343.4\_-33.1 & 343.904 & -32.957 & 0.47 & G    \\
68 & KIDS\_176.0\_0.5   & 175.596 & 0.338   & 0.39 & G    & 165 & KIDS\_344.9\_-31.2 & 344.324 & -31.267 & 0.6  & G    \\
69 & KIDS\_176.5\_-2.5  & 176.753 & -2.021  & 0.46 & G    & 166 & KIDS\_345.0\_-30.2 & 344.802 & -29.785 & 0.41 & G    \\
70 & KIDS\_177.0\_-1.5  & 176.588 & -1.681  & 0.44 & G    & 167 & KIDS\_345.7\_-33.1 & 346.22  & -32.868 & 0.5  & G    \\
71 & KIDS\_178.0\_-1.5  & 178.308 & -1.041  & 0.32 & G    & 168 & KIDS\_345.7\_-33.1 & 345.194 & -33.106 & 0.43 & G    \\
72 & KIDS\_178.0\_-1.5  & 178.1   & -1.945  & 0.66 & G    & 169 & KIDS\_345.9\_-32.1 & 345.527 & -32.611 & 0.3  & G    \\
73 & KIDS\_178.0\_-1.5  & 178.32  & -1.748  & 0.53 & G    & 170 & KIDS\_345.9\_-32.1 & 345.471 & -32.117 & 0.31 & G    \\
74 & KIDS\_179.0\_1.5   & 179.203 & 1.179   & 0.52 & G    & 171 & KIDS\_346.0\_-31.2 & 346.365 & -31.617 & 0.32 & G    \\
75 & KIDS\_179.0\_1.5   & 179.339 & 1.952   & 0.46 & G    & 172 & KIDS\_347.1\_-32.1 & 347.461 & -32.389 & 0.47 & G    \\
76 & KIDS\_179.0\_1.5   & 178.549 & 1.851   & 0.79 & G    & 173 & KIDS\_347.2\_-31.2 & 346.8   & -31.164 & 0.46 & G    \\
77 & KIDS\_179.0\_1.5   & 178.875 & 1.887   & 0.47 & G    & 174 & KIDS\_347.3\_-30.2 & 346.92  & -29.85  & 0.38 & G    \\
78 & KIDS\_179.0\_1.5   & 178.915 & 1.14    & 0.43 & G    & 175 & KIDS\_347.3\_-30.2 & 347.539 & -30.523 & 0.33 & G    \\
79 & KIDS\_179.5\_-2.5  & 179.245 & -2.455  & 0.41 & G    & 176 & KIDS\_348.0\_-34.1 & 348.148 & -34.493 & 0.32 & G    \\
80 & KIDS\_179.5\_-2.5  & 179.73  & -2.519  & 0.21 & G    & 177 & KIDS\_348.0\_-34.1 & 347.972 & -34.541 & 0.63 & G    \\
81 & KIDS\_179.5\_-2.5  & 179.429 & -2.409  & 0.47 & G    & 178 & KIDS\_348.0\_-34.1 & 348.097 & -34.526 & 0.04 & G    \\
82 & KIDS\_180.0\_-0.5  & 179.68  & -0.116  & 0.39 & G    & 179 & KIDS\_348.0\_-34.1 & 347.807 & -33.808 & 0.25 & G    \\
83 & KIDS\_180.0\_1.5   & 180.353 & 1.362   & 0.47 & G    & 180 & KIDS\_348.0\_-34.1 & 347.89  & -34.323 & 0.62 & G    \\
84 & KIDS\_180.5\_-2.5  & 180.849 & -2.068  & 0.47 & G    & 181 & KIDS\_348.0\_-34.1 & 348.133 & -34.33  & 0.72 & G    \\
85 & KIDS\_181.0\_-1.5  & 180.702 & -1.484  & 0.43 & G    & 182 & KIDS\_348.0\_-34.1 & 347.916 & -34.33  & 0.46 & G    \\
86 & KIDS\_181.0\_-1.5  & 181.015 & -1.645  & 0.45 & G    & 183 & KIDS\_348.0\_-34.1 & 347.464 & -34.469 & 0.2  & G    \\
87 & KIDS\_181.0\_-1.5  & 181.298 & -1.355  & 0.71 & G    & 184 & KIDS\_348.0\_-34.1 & 347.884 & -34.399 & 0.33 & G    \\
88 & KIDS\_181.0\_-1.5  & 180.733 & -1.379  & 0.42 & G    & 185 & KIDS\_348.0\_-34.1 & 348.079 & -34.183 & 0.52 & G    \\
89 & KIDS\_181.0\_1.5   & 180.706 & 1.454   & 0.33 & G    & 186 & KIDS\_348.0\_-34.1 & 347.94  & -33.73  & 0.29 & G    \\
90 & KIDS\_182.0\_-0.5  & 181.537 & -0.309  & 0.42 & G    & 187 & KIDS\_348.2\_-32.1 & 348.458 & -32.005 & 0.45 & G    \\
91 & KIDS\_182.0\_0.5   & 181.647 & 0.222   & 0.57 & G    & 188 & KIDS\_349.2\_-34.1 & 349.653 & -34.221 & 0.5  & G    \\
92 & KIDS\_182.0\_0.5   & 181.592 & 0.174   & 0.53 & G    & 189 & KIDS\_349.4\_-32.1 & 349.768 & -32.189 & 0.49 & G    \\
93 & KIDS\_182.0\_0.5   & 181.756 & 0.354   & 0.47 & G    & 190 & KIDS\_350.6\_-32.1 & 350.693 & -32.064 & 0.26 & G    \\
94 & KIDS\_182.0\_0.5   & 182.001 & 0.936   & 0.52 & G    & 191 & KIDS\_350.6\_-32.1 & 351.072 & -32.171 & 0.4  & G    \\
95 & KIDS\_182.5\_-2.5  & 182.545 & -2.261  & 0.22 & G    & 192 & KIDS\_350.7\_-31.2 & 350.158 & -31.042 & 0.21 & G    \\
96 & KIDS\_182.5\_-2.5  & 182.574 & -2.504  & 0.47 & G    \\
\hline
\end{tabular}
\end{adjustbox}
\end{table*}

\begin{figure*}[!htbp]
\includegraphics[width=\textwidth]{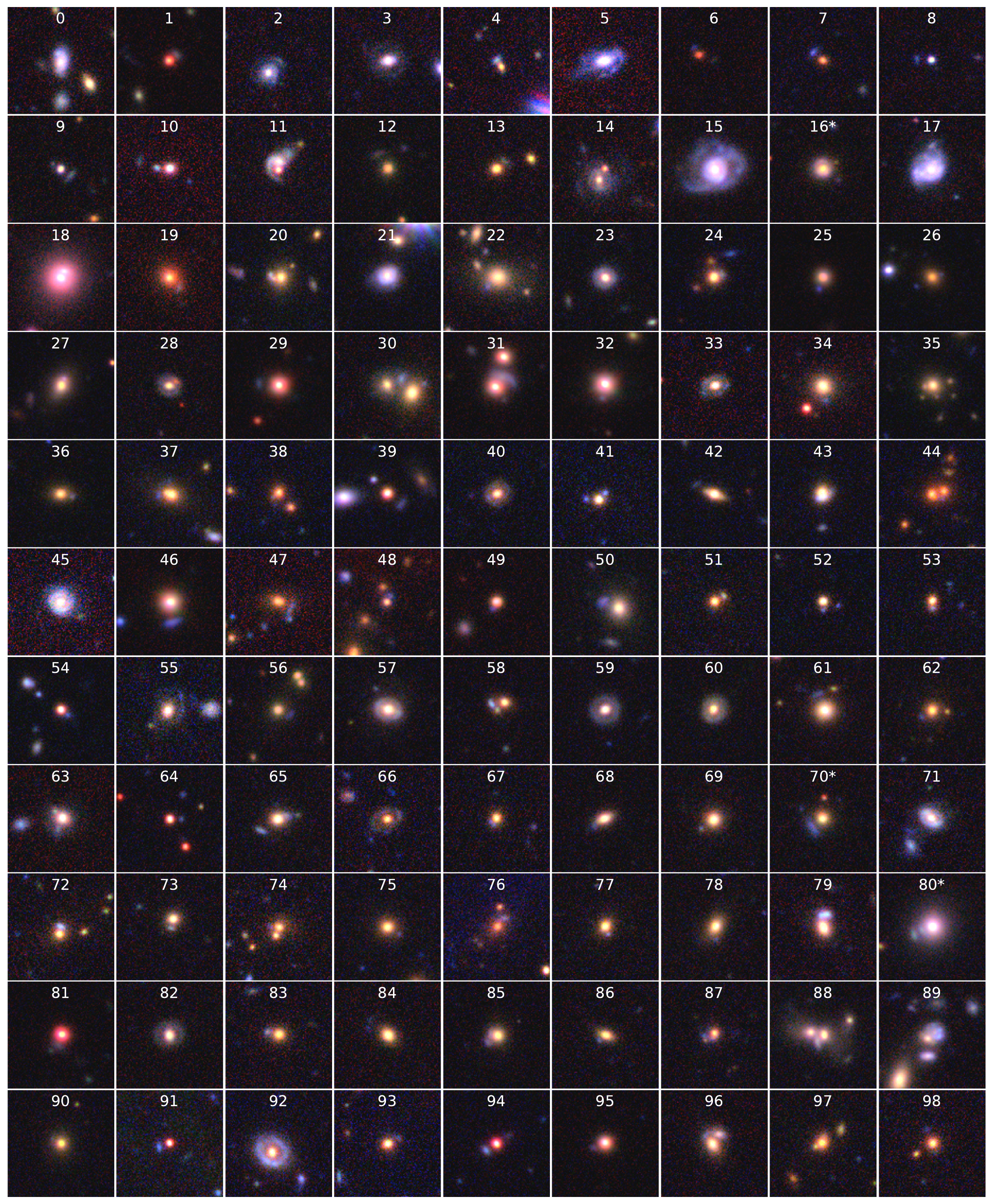}
\centering
\caption{TEGLIE grade 2 objects. These candidates exhibit some lensing features such as a distorted arc or multiple images. However, these features are not definitive enough to confirm the lensing hypothesis. The images marked with an asterisk are high-quality lens candidates already identified by the KiDS collaboration.}
\label{fig:grade_2_0}
\end{figure*}

\addtocounter{figure}{-1}
\begin{figure*}[!htbp]
\includegraphics[width=\textwidth]{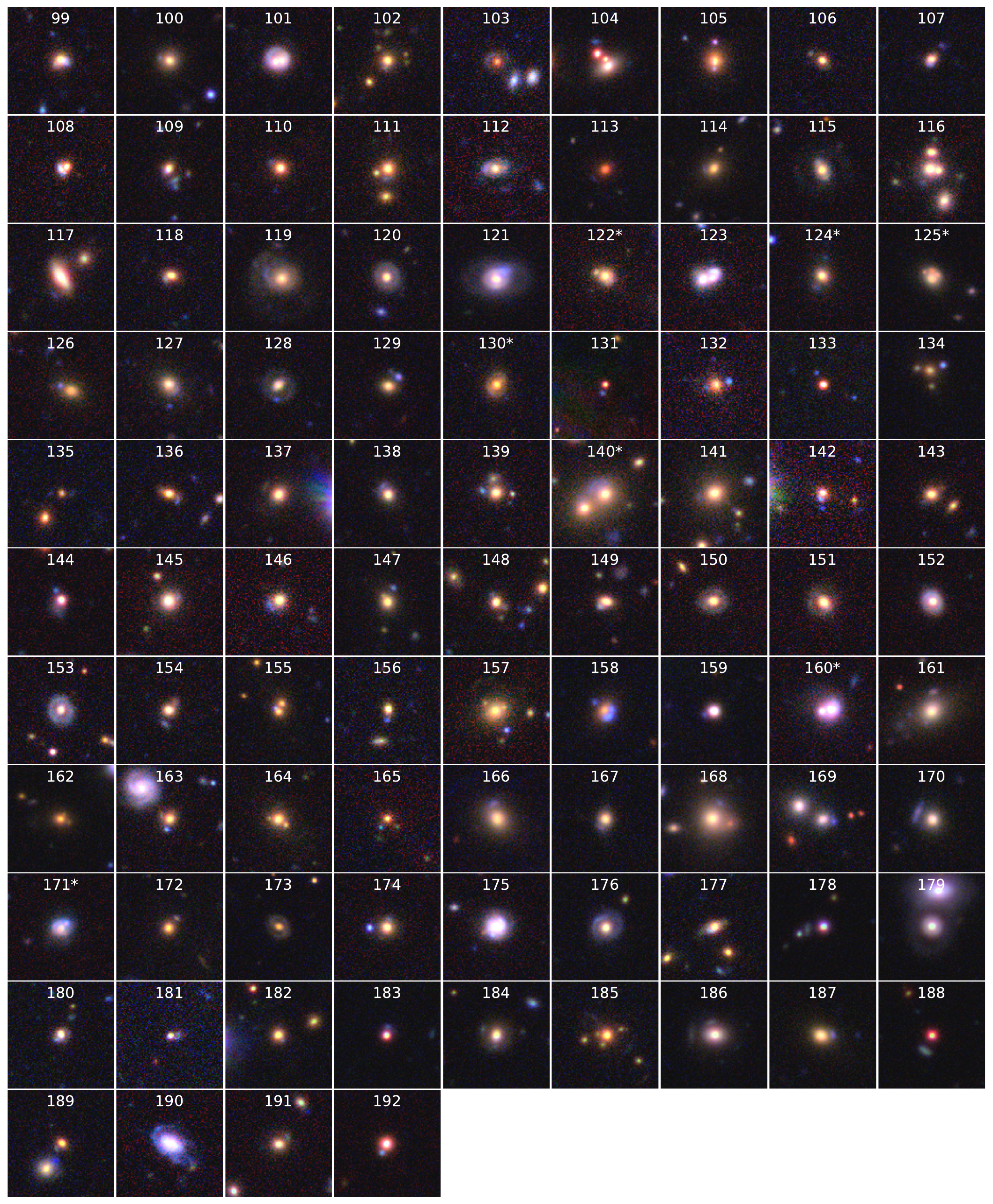}
\centering
\caption{(Continued)}
\label{fig:grade_2_1}
\end{figure*}

\clearpage
\section{Grade 3 candidates}

\begin{table*}[!htbp]
\centering
 \footnotesize
\caption{Candidates with grade 3, found during the search within the GAMA footprint.
`ID' is the index of the object in Fig. \ref{fig:grade_3_0}. 
`KiDS ID' is the object's ID in the KiDS catalog where we omitted 'KiDSDR4 ' present in every ID before the ICRS coordinates. `R.A.' and `Dec.' are the object's coordinates in degrees and  z$_{phot}$ is the photometric redshift as reported in the KiDS catalog. All the candidates have been found during the search in the GAMA footprint.}
\label{tab:grade_3}
\begin{adjustbox}{width=0.92\textwidth}
\begin{tabular}{lllll|lllll}
\hline\hline\noalign{\smallskip}
ID & KiDS ID & R.A. J2000 & Dec. J2000 & $z_{phot}$ & ID & KiDS ID  & R.A. J2000 & Dec. J2000 & $z_{phot}$ \\
\hline\noalign{\smallskip}
0  & J083729.272+020108.08 & 129.372 & 2.019  & 0.35 & 61  & J120945.192+003615.67 & 182.438 & 0.604   & 0.62 \\
1  & J083541.903+021003.79 & 128.925 & 2.168  & 0.04 & 62  & J120848.824-020651.40 & 182.203 & -2.114  & 0.18 \\
2  & J083948.242-000212.84 & 129.951 & -0.037 & 0.14 & 63  & J121307.317-001623.23 & 183.28  & -0.273  & 0.41 \\
3  & J084048.339-013506.83 & 130.201 & -1.585 & 0.12 & 64  & J121158.790-002308.55 & 182.995 & -0.386  & 0.47 \\
4  & J084003.763+024947.07 & 130.016 & 2.83   & 0.47 & 65  & J121356.270-015439.59 & 183.484 & -1.911  & 0.4  \\
5  & J084359.462+005601.49 & 130.998 & 0.934  & 0.25 & 66  & J121035.452-015614.83 & 182.648 & -1.937  & 0.33 \\
6  & J084320.446+001706.78 & 130.835 & 0.285  & 0.28 & 67  & J121228.277+000928.45 & 183.118 & 0.158   & 0.15 \\
7  & J084434.904+011107.28 & 131.145 & 1.185  & 0.6  & 68  & J121446.301+015823.64 & 183.693 & 1.973   & 0.28 \\
8  & J084445.884+022558.73 & 131.191 & 2.433  & 0.58 & 69  & J121434.349+013911.22 & 183.643 & 1.653   & 0.37 \\
9  & J084632.804-015420.40 & 131.637 & -1.906 & 0.06 & 70  & J121935.683-020829.23 & 184.899 & -2.141  & 0.32 \\
10 & J084920.737-012656.40 & 132.336 & -1.449 & 0.28 & 71  & J122139.692-002604.01 & 185.415 & -0.434  & 0.41 \\
11 & J084729.163+005604.58 & 131.872 & 0.935  & 0.27 & 72  & J122151.291-000032.18 & 185.464 & -0.009  & 0.43 \\
12 & J085353.059-000028.68 & 133.471 & -0.008 & 0.33 & 73  & J121851.994-015434.23 & 184.717 & -1.91   & 0.31 \\
13 & J085309.469-013527.60 & 133.289 & -1.591 & 0.31 & 74  & J121920.933+003435.85 & 184.837 & 0.577   & 0.61 \\
14 & J085110.857+004811.05 & 132.795 & 0.803  & 0.15 & 75  & J122009.474+002125.63 & 185.039 & 0.357   & 0.29 \\
15 & J085344.541+011923.09 & 133.436 & 1.323  & 0.42 & 76  & J122130.469+013811.98 & 185.377 & 1.637   & 0.5  \\
16 & J085101.021+014207.89 & 132.754 & 1.702  & 0.44 & 77  & J140756.219-002645.37 & 211.984 & -0.446  & 0.52 \\
17 & J085506.549+021401.93 & 133.777 & 2.234  & 0.5  & 78  & J141445.391-014051.72 & 213.689 & -1.681  & 0.17 \\
18 & J085727.872-004150.70 & 134.366 & -0.697 & 0.22 & 79  & J141824.188+020911.32 & 214.601 & 2.153   & 0.66 \\
19 & J085842.737+000354.65 & 134.678 & 0.065  & 0.33 & 80  & J142159.038-003525.18 & 215.496 & -0.59   & 0.16 \\
20 & J090240.270-000814.29 & 135.668 & -0.137 & 0.33 & 81  & J142320.449-003213.92 & 215.835 & -0.537  & 0.11 \\
21 & J090251.816-000815.43 & 135.716 & -0.138 & 0.22 & 82  & J142251.341-010308.43 & 215.714 & -1.052  & 0.48 \\
22 & J090249.449+004901.71 & 135.706 & 0.817  & 0.75 & 83  & J142217.324+005113.48 & 215.572 & 0.854   & 0.31 \\
23 & J090639.021+020905.25 & 136.663 & 2.151  & 0.16 & 84  & J142933.367-002655.80 & 217.389 & -0.449  & 0.21 \\
24 & J090547.790+021913.38 & 136.449 & 2.32   & 0.21 & 85  & J142727.154-005844.29 & 216.863 & -0.979  & 0.12 \\
25 & J090822.399-012043.89 & 137.093 & -1.346 & 0.7  & 86  & J142724.520-013720.01 & 216.852 & -1.622  & 0.21 \\
26 & J090705.974+012107.84 & 136.775 & 1.352  & 0.33 & 87  & J142913.561+013519.06 & 217.307 & 1.589   & 0.38 \\
27 & J090750.075+010141.07 & 136.959 & 1.028  & 0.2  & 88  & J143310.615-000807.05 & 218.294 & -0.135  & 0.21 \\
28 & J091321.371+025802.38 & 138.339 & 2.967  & 0.55 & 89  & J143134.344-003421.89 & 217.893 & -0.573  & 0.61 \\
29 & J091246.457+024038.15 & 138.194 & 2.677  & 0.19 & 90  & J143543.436+000223.51 & 218.931 & 0.04    & 0.18 \\
30 & J091440.317-010640.69 & 138.668 & -1.111 & 0.6  & 91  & J144031.305-011748.97 & 220.13  & -1.297  & 0.23 \\
31 & J091946.084-010642.10 & 139.942 & -1.112 & 0.27 & 92  & J144139.063-015500.93 & 220.413 & -1.917  & 0.37 \\
32 & J091906.528-012952.18 & 139.777 & -1.498 & 0.63 & 93  & J143927.970-012348.56 & 219.867 & -1.397  & 0.26 \\
33 & J092112.702-014449.18 & 140.303 & -1.747 & 0.21 & 94  & J144026.814+020430.29 & 220.112 & 2.075   & 0.2  \\
34 & J092125.355-015809.91 & 140.356 & -1.969 & 0.32 & 95  & J144426.730+023143.37 & 221.111 & 2.529   & 0.3  \\
35 & J091903.838+003900.49 & 139.766 & 0.65   & 0.57 & 96  & J144303.020-015633.92 & 220.763 & -1.943  & 0.7  \\
36 & J091900.407+012229.83 & 139.752 & 1.375  & 0.45 & 97  & J144357.047-011030.17 & 220.988 & -1.175  & 0.36 \\
37 & J092207.693+021235.87 & 140.532 & 2.21   & 0.32 & 98  & J144451.291-011350.36 & 221.214 & -1.231  & 0.16 \\
38 & J114141.904-012007.68 & 175.425 & -1.335 & 0.19 & 99  & J144515.098+003906.93 & 221.313 & 0.652   & 0.69 \\
39 & J114040.003-012230.65 & 175.167 & -1.375 & 0.4  & 100 & J144711.141-015848.48 & 221.796 & -1.98   & 0.17 \\
40 & J114007.294-025136.20 & 175.03  & -2.86  & 0.12 & 101 & J144937.952-015619.16 & 222.408 & -1.939  & 0.32 \\
41 & J114518.853-010942.67 & 176.329 & -1.162 & 0.26 & 102 & J144612.399-014832.51 & 221.552 & -1.809  & 0.12 \\
42 & J114304.221-010806.88 & 175.768 & -1.135 & 0.11 & 103 & J224051.563-303256.88 & 340.215 & -30.549 & 0.17 \\
43 & J114502.682-023801.98 & 176.261 & -2.634 & 0.09 & 104 & J224029.645-303533.23 & 340.124 & -30.593 & 0.67 \\
44 & J114654.701-021710.40 & 176.728 & -2.286 & 0.58 & 105 & J224059.474-343241.63 & 340.248 & -34.545 & 0.43 \\
45 & J114711.547-023458.84 & 176.798 & -2.583 & 0.27 & 106 & J224805.958-341349.56 & 342.025 & -34.23  & 0.45 \\
46 & J114713.170-013912.79 & 176.805 & -1.654 & 0.67 & 107 & J224809.748-304517.86 & 342.041 & -30.755 & 0.41 \\
47 & J114900.031+001859.47 & 177.25  & 0.317  & 0.61 & 108 & J224855.698-301405.29 & 342.232 & -30.235 & 0.22 \\
48 & J114843.457-023820.16 & 177.181 & -2.639 & 0.66 & 109 & J225332.623-334507.31 & 343.386 & -33.752 & 0.43 \\
49 & J115238.093-010015.67 & 178.159 & -1.004 & 0.34 & 110 & J230433.400-334831.67 & 346.139 & -33.809 & 0.29 \\
50 & J115232.387-011822.22 & 178.135 & -1.306 & 0.33 & 111 & J230232.312-331138.64 & 345.635 & -33.194 & 0.33 \\
51 & J115137.253+000000.24 & 177.905 & 0.0    & 0.24 & 112 & J230229.579-322633.51 & 345.623 & -32.443 & 0.67 \\
52 & J115530.540-002402.93 & 178.877 & -0.401 & 0.48 & 113 & J230258.316-301254.92 & 345.743 & -30.215 & 0.32 \\
53 & J115612.668-000906.27 & 179.053 & -0.152 & 0.48 & 114 & J230647.024-343130.10 & 346.696 & -34.525 & 0.25 \\
54 & J115752.611-000156.17 & 179.469 & -0.032 & 0.29 & 115 & J230804.894-301434.11 & 347.02  & -30.243 & 0.39 \\
55 & J115953.579-021438.14 & 179.973 & -2.244 & 0.68 & 116 & J231028.404-334913.99 & 347.618 & -33.821 & 0.25 \\
56 & J115742.900-024725.74 & 179.429 & -2.79  & 0.24 & 117 & J231208.906-341513.12 & 348.037 & -34.254 & 0.52 \\
57 & J115920.987-025436.03 & 179.837 & -2.91  & 0.32 & 118 & J231024.308-342650.18 & 347.601 & -34.447 & 0.13 \\
58 & J115841.260-001657.42 & 179.672 & -0.283 & 0.19 & 119 & J232404.432-321557.06 & 351.018 & -32.266 & 0.35 \\
59 & J120204.037+011733.51 & 180.517 & 1.293  & 0.76 & 120 & J232307.980-301420.63 & 350.783 & -30.239 & 0.21 \\
60 & J120654.110-021819.65 & 181.725 & -2.305 & 0.3  & \\              \hline
\end{tabular}
\end{adjustbox}
\end{table*}

\clearpage
\section{Blended spectra search}

\begin{figure*}[!htbp]
\includegraphics[width=\textwidth]{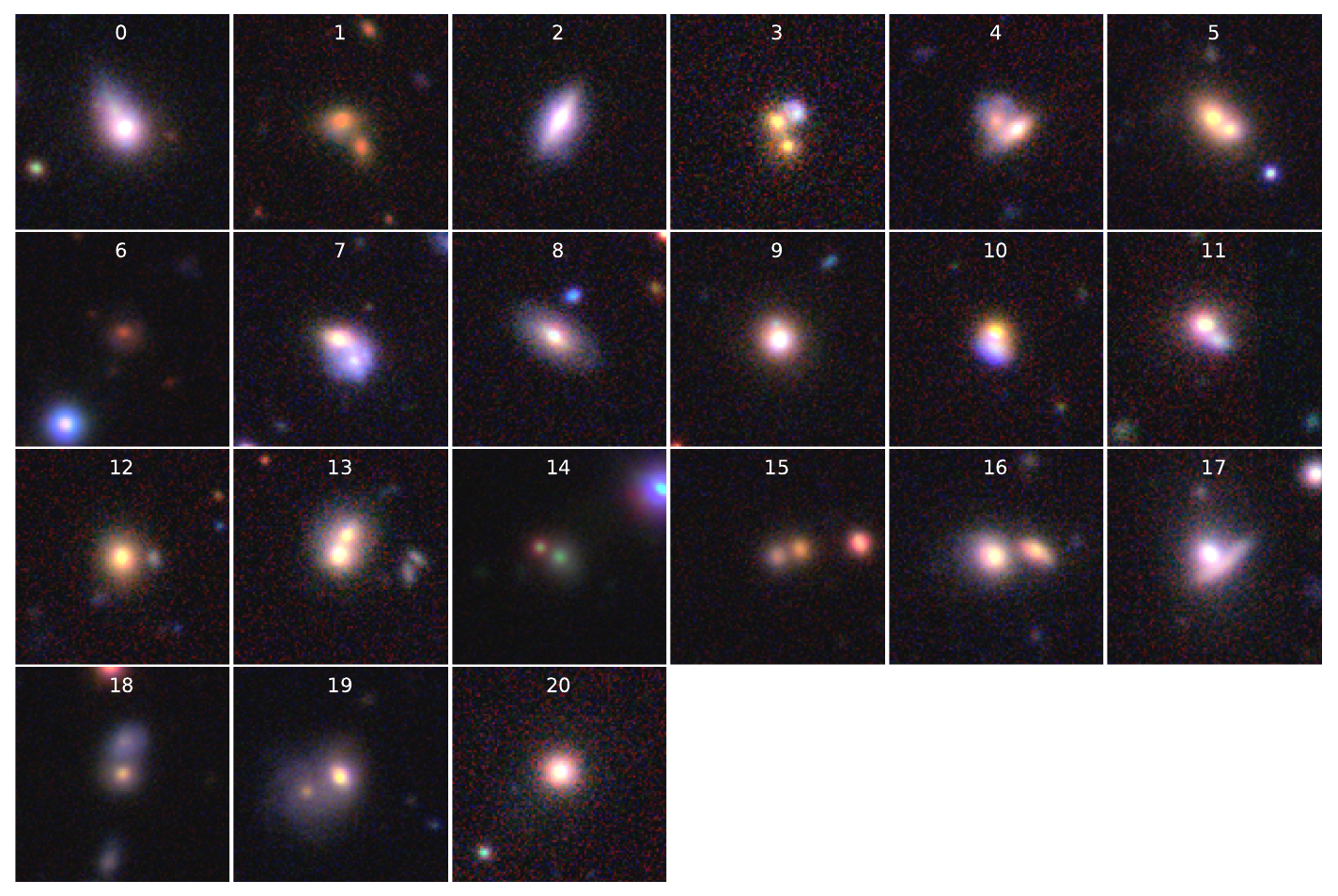}
\centering
\caption{Strong lens candidates found through the blended spectra search by \citet{Holwerda_2015} and labeled as not lenses during the visual inspection.}
\label{fig:pg_elg_0}
\end{figure*}

\end{appendix}
\end{document}